\documentclass[twocolumn,epjc3]{svjour3}
\smartqed  
\RequirePackage{lmodern}
\RequirePackage{lscape}
\RequirePackage{graphicx}
\RequirePackage{mathrsfs}
\RequirePackage{doi}
\RequirePackage[numbers,sort&compress]{natbib}

\journalname{Eur. Phys. J. C}

\begin{document}

\title{TREX-DM: a low-background Micromegas-based TPC for low-mass WIMP detection}
\titlerunning{TREX-DM}        

\author{F.J.~Iguaz\thanksref{e1} \and J.G.~Garza\thanksref{e2} \and F.~Aznar\thanksref{e3} \and
  J.F.~Castel \and S.~Cebri\'an \and T.~Dafni \and
  J.A.~Garc\'ia \and I.G.~Irastorza \and A.~Lagraba \and
  G.~Luz\'on \and A.~Peir\'o
}

\thankstext{e1}{e-mail: iguaz@unizar.es}
\thankstext{e2}{e-mail: jgraciag@unizar.es}
\thankstext{e3}{\emph{Present Address:} Centro Universitario de la Defensa, Universidad de Zaragoza,
Crta. de Huesca s/n, 50090, Zaragoza, Spain}

\institute{Grupo de F\'isica Nuclear y Astropart\'iculas,
Universidad de Zaragoza, C/ Pedro Cerbuna 12, 50009, Zaragoza, Spain.}

\date{Received: date / Accepted: date}

\maketitle

\begin{abstract}
If Dark Matter is made of Weakly Interacting Massive Particles (WIMPs) with masses below $\sim$20 GeV,
the corresponding nuclear recoils in mainstream WIMP experiments are of energies too close,
or below, the experimental threshold. 
Gas Time Projection Chambers (TPCs) can be operated with a variety of target elements,
offer good tracking capabilities and, on account of the amplification in gas,
very low thresholds are achievable.
Recent advances in electronics and in novel radiopure TPC readouts,
especially micro-mesh gas structure (Micromegas),
are improving the scalability and low-background prospects of gaseous TPCs.
Here we present TREX-DM, a prototype to test the concept of a Micromegas-based TPC to search for low-mass WIMPs.
The detector is designed to host an active mass of $\sim$0.300 kg of Ar at 10 bar,
or alternatively $\sim$0.160 kg of Ne at 10 bar, with an energy threshold below 0.4~keVee,
and is fully built with radiopure materials.
We will describe the detector in detail, the results from the commissioning phase on surface,
as well as a preliminary background model.
The anticipated sensitivity of this technique may go beyond current experimental limits for WIMPs of masses of 2--8 GeV.
\end{abstract}

\section{Introduction}
\label{sec:intro}
There is compelling evidence now, from cosmology and astrophysics,
that most of the matter of the Universe is in the form of non-baryonic cold Dark Matter (DM)~\cite{Bertone:2004pz}.
The particle physics nature of this matter, however, remains a mystery.
The Weakly Interacting Massive Particle (WIMP) is a good generic candidate to compose the DM.
In addition, WIMPs appear naturally in well-motivated extensions of the Standard Model,
in particular those including SuperSymmetry (SUSY)~\cite{Jungman:1995df}.

If our galactic DM halo is made of WIMPs, they could interact with nuclei
and produce detectable nuclear recoils in the target material of underground terrestrial experiments.
Due to the extreme low rate and low energy of such events, the experimental challenge in terms of background rate,
threshold and target mass is formidable.
During the last 30 years an ever growing experimental activity has been devoted to the development of detection techniques
that have achieved increasingly larger target masses and lower levels of background,
in the quest of reaching higher sensitivity to DM WIMPs.
At the moment, the leading experiments in the ``WIMP race'' are those using relatively heavy target nuclei (e.g. Xe or Ge)
--to exploit the $A^2$ dependence of the coherent WIMP-nucleus interaction--
and using detection techniques that provide nuclear recoil discrimination.
This is the case, e.g. of liquid Xe double-phase detectors (e.g. LUX~\cite{Akerib:2013tjd} or XENON~\cite{Aprile:2012nq})
or hybrid Ge bolometers (like CDMS/SuperCDMS~\cite{Ahmed:2009zw,Agnese:2015ywx,Agnese:2014aze}).
These experiments are currently operating at target masses of order 100 kg,
with background levels of a few counts per year.
As illustrative examples, SuperCDMS~\cite{Agnese:2014aze} has operated $\sim$9~kg of Ge target mass
observing 11 nuclear-recoil candidate events in 577~kg-days,
with plans for the 100~kg scale are ongoing;
LUX~\cite{Akerib:2013tjd} has operated 118 kg of liquid Xe fiducial mass, observing a background level that effectively limits a possible WIMP nuclear recoil signal to 2-5 events (depending on the mass) in a run of 85.3 live days. This corresponding to the current most stringent upper limit on the WIMP-nucleon cross-section of $7.6 \times 10^{46}$ cm$^2$ at a WIMP mass of 33~GeV.
Such impressive numbers are obtained as a result of the availability of discrimination techniques
that allow distinguishing --with some efficiency-- electron recoils (produced e.g. by gammas)
from the signal-like nuclear recoils. This happens because the different ionization density of nuclear
and electron types of events leads to a different yield-ratio in the detection medium
(ionization/scintillation in the case of noble liquids, and ionization/phonon in case of hybrid Ge bolometers).
However, this discrimination capability is energy-dependent and for very low energies (typically few~keV)
it disappears, setting the effective threshold of the experiment.

WIMP searches are conventionally and somewhat simplistically expressed
in the two-dimensional effective parameter space ($\sigma_N$,$M_W$),
where $\sigma_N$ is the WIMP-nucleon interaction cross section
and $M_W$ is the WIMP mass. This representation usually comes with a number of additional oversimplifying assumptions, e.g.,
that the velocity distribution of WIMPs in the galactic halo follows a Maxwellian distribution,
or that WIMPs interact exclusively (or mainly) with nuclei via elastic coherent spin-independent scattering.
Although this conventional scenario is appealing to set a common ground for inter-comparison of experimental sensitivities,
one has to keep in mind the implied assumptions.

The large majority of the experimental effort so far has focused on the search for WIMPs
of relatively large masses (of around 50 GeV and larger).
This is mainly because of theoretical considerations set in the early days of WIMP searches,
that identified the WIMP with the neutralino of (minimal) SUSY extensions of the Standard Model,
and interpreted the early accelerator limits in light of these models.
The establishment of this ``WIMP orthodoxy'' (as it is called in~\cite{DRUKIER:2013lva})
was facilitated by the fact that the best WIMP detection techniques available
were already well suited for this mass range.
Indeed, mainstream experiments show the best sensitivity for $M_W \sim 50$~GeV,
partially due to the kinematical matching between the WIMP and the nuclear mass.
For higher masses the sensitivity to $\sigma_N$ slowly decreases,
while for lower masses it gets sharply reduced mostly because of the effect of the energy threshold.

Despite the improvement in sensitivity to $\sigma_N$ of more than 4 orders of magnitude over the past 15 years,
no conclusive WIMP signal has been observed.
This fact has triggered the revision of the mentioned assumptions and
the study of more generic phenomenological WIMP frameworks, e.g. different WIMP interactions~\cite{Hill:2013hoa} or different WIMP velocity distributions (see~\cite{Green:2010gw} and references therein). In addition, the non-observation of signals of SUSY thus far~\cite{Chatrchyan:2012jx,Aad:2014wea} in the Large Hadron Collider (LHC)
calls also for adopting more open-minded views of the theoretical frameworks of dark matter candidates.
With these facts in mind, recent theoretical and phenomenological efforts
have focused on the study of less conventional SUSY models,
or even non-SUSY WIMP models (like e.g. Asymmetric DM models~\cite{Zurek:2013wia}).

\subsection{Low-mass WIMPs}

As part of this view of going beyond the WIMP orthodoxy,
some recent experimental and phenomenological efforts have been focused
on the study of WIMPs in the low-mass range (i.e. $M_W < 10-20$~GeV).
The interest on this region of the parameter space, traditionally out of reach of mainstream experiments,
was increased by the appearance of a number of hints that could be interpreted
as collisions of low-mass WIMPs~\cite{Bernabei:2013rb,Aalseth:2012if,Angloher:2011uu}
(although those interpretations have weakened over time~\cite{Agnese:2014aze,Angloher:2014myn}).
In addition, the well-known and persistent DAMA/LIBRA claim~\cite{Bernabei:2013rb},
incompatible with results from other experiments in conventional scenarios,
might be reconciled only within very non-standard model assumptions,
some of them invoking low-mass WIMPs~\cite{Savage:2008er}.
In any case, it is as important to extend WIMP search sensitivities to lower WIMP masses as it is to lower cross section values.

Sensitivity to low-mass WIMPs poses particular experimental challenges.
As mentioned above, mainstream experiments are severely limited at low-masses
due to the threshold requirements for nuclear recoil discrimination.
Sensitivity projections for low WIMP masses should be treated with great caution because such low-mass WIMP interactions produce recoil energy deposits that are mostly below the energy threshold of experiments based on heavy target nuclei like Xe or Ge. This means that the exclusion limits derived for low-mass WIMPs by these experiments
rely on detecting the interactions of a very small (1\% or lower) fraction of the incident WIMP velocity distribution,
corresponding to the WIMPs with kinetic energies high enough to produce a nuclear recoil above the detector energy threshold.
But precisely this part of the distribution is the most uncertain~\cite{Vogelsberger:2008qb},
and in some plausible galactic halo models (i.e. those with lower maximum WIMP velocity) it can altogether disappear~\cite{Green:2010gw}.

It is clear that to tackle the low-mass WIMP region,
specific experiments optimized for this mass range are needed.
A robust detection or exclusion requires that a substantial fraction (of order 50\%) of the WIMP spectrum
is above the experimental threshold.
To achieve this the use of light target nuclei is preferred (to kinematically reach higher recoil energies),
as well as techniques with intrinsically low energy detection threshold.
These requirements are incompatible with the discrimination between nuclear and electron recoils, as the yield-ratio method employed lose power at low energies.
Some conventional experiments, like e.g. CDMS~\cite{Agnese:2015ywx} and XENON~\cite{Angle:2011th},
have developed analyses specifically for low energy data,
bypassing their nuclear/electron discrimination and going to lower thresholds.
More relevantly, the first experiments specifically focused on the new low-mass WIMP paradigm
are appearing, for example, DAMIC~\cite{Chavarria:2014ika}, CDEX~\cite{Yue:2014qdu}, or CDMSlite~\cite{Agnese:2015nto}.
As the background levels in these experiments must rely on more conventional strategies like e.g.
ultra-high levels of radiopurity of the detector components,
the scale of these experiments remain so far at a relatively modest scale (still below the kg level of target mass).

\subsection{High pressure TPCs to search for low-mass WIMPs}
In this paper we propose the use of gas Time Projection Chambers (TPCs)
with novel Micromegas readouts to search for low-mass WIMPs.
Being gaseous detectors, the scaling-up prospects of gas TPCs are typically considered modest.
However, advances in electronics and novel micro-pattern gas readout planes (especially Micromegas)
are changing this view (see~\cite{Irastorza:2015dcb,Irastorza:2015geo} and references therein). The objective of the T-REX project~\cite{Irastorza:2011hh,Dafni:2012fi} has been to study
the applicability of Micromegas readouts TPCs to rare event searches
(not just to WIMP searches, but also axions~\cite{Aune:2013pna} and double beta decay~\cite{Cebrian:2010nw}).
The T-REX activity\footnote{T-REX webpage: \url{http://gifna.unizar.es/trex/}} during the last years has included
the study and characterization of novel Micromegas readouts~\cite{Giomataris:1995fq},
especially those of microbulk type~\cite{Andriamonje:2010zz},
study and improvement of their radiopurity~\cite{Cebrian:2010ta},
simulation and development of discrimination algorithms~\cite{Cebrian:2013mza}, and the construction
and test of prototypes~\cite{Aune:2013pna,Aznar:2015iia,Alvarez:2013oha,Alvarez:2013kqa,Gonzalez-Diaz:2015oba}.
It is our claim here that gaseous detectors are very promising options for low-mass WIMP detection for many reasons.  The charge amplification inherent to gaseous detectors yields an appropriately low energy threshold for low-mass WIMP searches.  The aforementioned advances in radiopurity and general simplification of these detectors increase the feasibility of scaling-up such detectors.  Further, there is flexibility in the choice of target gas and pressure

As part of the T-REX project, a prototype to assess the feasibility of a low-mass WIMP detector
with this technique has been developed: TREX-DM.
This paper constitutes the first detailed presentation of this activity, its current status and prospects.
In Sec.~\ref{sec:detector} a technical description of the TREX-DM prototype is made.
Section~\ref{sec:characterization} is devoted to the first experimental results of the commissioning on surface,
focused on performance results of the Micromegas readout planes.
In Sec.~\ref{sec:radiopurity} we review the radiopurity results of the detector components,
a very important aspect of the project.
Based on these, in Sec.~\ref{sec:backmodel} we introduce a preliminary background model for the detector,
with which we tentatively assess the physics prospects in Sec.~\ref{sec:sensitivity}.
The conclusions and the outlook (Sec.~\ref{sec:con}) complete this paper.

We must note here that another important reason why gas TPCs are being considered as WIMP detectors is because
they could provide access to the imaging of the nuclear recoils,
and therefore to the WIMP incoming direction~\cite{Ahlen:2009ev}.
WIMP directionality is considered the ultimate signature to unambiguously identify
the extraterrestrial origin of a putative signal.
The experimental challenge is big, due to the tiny size of nuclear recoils,
and it requires working at very low pressure and with very high granularity readouts.
Apart from the pioneer DRIFT experiment~\cite{Daw:2010ud}, a number of more recent initiatives
are ongoing to demonstrate directional sensitivity with a number of different TPC prototypes,
like MIMAC~\cite{Santos:2011kf}, NEWAGE~\cite{Nakamura:2015iza}, DMTPC~\cite{Ahlen:2010ub} and others.
Although we acknowledge the importance of this goal as a motivation to develop gas TPCs for WIMP searches,
TREX-DM is focused on the non-directional detection of WIMPs. This allows
operation at high pressure in order to increase target mass.

\section{Description of the experimental setup}
\label{sec:detector}
The TREX-DM detector is conceived to host 0.3 kg of Ar target mass at 10 bar (or, alternatively, 0.16 kg of Ne).
In some respects, the detector is a scaled-up version of the low-background Micromegas x-ray detectors
developed for axion research~\cite{Aune:2013pna}, but with a $10^3$ times larger active mass.
The detector is built taking into account state-of-the-art radiopurity specifications,
for which a dedicated campaign of material identification
and measurements has been carried out (see Sec.~\ref{sec:radiopurity}).
A few components of the detector described here will be replaced to improve radiopurity
for the physics run underground, which is discussed in Sec.~\ref{sec:con}.

\subsection{Vessel and shielding}
The vessel is composed of a forged and machined Electrolytic Tough Pitch Copper (ETP Cu) sleeve,
with a 0.5~m diameter and 0.5~m length
and two 6~cm thick Oxygen Free Electronic Copper (OFE Cu) machined flat end caps. Its thickness (6 cm) is enough
to both hold pressures up to 12 bar and be part of the passive shielding of external backgrounds.
The vessel is supported by an aluminum frame composed of three independent parts:
a central one to keep the central body and two others for the end-caps.
This configuration allows the separation of the two end-caps from the central body,
so as the readout planes (Fig.~\ref{fig:Setup}, (f)), which are bolted to the end-caps,
and the drift cage (Fig.~\ref{fig:Setup}, (a)) could be independently repaired or replaced.

\subsection{Drift cage and mechanical support}
\begin{figure*}[htb!]
\centering
\includegraphics[width=0.9\textwidth]{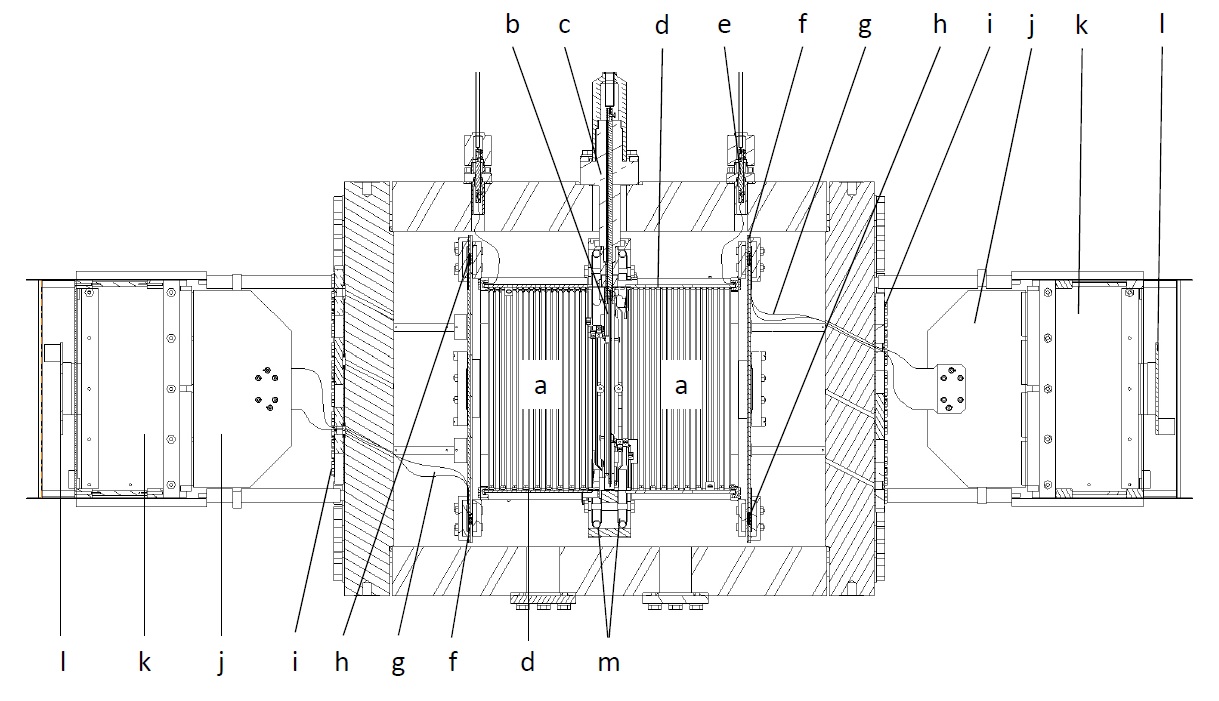}
\caption{Schema of the experimental setup. The different components are described in detail in the text:
active volumes (a), central cathode (b), high voltage feedthrough (c), field cage (d), \emph{last ring} feedthrough (e),
Micromegas readout planes (f), flat cable (g), Samtec connectors (h), signal feedthroughs (i), interface card (j),
AFTER-based FEC (k) and FEM boards (l), and calibration tube (m).}
\label{fig:Setup}
\end{figure*}

\begin{figure}[htb!]
\centering
\includegraphics[width=0.49\textwidth]{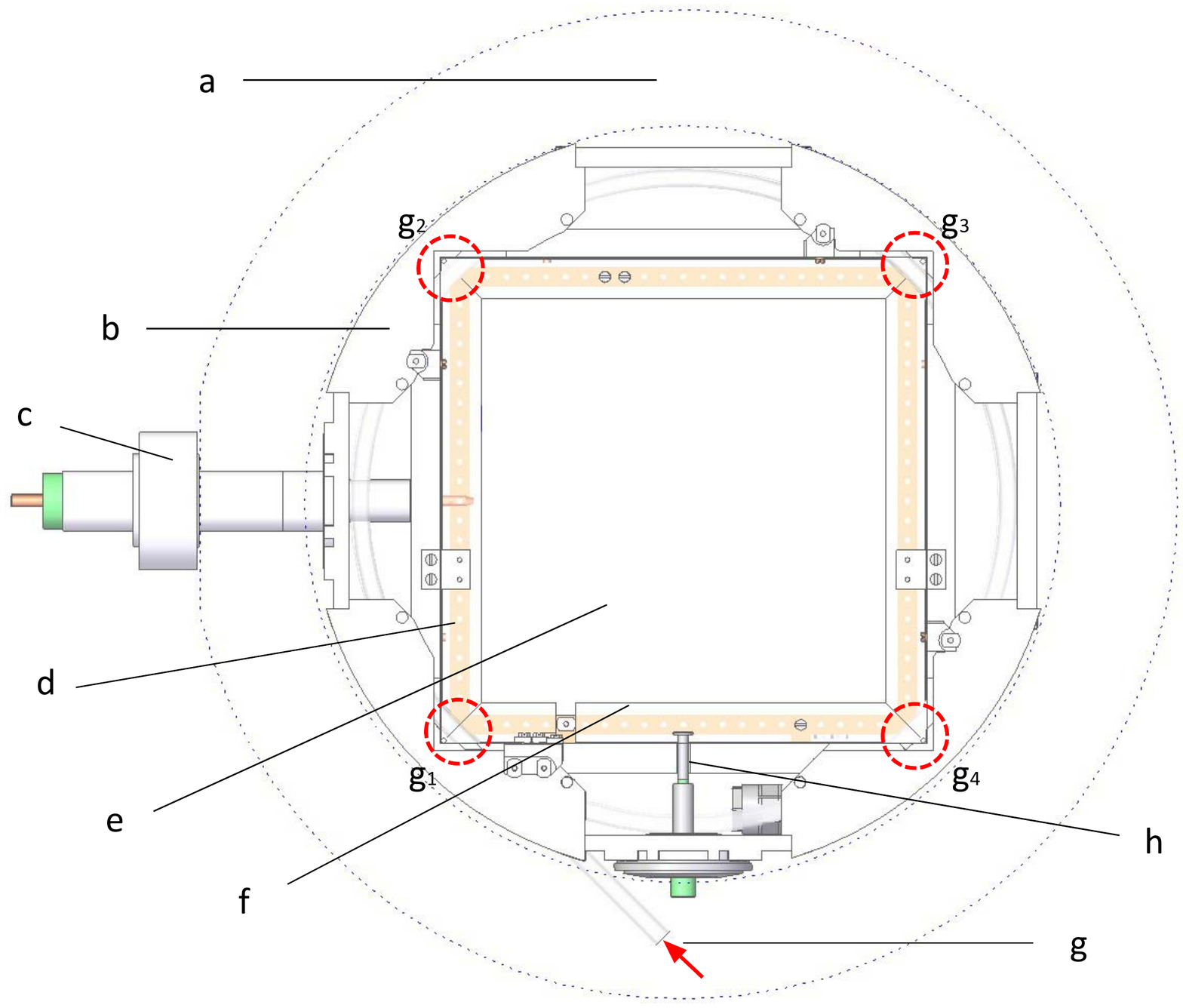}
\caption{Section of the experimental setup. The different components are described in detail in the text:
copper vessel contour (a), PTFE cartridge (b), radiopure HV feedthrough (c), cathode copper frame (d),
cathode mylar foil (e), cathode PTFE cassette (f), calibration tube with four source calibration positions (g),
and gas inlet (h).}
\label{fig:DesignCathode}
\end{figure}

The inner volume of the vessel is divided into two active volumes (Fig.~\ref{fig:Setup}, (a)),
separated by a central cathode (Fig.~\ref{fig:Setup}, (b)).
The cathode assembly consists of a squared copper frame (243~mm side length,
10~mm width and 1,5~mm thickness) (Fig.~\ref{fig:DesignCathode}, (d))
with an aluminized mylar foil glued on it and electrically connected (Fig.~\ref{fig:DesignCathode} (e)).
A PTFE cassette (Fig.~\ref{fig:DesignCathode}, (f)) covers the copper frame to reduce the copper fluorescence (at 8~keV)
induced by background events.
The cathode assembly is electrically isolated from the vessel
by a cylindrical Teflon cassette (190~mm radius; Fig.~\ref{fig:DesignCathode}, (b)),
which surrounds it and prevents any spark at voltages up to 40 kV.
The cathode is connected to a tailor-made high voltage feedthrough
(Fig.~\ref{fig:Setup}, (c); Fig.~\ref{fig:DesignCathode}, (c)),
composed by a ETP Cu round bar inserted in a machined teflon
rod that also works as gasket for sealing purposes.
Around each active volume, there is a 19 cm long and 25 cm side square sectioned field cage (Fig.~\ref{fig:Setup}, (d)),
composed of a copper-kapton printed circuit.
Each circuit is screwed to four Teflon walls with two purposes: the electrical isolation
and the suppression of the copper fluorescence emitted from the vessel walls.
The copper strips are 1~mm thick, are separated 7~mm and are
electrically linked one after the other by 10~M$\Omega$ resistors\footnote{SM5D resistor, produced by Finechem.}.
The inner drift chain ends at each side at a 1 mm thick copper squared ring,
also covered by a teflon gasket to prevent from sparks damaging the readout plane frame.
This last ring is connected via a cable\footnote{AWG 18/19/30, produced by Druflon.}
to a customized high voltage feedthrough,
made of a copper bar glued to a copper flange by epoxy Hysol\footnote{Hysol RE2039, produced by Henkel.} (Fig.~\ref{fig:Setup}, (e)).
Its voltage is adjusted by an external variable resistor (connected to ground) in order to have an
homogeneous drift field independently on the voltage applied
to the central cathode\footnote{Powered by a Spellman SL50N30/230/LR/SIC module.}
and the Micromegas mesh\footnote{Powered by a CAEN N470A module.}.
The electronics, described in Sec.~\ref{sec:DetElec}, sets the Micromegas strip pads to ground
while the central cathode and the Micromegas mesh are set to a negative voltage.
A diagram of the field cage is shown in Fig.~\ref{fig:SchemaElec} (left),
while the voltages used during the data-taking are detailed in Sec.~\ref{sec:sensitivity}.

\subsection{Micromegas readout planes}
\begin{figure}[htb!]
\centering
\includegraphics[width=0.49\textwidth]{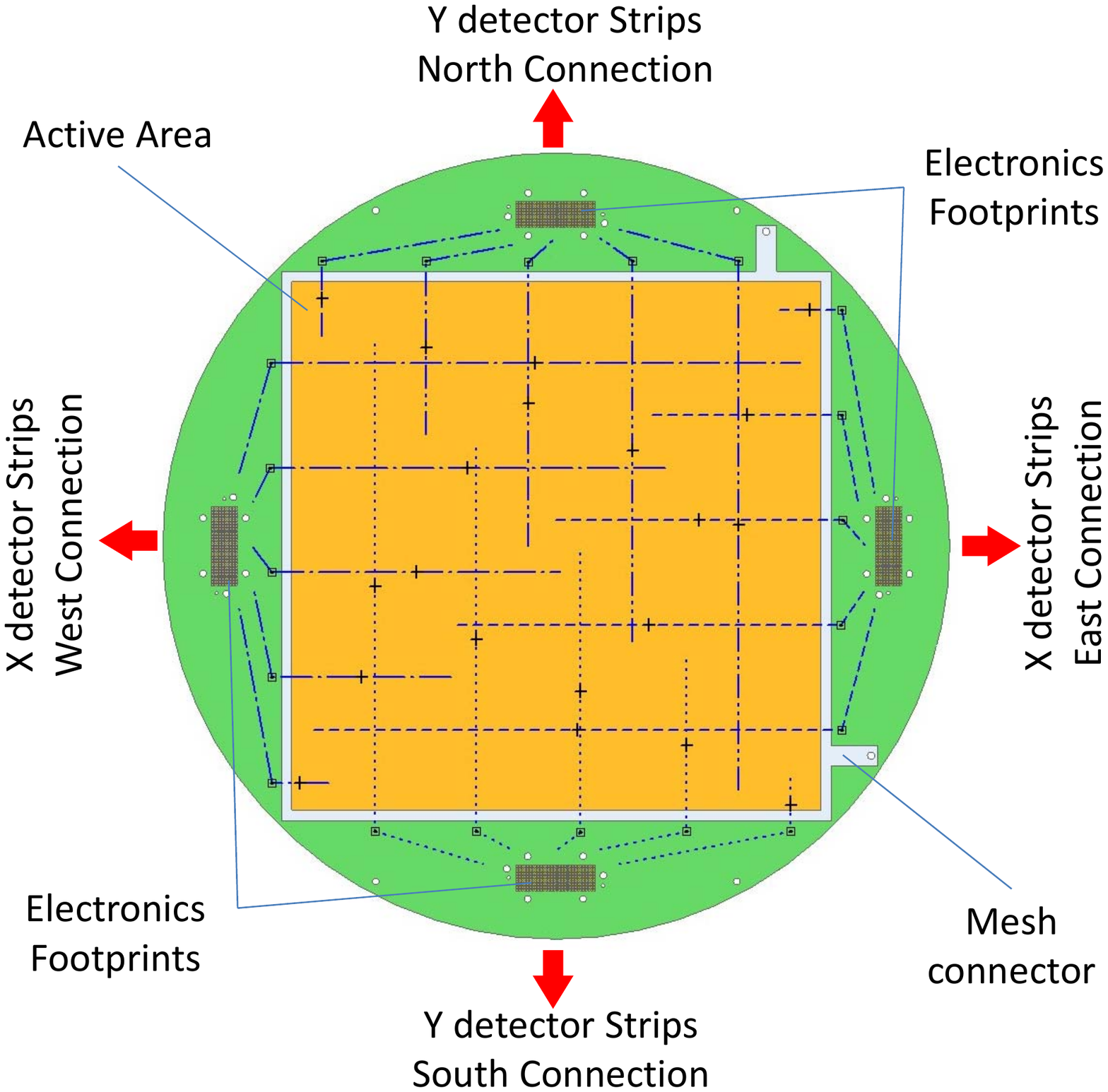}
\includegraphics[width=0.49\textwidth]{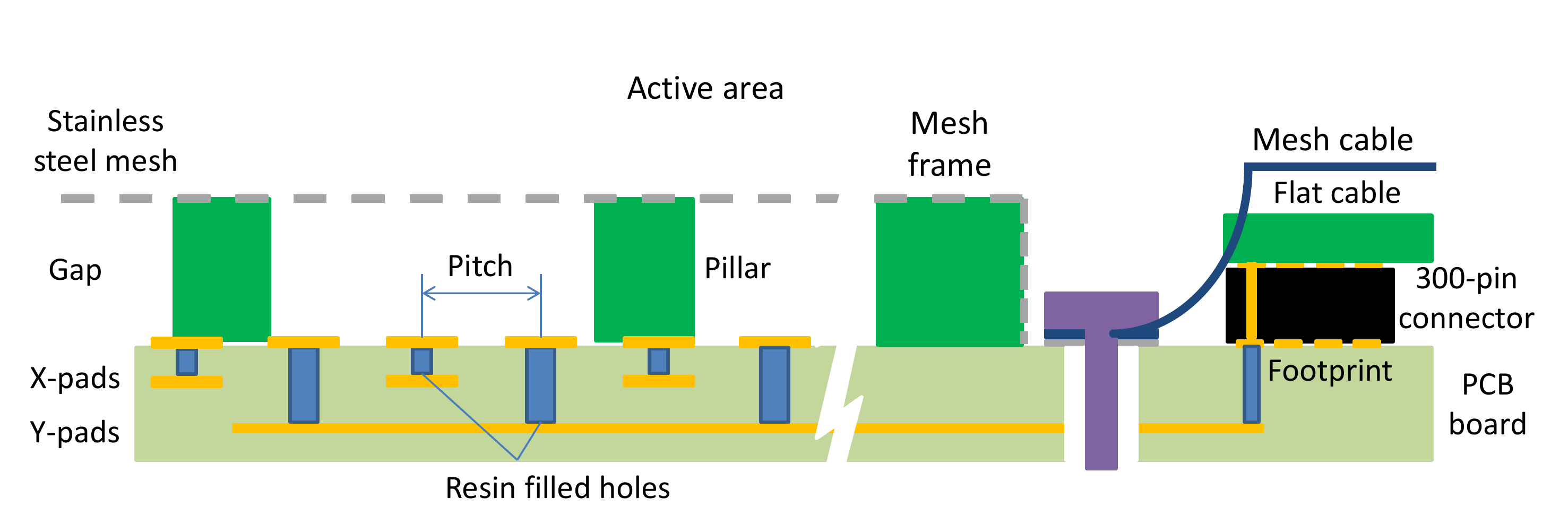}
\caption{Top view and section of a TREX-DM Micromegas readout,
described in detail in the text.
In these designs, the scale of some components has been exaggerated.}
\label{fig:DesignMMDetector}
\end{figure}

\begin{figure}[htb!]
\centering
\includegraphics[width=0.40\textwidth]{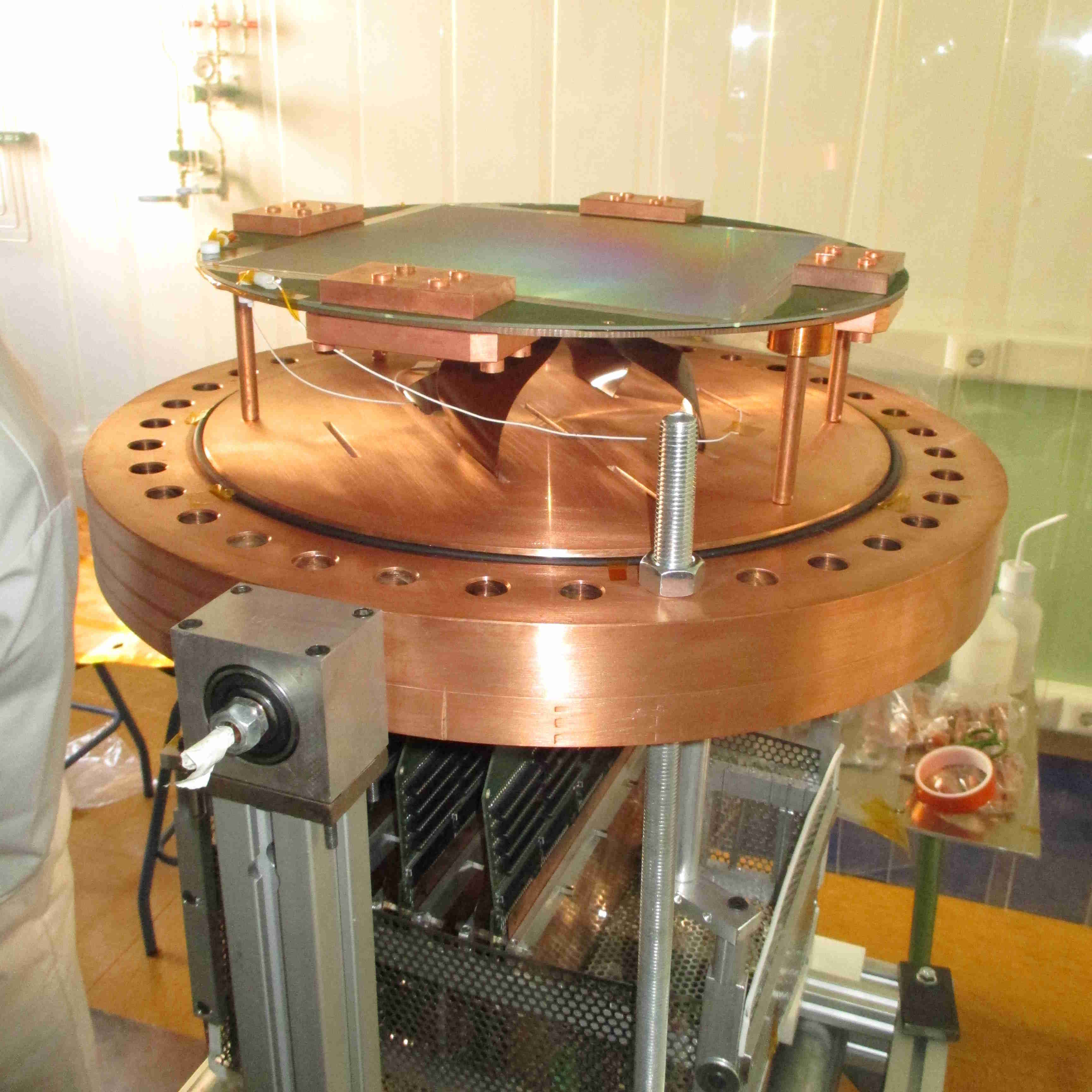}
\caption{View of one of the vessel's caps,
where several components described in detail in the text are shown:
a readout plane, its copper shielding pieces, its four flat cables
and part of the electronics: the interface cards and the FEC cards.}
\label{fig:MMDetector}
\end{figure}

The Micromegas anode planes (Fig.~\ref{fig:Setup}, (f); and Fig.~\ref{fig:MMDetector})
are a modified version of those used in CAST~\cite{Aune:2013pna}.
Each readout plane is on a circular Printed Circuit Board (PCB, made by Somacis) of 375~mm diameter and 1.6~mm thickness,
whose core materials are FR4/phenolic and copper (17~$\mu$m of thickness).
The active surface (Fig.~\ref{fig:DesignMMDetector}) is $25.2 \times 25.2$ cm$^2$
and is divided in squared pads of 332~$\mu$m length with a pitch of 583~$\mu$m.

Pads are alternatively interconnected following $X$ and $Y$ axis to 432~strips per direction,
as shown in Fig.~\ref{fig:DesignMMDetector}.
This connection is made through resin filled holes ($\sim$120$\mu$m diameter).
Routing strips lie in two different circuit layers
and finish at four rectangular connectors prints at the PCB sides, two per direction.
A connector print contains 300~pads, but not all of them are connected to a strip pad:
one print is connected to 288~strip pads (two thirds) and the other to 144 (one third).
The other print pads are connected to the readout ground.
A stainless-steel mesh was laminated on the PCB (creating an amplification gap of 128~$\mu$m)
at the Saclay workshop using the bulk technology~\cite{Giomataris:2004aa}.

Each PCB is fixed to a circular copper base, which is then fixed to the respective cap by four copper columns.
A flat cable (Fig.~\ref{fig:Setup}, (g)) links each readout footprint to the electronics, as described in Sec.~\ref{sec:DetElec},
by means of a commercial 300-pin solderless connector\footnote{GFZ300, produced by SAMTEC.} (Fig.~\ref{fig:Setup}, (h)).
The connectivity is assured by four screws, which also join two 0.5~cm thick lead covers and two 0.5~cm thick copper containers.
These pieces are conceived to partially shield the intrinsic radioactivity of the connectors.
Each flat cable goes out from the vessel through a slit
at the corresponding end cap that ends in a copper feedthrough (Fig.~\ref{fig:Setup}, (i)).
The flat cable is fixed to this piece by a teflon gasket, which is then glued by epoxy Hysol.
The copper feedthrough is then screwed to the end cap and its leak-tightness is assured by a teflon o-ring.

\subsection{Readout electronics}
\label{sec:DetElec}
\begin{figure*}[htb!]
\centering
\includegraphics[width=0.65\textwidth]{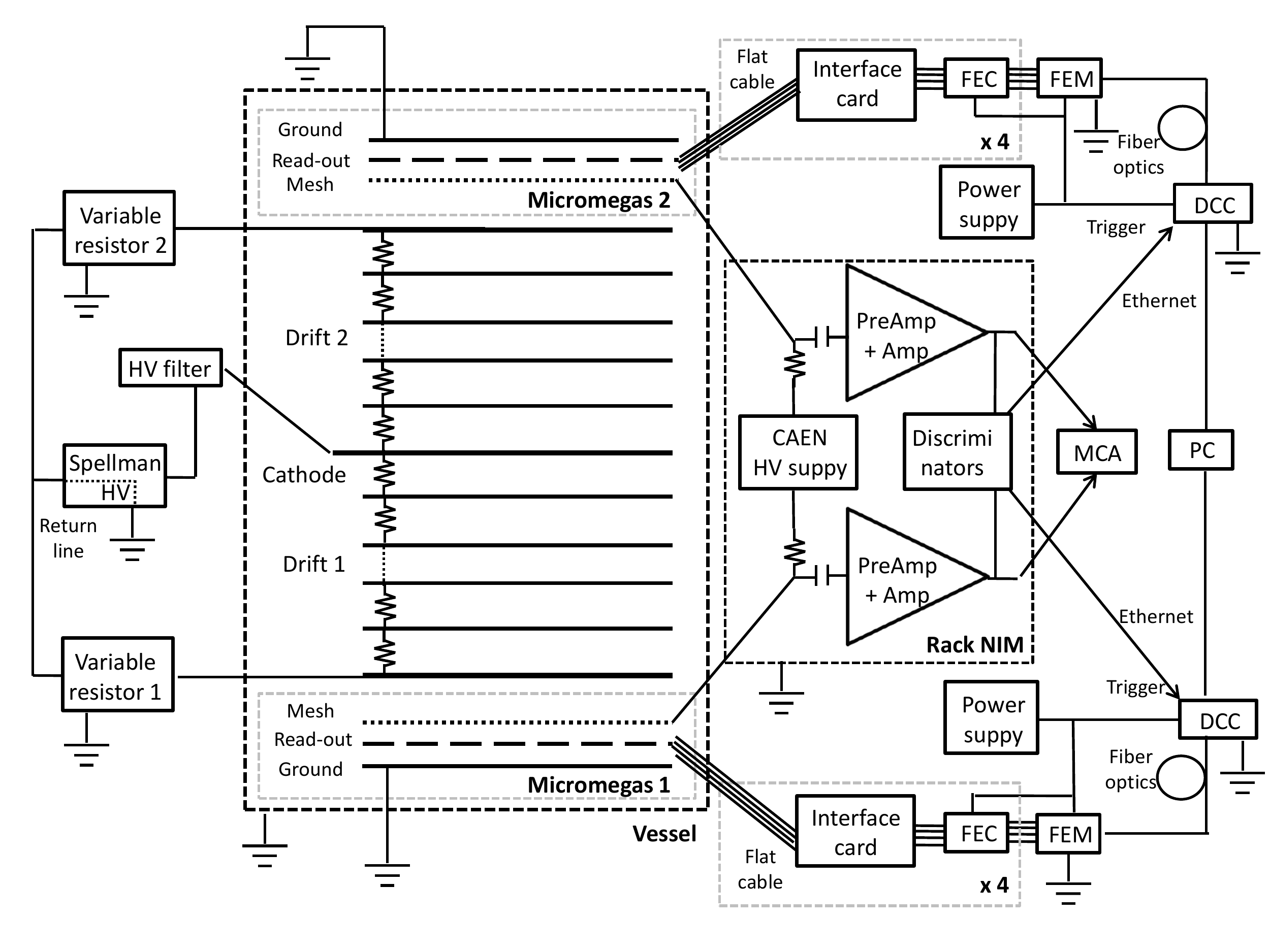}
\includegraphics[width=0.34\textwidth]{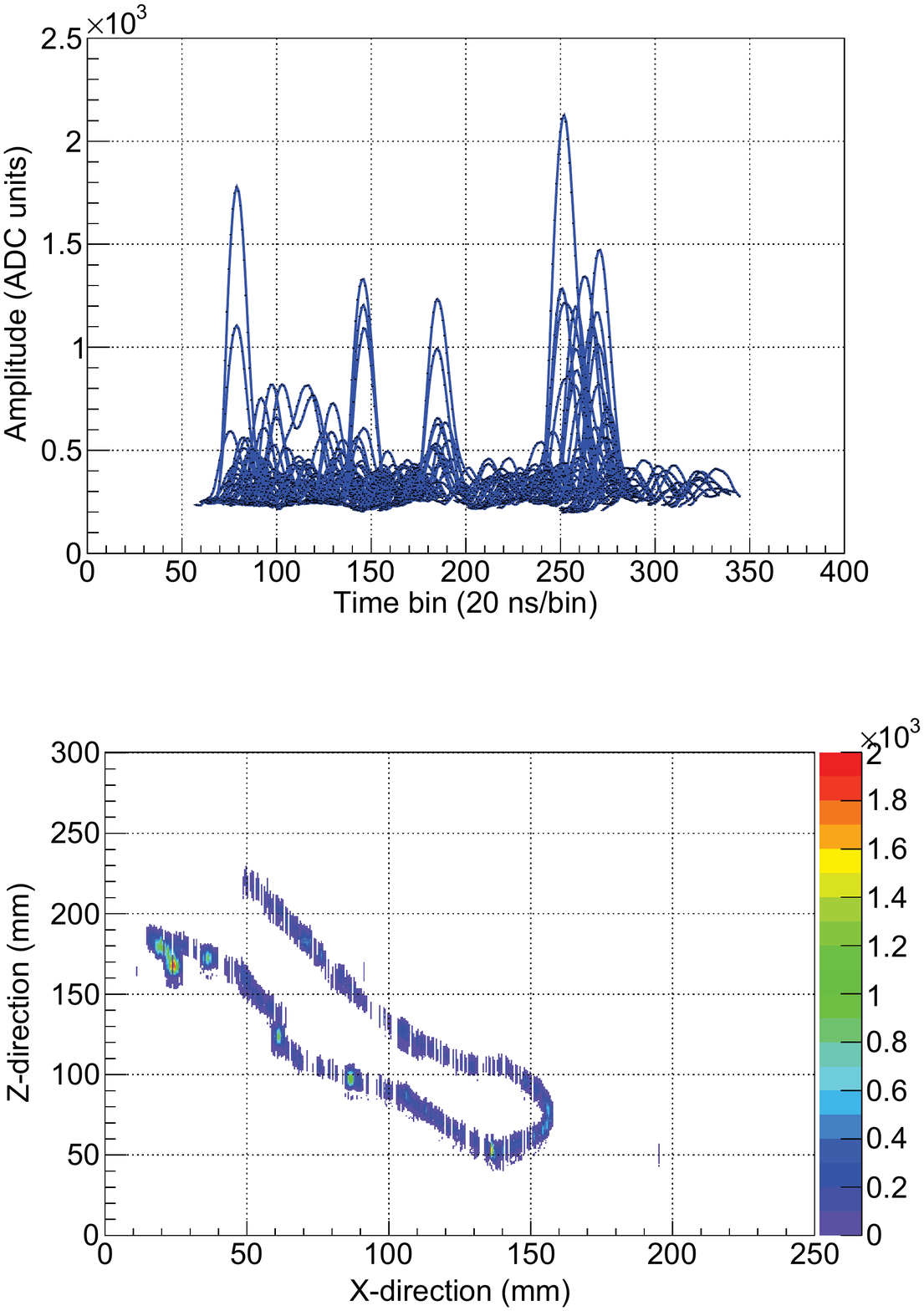}
\caption{Left: A diagram of the electronic chain, which is described in detail in text.
Top right: The strip pulses of an event as they are recorded in a FEC card.
Bottom right: The $XZ$ view of a reconstructed event, as obtained from previous pulses.
It corresponds to an electron with a long twisted track and a final big energy deposition or blob.}
\label{fig:SchemaElec}
\end{figure*}

An event interacting in either of the active volumes releases electrons, which drift towards the Micromegas readout planes.
These primary electrons are then amplified in the gap and the charge movement induces signals both at the mesh and the strips.
Both signals are processed by two different electronic chains, whose schema are shown in Fig.~\ref{fig:SchemaElec} (left).
The mesh signal is extracted from the vessel by a coaxial low noise cable\footnote{SML50SCA, produced by AXON.}
and a feedthrough (Fig.~\ref{fig:Setup}, (i)) similar to the field cage ones.
The signal is decoupled from the high voltage by a filter, whose characteristic $RC$ constant
minimizes the recovery time after a possible current excursion produced by a spark at the amplification gap.
The signal is afterwards processed by a preamplifier\footnote{Model 2004 by Canberra.},
a spectroscopy amplifier\footnote{Model 2021 by Canberra.},
and is subsequently recorded by a Multichannel Analyzer (MCA)\footnote{Amptek MCA8000A.}.
In parallel, strip pulses are routed to the four readout footprints and
go through four flat cables that come out from the vessel.
Each cable is connected to the so-called {\it interface card} (Fig.~\ref{fig:Setup}, (j)),
that routes the signals to the ERNI connectors of
an AFTER (ASIC For TPC Electronics Readout)-based front-end card (FEC) board (Fig.~\ref{fig:Setup}, (k))
\cite{Baron:2008zza,Baron:2010zz}.
The interface card includes a jumper for each strip signal path
to isolate it from the electronics if a spark connects it with the mesh.
Each FEC board has four AFTER ASICs that amplify
and sample the strip signals continuously at 50~MHz in 512~samples,
corresponding to a time window of $\sim$10~$\mu$s,
which is longer that the maximum drift time (5.7~$\mu$s) of an event in an active volume.
The electronics is triggered by the negative component of the mesh's amplified bipolar pulse, which passes through
a discriminator\footnote{Lecroy 623A.}
and a NIM-to-TTL adapter\footnote{Lecroy 688.},
and is fed to a Data Concentrator Card (DCC)\footnote{Xilinx ML-405.}.
If a trigger arrives, the analog data from all channels is digitized by the ASICs. Then, a pure digital electronics
card, the front-end mezzanine (FEM) board (Fig.~\ref{fig:Setup}, (l)),
gathers all digital data, performs a pedestal subtraction
and sends it to the DCC card via optical fiber, 
which is connected to the computer by means of a standard network cable.
The electronics has two modes of operation: non-compressed and compressed one.
In the first one, the 512 digitized samples are recorded for each strip channel.
In the second one, only the samples whose height is bigger than a {\it strip threshold} are recorded.
This threshold is calculated for each channel during a pedestal run, which is made before the data acquisition,
and is equal to $4.0 \times \sigma$ adc units over the baseline level,
where $\sigma$ is the baseline fluctuation of the channel.
The second mode has been used for all data presented in this article,
except for some data-sets taken to evaluate the noise prospects of the experiment (see Sec.~\ref{sec:enerthers} for more details).
The $XZ$ and $YZ$ views of an event are reconstructed
combining the strip pulses, whose temporal position gives the relative $z$ position,
and the routing of both the readout plane and the interface card.
An example of the pulses acquired by the electronics and the corresponding reconstructed event
is shown in Fig.~\ref{fig:SchemaElec} (right).

As a low energy threshold is one of the main goals of the experiment, special care has been given to grounding
in the electronics design: each high voltage line has a dedicated low-frequency filter to dim the signal ripples
from high voltage sources; coaxial cables are used for mesh and ground connexions; signal paths are
surrounded by a ground layer both at the readout plane, the flat cables and the interface cards to avoid any coupling;
the AFTER-based cards (FEM and FECs) of each side are inside a Faraday cage to minimize induced noises.
These Faraday cages, which were initially isolated from the vessel, were found the origin of a MHz-frequency noise.
This noise was removed by covering with aluminum foil both the flat cables and the interface cards,
and connecting simultaneously the Faraday cages and the vessel.
This fact points to either a design issue of the interface cards
or an intrinsic noise source at the level of the AFTER-based electronics, which should be solved in near-term upgrades.
As described in Sec.~\ref{sec:enerthers}, the noise level is equivalent to an energy threshold of 0.6~keV
for a readout gain of $10^3$, i.e., $2.3 \times 10^4$ electrons.
These values are limited by the electronic noise of the mesh channel and the readouts gain.

\subsection{Calibration, gas and pumping systems}
The calibration source consists of a cylindrical container of thin aluminum wall
closed at one end and with a deposition of $^{109}$Cd inside.
This radioactive source emits x-rays of 22.1 (K$_\alpha$) and 24.9~keV (K$_\beta$).
This holder is screwed to a 3~mm diameter nylon wire pushed forwards (or pulled backwards) 
through a Teflon tube located inside the vessel and around the cathode plane
(Fig.~\ref{fig:Setup}, (m); Fig.~\ref{fig:DesignCathode}, (g)).
The wire can be manually moved to eight calibration points (four per active volume), situated
at the corners of the central cathode (Fig.~\ref{fig:DesignCathode}, (g$_x$)),
where the source illuminates directly an active volume. The source can be retracted
outside the vessel inner volume to a parking position situated at the bottom port.
The $^{109}$C source was chosen because its x-rays can go through the Teflon tube with small loses,
which is not the case of $^{55}$Fe.
However, an extra x-ray line in the 1-10 keV energy range is needed for the analysis,
as discussed in Sec.~\ref{sec:BackXraysAna}.
Several options are being studied for the LSC setup, which are described in Sec.~\ref{sec:con}.

The gas system consists of two ports situated at the bottom (inlet) and the top (outlet),
where gas enters and comes out from the vessel.
The gas comes from a premixed bottle,
whose pressure is adjusted by a pressure transducer\footnote{F-702CV-AGD-33-V by Bronkhorst.}
and whose flow is set by a mass flowmeter\footnote{F-201CV-AGD-33-V by Bronkhorst.}.
These two components, three temperature sensors, a pressure sensor\footnote{PTU-F-AC15-33AG by Swagelok.}
and the HV sources are continuously monitored by a slow control,
programmed in Python and based on Arduino cards~\cite{PeiroVal:12149}.

The vessel has a stainless steel CF40 flange through which it can be pumped before its operation
to reduce the release of trapped air or other impurities from elements inside the vessel.
After $\sim$96~hours of continuous pumping, a level of $3.0 \times 10^{-4}$~mbar was achieved,
while the outgassing/leak rate was $3.0 \times 10^{-4}$~mbar$\cdot$l/s.
We think that these numbers are limited by the outgassing of the inner plastic components,
as feedthroughs and unions show leak rates below $\sim 10^{-6}$~mbar$\cdot$l/s.
As no attachment effect has been observed during the characterization,
the actual detector can work with a continuous gas flow.
For close regimes (static or recirculation mode),
the outgassing rate may not be enough and should be probably reduced by a bake-out system.


\section{Detector characterization at low energy}
\label{sec:characterization}
This section describes the studies undertaken to characterize
the performance of the Micromegas readout planes in argon- and neon-based mixtures
at high pressure using a $^{109}$Cd source.
The aim is to find the optimum point of operation,
in terms of general performance and energy threshold, for a physics run.
This optimization is a challenging task as it depends on many factors
(the base gas, the quencher and its quantity; and the pressure)
and there are few studies in literature on gas properties.

For simplicity, we have started the characterization with Ar+2\%iC$_{4}$H$_{10}$,
the gas normally used in CAST Micromegas detectors~\cite{Aune:2013pna}.
This initial choice is reasonable: isobutane gives high gain (up to $10^4$)
and good energy resolution (13\% FWHM at 5.9 keV) in argon-based mixtures~\cite{Iguaz:2012ur}
for microbulk Micromegas detectors.
Even if 2\% of isobutane may be a too small quencher concentration
for atmospheric pressure~\cite{Iguaz:2012ur},
it is known that optimal relative quencher concentrations decrease with pressure,
e.g. for Xe-TMA mixtures~\cite{Cebrian:2012sp}.
The results of this characterization will be presented and discussed here.
We will also present some data taken in Ar+5\%iC$_4$H$_{10}$ at 1.2 bar
in the best noise conditions of the actual experiment.
These data-sets have been used to estimate the actual energy threshold of TREX-DM
and its future prospects (see Sec.~\ref{sec:enerthers}).

\subsection{The experimental procedure}
The experimental procedure starts by a several-day long pump-down of the vessel.
Then, the leak-tightness of all the vessel feedthroughs and unions is checked by means of a Helium leak detector.
All components must show a value lower than $\sim 10^{-6}$~mbar$\cdot$l/s in order to start the characterization.
Once the leak-tightness is verified, gas is injected into the vessel at an adjustable flow
and high-voltage tests are performed to verify the connectivity and the spark protection.
A gas flow of of 3--5~l/h is kept during all the measurements.

\begin{figure}[htb!]
\centering
\includegraphics[width=0.48\textwidth]{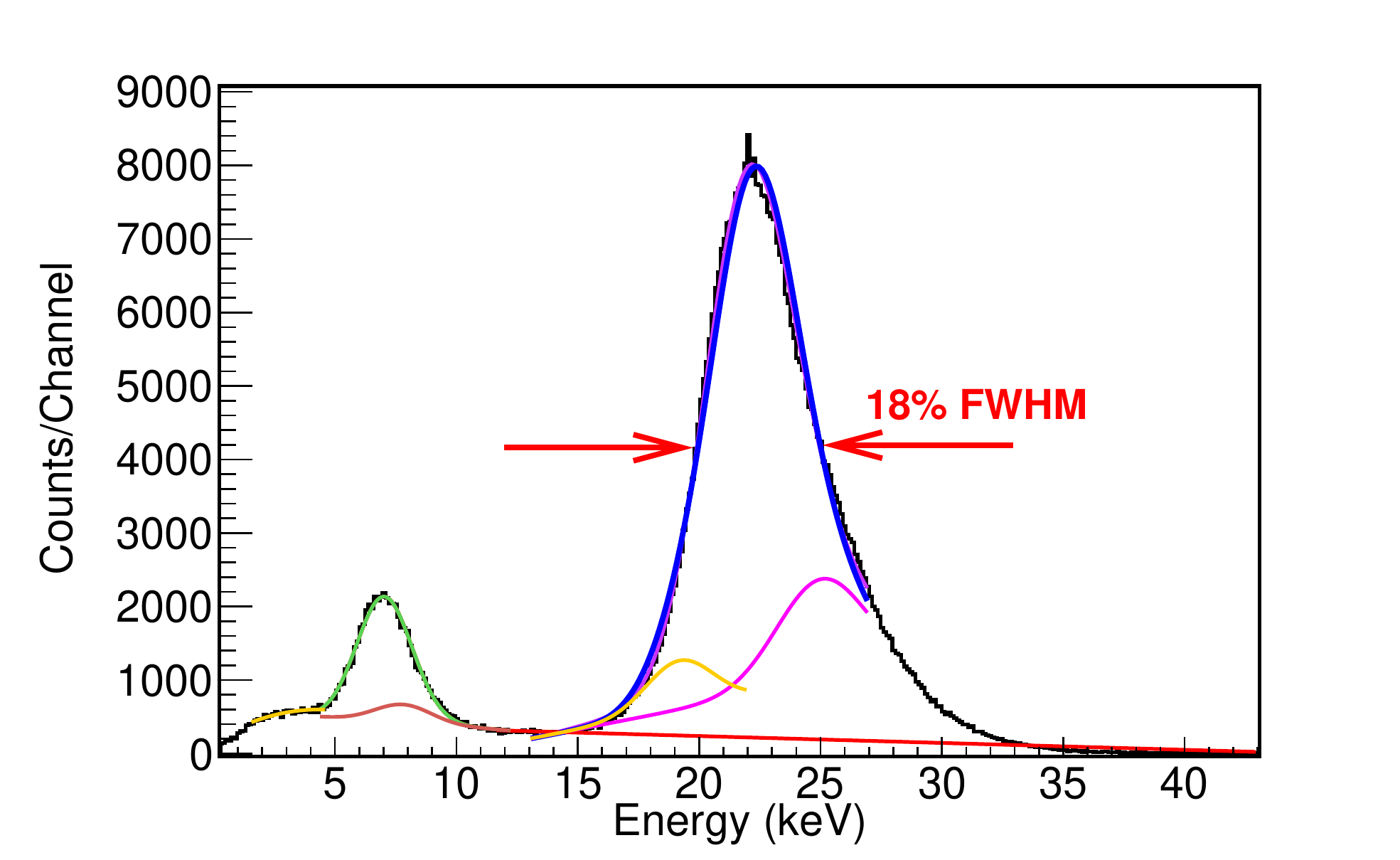}
\caption{Energy spectrum  generated by the mesh signals when one of the Micromegas readouts
is irradiated by a $^{109}$Cd source in Ar+2\%iC$_{4}$H$_{10}$ at 2~bar.
The spectral parameters are defined through an iterative multi-Gaussian fit
corresponding to the K$_{\alpha}$ (22.1~keV, blue line)
and K$_{\beta}$ (24.9~keV, magenta line) emission lines of the source
and their escape peaks (located at 19.1 and 21.9~keV, orange line).
The fluorescence lines of iron (at 6.4~keV, emitted from the mesh)
and copper (8~keV, from the vessel or the field cage strips)
are also present (green and brown lines, respectively).}
\label{fig:fit}
\end{figure}

The two Micromegas planes (\emph{MM1} and \emph{MM2}) are characterized
in terms of electron transmission, gain,
gain uniformity and energy threshold,
over a wide range of operating pressures (and therefore operating voltages).
The gas mixture used was Ar+2\%iC$_{4}$H$_{10}$ for pressures between 1.2 and 10~bar, in steps of 1~bar.
For this purpose, the two readouts were calibrated at low energies
by a $^{109}$Cd $\gamma$-source.
The calibration spectra are characterized by the K-peaks and the fluorescence emissions at 6.4 and 8~keV
from the iron and copper components (see Fig.~\ref{fig:fit}).
The mean position and the width of the K$_{\alpha}$ is calculated through
an iterative multi-Gaussian fit, previously used in~\cite{Cebrian:2012sp},
including both the K$_{\alpha}$ and K$_{\beta}$ emission lines
and their escape peaks (at 19.1 and 21.9~keV).
A wide range of amplification and drift fields are scanned at each pressure,
which requires a bias ranging from $\sim$300~V at 1.2~bar to $\sim$900~V at 10~bar,
and from $\sim$1.5~kV to $\sim$30~kV, respectively. 

\subsection{Electron transmission and detector gain}
The electron transmission is the probability for primary electrons
to pass from the drift regions to the amplification gap through the mesh holes.
The measurement of the electron transmission depends, therefore,
on two different mechanisms that cannot be measured separately:
the electron attachment and recombination on the drift region,
and the so-called transparency of the mesh electrode.
This parameter informs about the electron collection efficiency of a Micromegas detector.

\begin{figure}[htb!]
\centering
\includegraphics[width=0.48\textwidth]{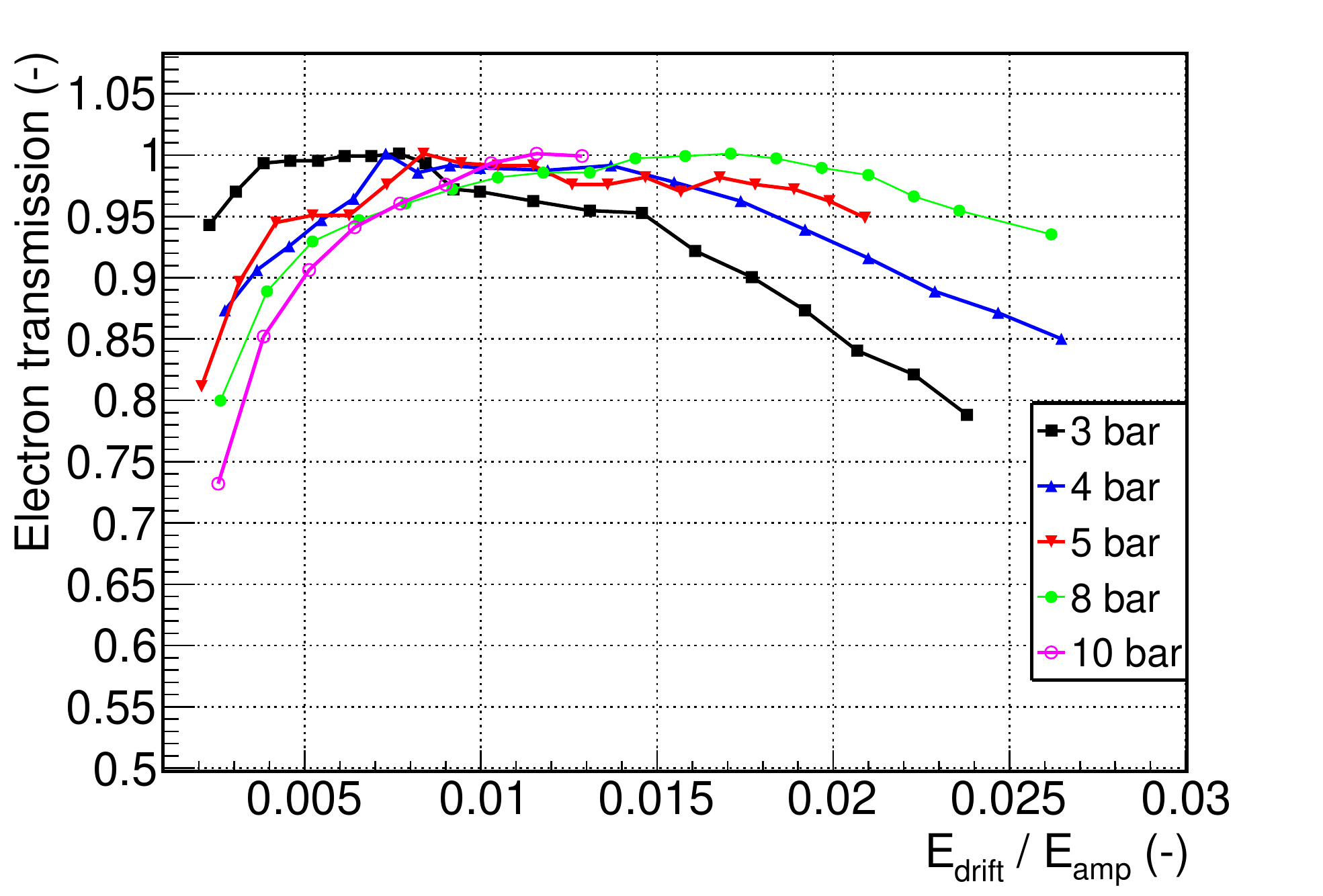}
\caption{Dependence of the electron transmission with the drift-to-amplification field ratio
for the \emph{MM2} readout in Ar+2\%iC$_{4}$H$_{10}$ at different gas pressures.
The peak positions have been normalized to the maximum of each series,
assuming that the full electron transmission is always achieved.}
\label{fig:transparency}
\end{figure}

The drift voltage is varied for a fixed mesh voltage
to obtain the dependence of the electron transmission
with the drift-to-amplification field ratio at each pressure (Fig.~\ref{fig:transparency}).
As expected, the readouts show a plateau of full electron transmission
for a wide range of drift-to-amplification field ratios at all pressures.
Although this is not an absolute measurement of the electron transmission,
the fact that the signal height becomes independent of the field ratios
suggest that the mesh transmission is close to 100\% in the plateau range,
allowing to normalize to the maximum value of the signal height.
If no plateau is observed, however,
the identification of the maximum with 100\% electron transmission 
would not be supported and the normalization would not be justified.

The electron transmission drops at very low reduced electric fields
in the drift regions due to electron attachment
and recombination of the primary electrons generated in the conversion volume.
In these measurements the plateau of full electron transmission starts
at higher values of drift field ($\sim$50~V/cm/bar)
than those observed in \cite{pacotesis} ($\sim$ 20 V/cm/bar) for microbulk Micromegas detectors.
This effect has been attributed to a ballistic deficit that appears
when the integration time of the amplifier is lower than the collection time,
favoured by lower drift velocities and larger longitudinal diffusion coefficients.
In fact, at 100 and 20 V/cm/bar the drift velocities are 3.3 and 1.1~cm/$\mu$s, respectively;
and the longitudinal diffusion coefficients are 405.6 and 914 $\mu$m/cm$^{1/2}$.

For high drift fields, the electron transmission is reduced
since the configuration of the field lines make that some primary electrons get trapped in the mesh electrode.
It is observed that the plateau extends to higher values of field ratios than those of \cite{pacotesis}.
On the other hand, it is also observed that the right edge of the plateau moves
to higher ratio of fields as the pressure increases,
an effect already observed in \cite{pacotesis} for microbulk readouts,
a fact attributed to the decreasing diffusion coefficient with pressure.
The reduction of the electron transmission also degrades the energy resolution.

The drift-to-amplification field ratio is set for every pressure
at the point where the mesh shows the maximum electron transmission,
typically at a reduced drift field of around 100~V/cm/bar.
Then, the available range of mesh voltages is scanned, from very low amplification fields
where the amplitude of the mesh signal is just above the noise threshold, up to the spark limit,
where micro-discharges between the mesh and the readout produce a current excursion.
If the current exceeds the HV current-limit of 300~nA,
it results in a HV trip that reduces the exposure time.
The spark rate is selected so that the overall exposure reduction is below $10^{-3}$.
In order to prevent that a high intensity discharge develops
a high conductivity path, the HV is ramped down if the HV current-limit is exceeded during more than 10 seconds.

The signal amplitude increases with the applied amplification field,
while the peak position moves to higher values in the energy spectra.
The peak position is used to calculate the absolute gain of the Micromegas readout planes,
defined as the ratio of the number of electrons after the avalanche $n$
and the number of primary electrons, $n_0$: $G = \frac{n}{n_0}$.
Determining $G$ requires the characterization of the electronic chain
in order to obtain the conversion factor between the peak position
registered by the MCA and the number of electrons
before the preamplifier $n$. As described in Sec.~\ref{sec:BackDetResponse},
the number of primary electrons $n_0$ is given by 22.1~keV/$W_{Ar}$,
where $W_{Ar}$ = 26.3~eV~\cite{Christophorou:1971}.
The presence of iC$_{4}$H$_{10}$ is disregarded in the gain calculation
since is low concentrations and similar W-value (23~eV~\cite{Meisels}) to that of Argon.

\begin{figure}[htb!]
\centering
\includegraphics[width=0.48\textwidth]{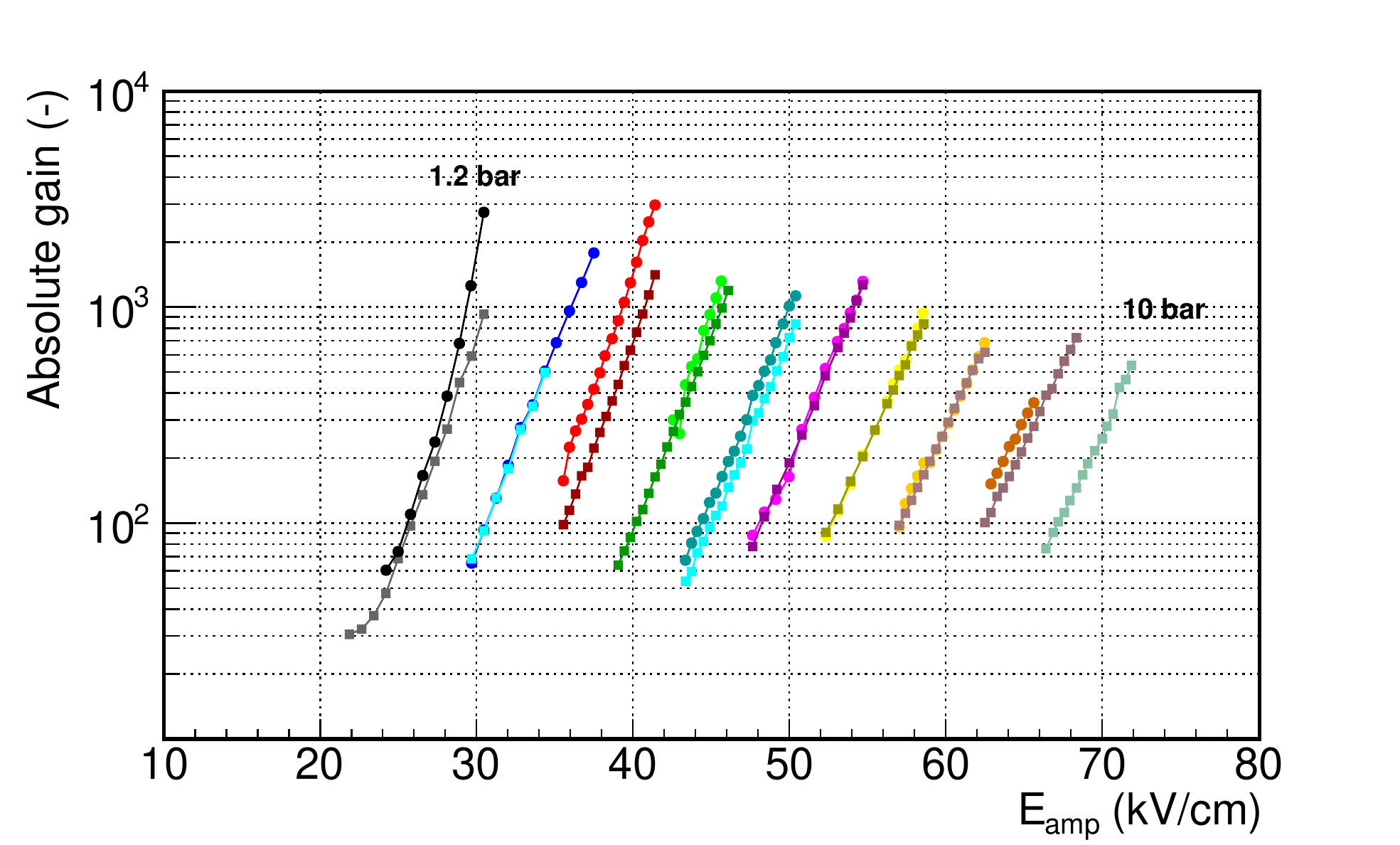}
\caption{Dependence of the absolute gain with the amplification field
in units of kV/cm in Ar+2\%iC$_{4}$H$_{10}$ between 1.2 and 10~bar (in steps of 1~bar)
for the \emph{MM1} (circles) and \emph{MM2} (squares) readouts.
The maximum gain of each curve is obtained just before the spark limit.}
\label{fig:gain}
\end{figure}

The gain curves obtained in Ar+2\%iC$_{4}$H$_{10}$ between 1.2 and 10~bar are shown in Fig.~\ref{fig:gain}.
The two readouts present a similar gain and, in both cases,
the maximum attainable gain before the spark limit decreases with the gas pressure,
from $3 \times 10^3$ at 1.2~bar down to $5 \times 10^2$ at 10~bar. 
Both planes reach gains higher than 10$^3$ for pressures up to $\sim$6~bar.
The dependence of the gain with pressure was also studied
for a triple-GEM gaseous detector in argon in reference~\cite{Bondar:2001mt}.
Much larger detector gains ($10^5$) than the ones reported in this work were found for pressures below 3 bar
but the gain per GEM plane was lower ($< 10^2$).
Moreover, but the maximum gain abruptly dropped at higher pressures,
reaching gains below $5 \times 10^2$ at 5 bar.

\begin{figure}[htb!]
\centering
\includegraphics[width=0.48\textwidth]{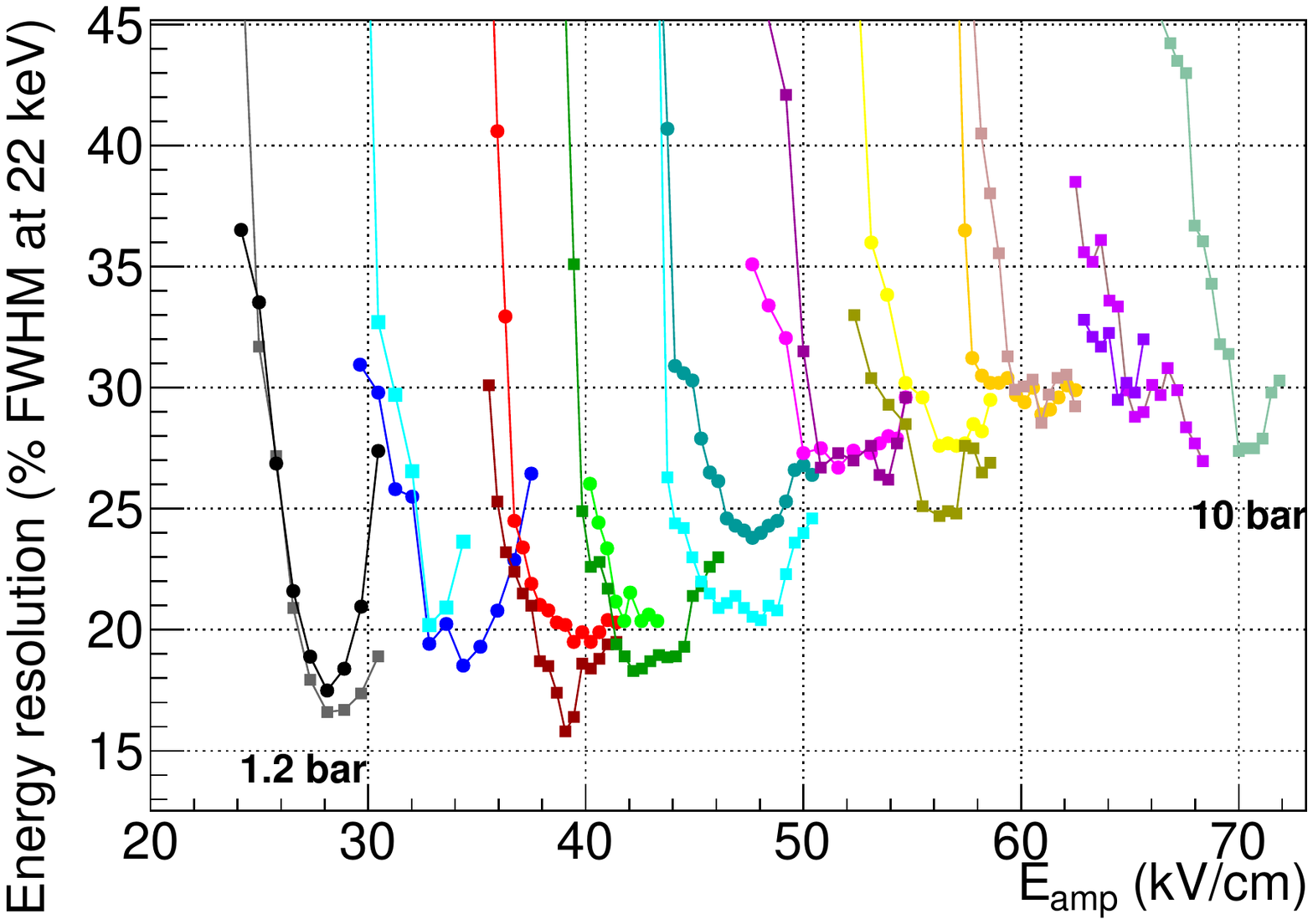}
\caption{Dependence of the energy resolution at 22~keV with the amplification field
for the \emph{MM1} (circles) and \emph{MM2} (squares) readout planes of TREX-DM
in Ar+2\%iC$_{4}$H$_{10}$ between 1.2 and 10~bar in steps of 1~bar.
The best achievable energy resolution degrades from 16\%~FWHM at 1.2~bar to 25\%~FWHM at 10 bar.
The statistical error of the energy resolution is up to 0.3\%~FWHM,
given by the error from the fit to the gaussian parameters. }
\label{fig:fwhm}
\end{figure}

The energy resolution (expressed in FWHM) of the readout planes
is obtained from the width of the gaussian fit
to the K$_{\alpha}$, K$_{\beta}$ and escape peaks.
The dependence of the energy resolution on the amplification field
for all the pressure settings is shown in Fig.~\ref{fig:fwhm}.
The statistical error of the energy resolution is less than 0.3\%~FWHM,
given by the error from the fit to the gaussian parameters.
There is also an uncertainty of $\pm$1~V in the high-voltage power-supply,
which produces a systematic error of about a 0.2\% in the amplification field determination.
At each pressure there is a range of amplification fields
for which the energy resolution is optimized.
At low gains, the energy resolution degrades
because the signal becomes comparable with the electronic noise.
At high fields, the resolution degrades due to the increase
in the gain fluctuations by the UV photons generated in the avalanche.

As it is shown, the best energy resolution degrades with pressure,
being 16\%~FWHM at 22.1~keV at 1.2~bar and 25\%~FWHM at 10~bar.
These values may be limited by the noise level and low quantity of quencher (2\%).
In fact, an energy resolution of 14\%~FWHM at 1.2~bar
was measured in Ar+5\%iC$_4$H$_{10}$ at 1.2 bar (as shown in Fig.~\ref{fig:EnerSpecThres}).
This energy resolution is closer to the best value measured by a 128~$\mu$m-gap bulk Micromegas readout
in an argon + 5\% isobutane mixture (11.9\% FWHM at 22.1~keV,
calculated from a 23\%~FWHM at 5.9~keV~\cite{Iguaz:2011yc}, supposing only an energy dependence).

\subsection{Gain homogeneity}
\begin{figure*}[htb!]
\centering
\includegraphics[width=0.48\textwidth]{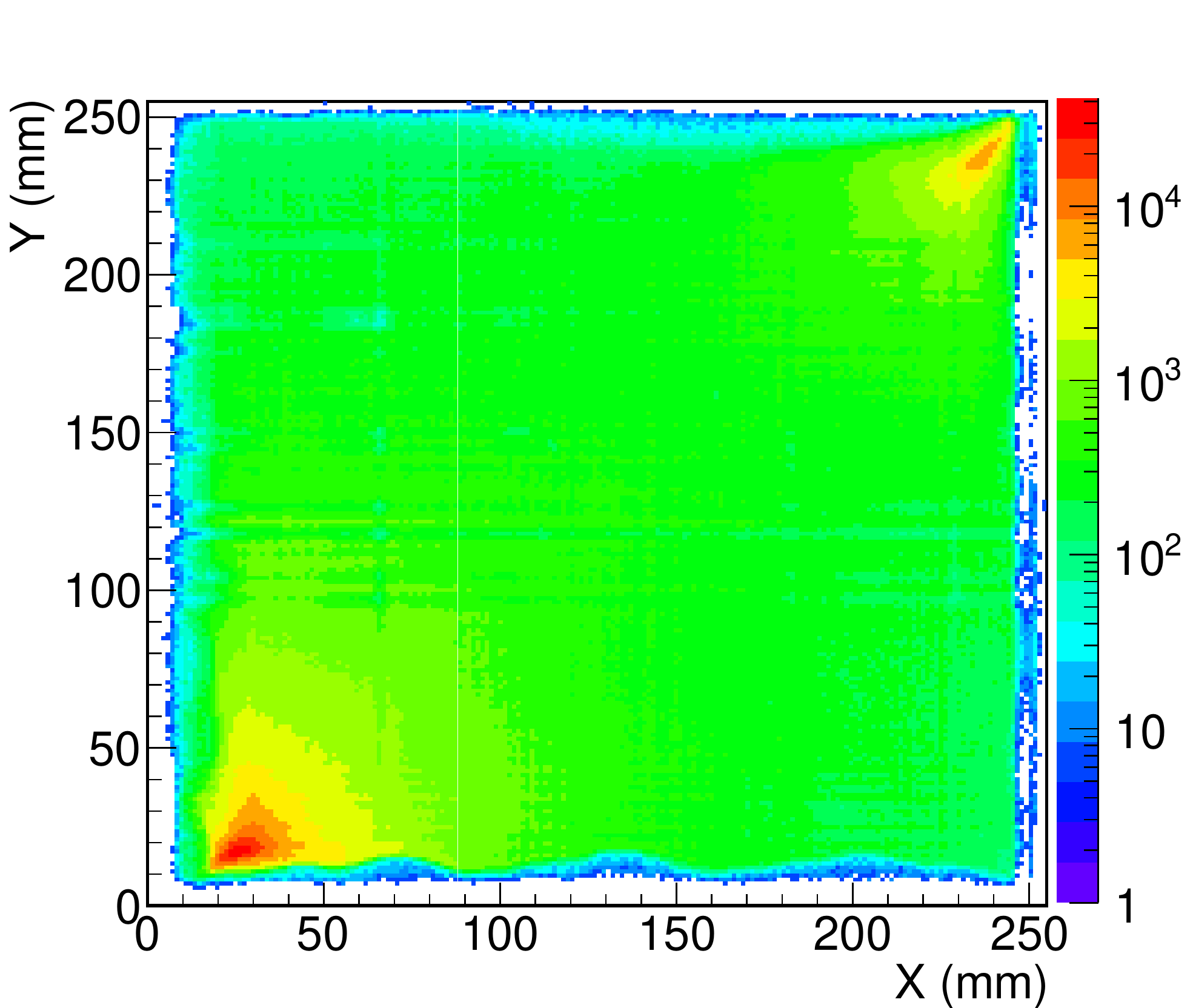}
\includegraphics[width=0.48\textwidth]{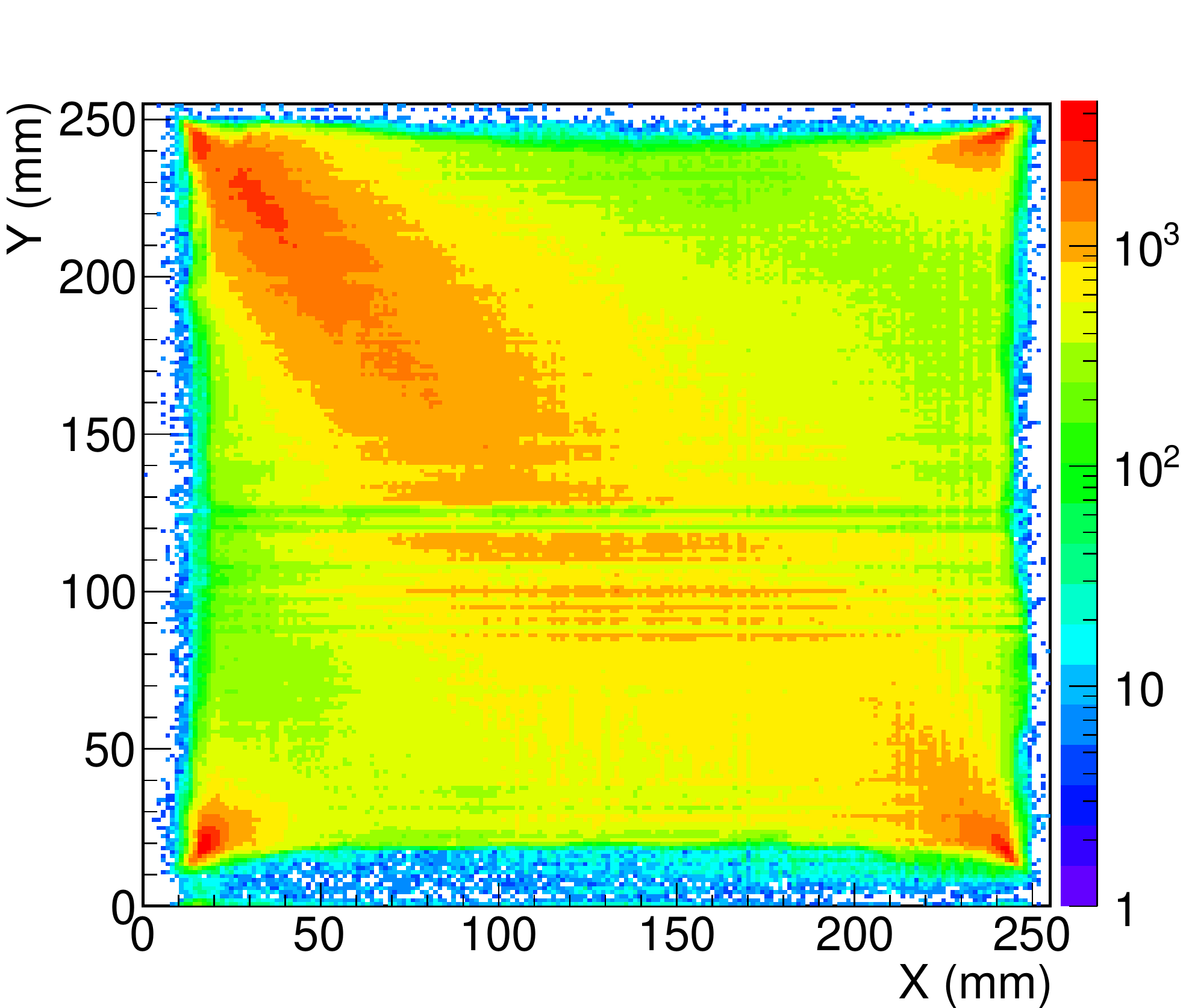}
\caption{Event distribution of the \emph{MM1} (left) \emph{MM2} (right)
Micromegas readout planes generated by $^{109}$Cd calibrations
when the vessel was filled with Ar+2\%iC$_{4}$H$_{10}$ at 2~bar.
There are more events at the corners, as these points are nearer to the calibration points.}
\label{fig:hitmap}
\end{figure*}

The response homogeneity of each readout was studied in Ar+2\%iC$_{4}$H$_{10}$ at 2~bar,
illuminating its surface with the $^{109}$Cd source at its four calibration points
and using the strip signals recorded by the AFTER-based electronics.
The sampling frequency was set to 50~MHz to get a temporal window of $\sim$10~$\mu$s.
In Ar+2\%iC$_4$H$_{10}$ at a reduced drift field of 100~V/cm/bar, the drift velocity is 3.33~cm/$\mu$s,
so ionization tracks as long as the active volume's length (19~cm) could be fully recorded.
The event distribution in each readout plane is shown in Fig.~\ref{fig:hitmap}. 

Before analyzing the data, the readout plane surface was binned into a 2D histogram of 216$\times$216 cells
(each readout has 432$\times$432 strips). Then, for each calibration event at the K$_{\alpha}$-line
($\pm$ 30\% of the energy), the mean position in $x$- and $y$-directions was calculated
and its energy recorded at the corresponding $(x,y)$ entry of the 2D histogram.
Although the event distribution is non-uniform,
we request to have more than 10 events per cell in order to compute the gain homogeneity in that cell.
Finally, each entry of the histogram was normalized by the number of x-rays registered in that cell.

\begin{figure*}[htb!]
\centering
\includegraphics[width=0.48\textwidth]{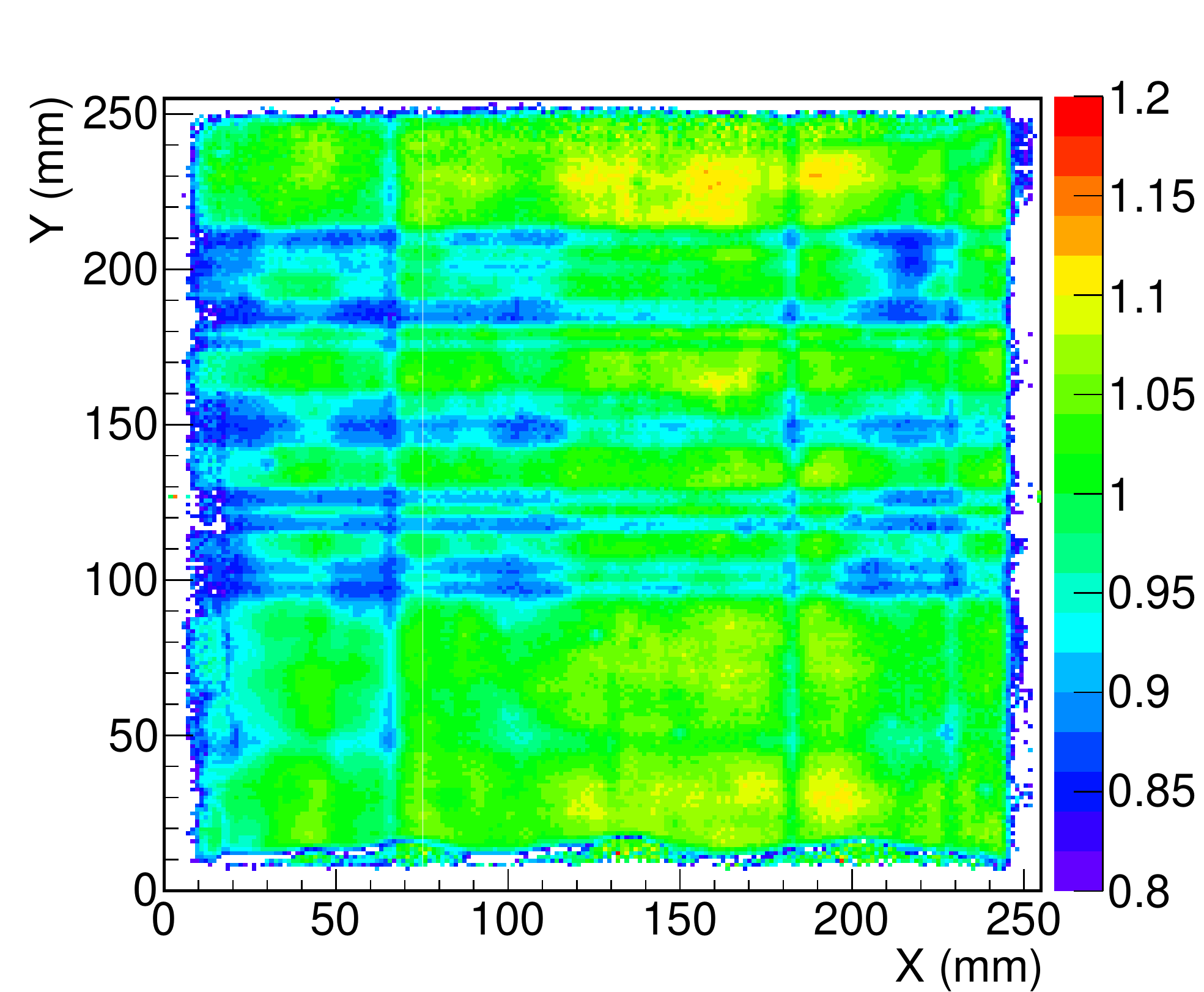}
\includegraphics[width=0.48\textwidth]{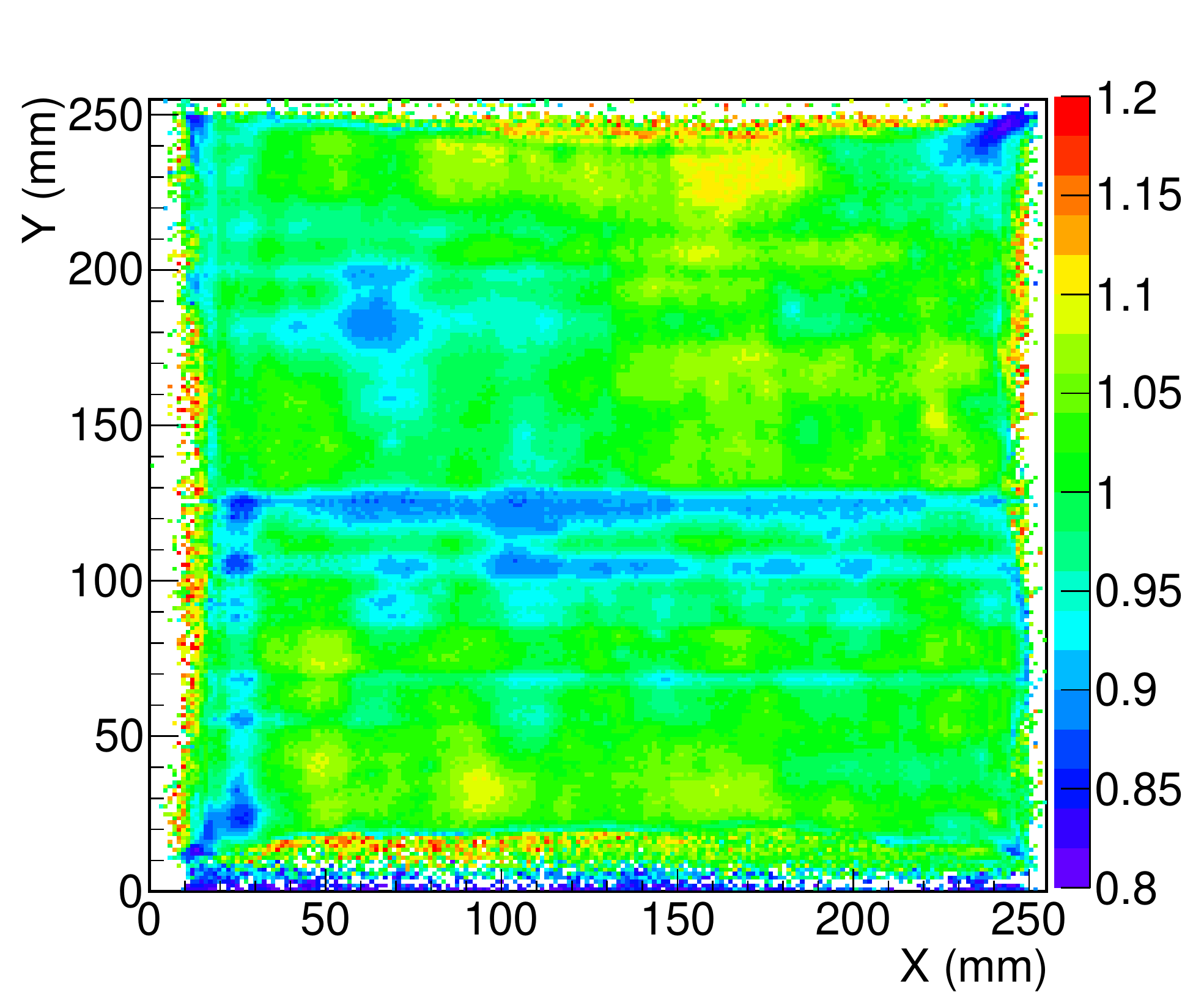}
\caption{Gain uniformity of the \emph{MM1} (left) \emph{MM2} (right) Micromegas readout planes,
generated by $^{109}$Cd calibrations
when the vessel was filled with Ar+2\%iC$_{4}$H$_{10}$ at 2~bar.
The readout cells without enough statistics (10 events or more) are shown in white.}
\label{fig:gainmap}
\end{figure*}

The resulting 2D histogram is the gain map,
which is shown for both readout planes in Fig.~\ref{fig:gainmap}.
For \emph{MM2}, the readout response's is uniform in almost all its surface,
except for two \emph{dead strips}\footnote{A dead strip is short-circuited with the mesh.
To recover the readout plane, the strip is electrically disconnected from the AFTER-based electronics
at the level of the interface card by removing a resistor.}
in $Y$-direction that reduce the effective gain at two lines.
The small dead area at the margins of the active area is caused by some wrinkles of the field-cage kapton PCB.
The \emph{MM1} readout's response has a similar behavior,
except for some more dead strips ($\sim$20)
that were accidentally caused by a bad isolation of the inner mesh cable.
The gain fluctuations over the readouts surface are better than 10\%,
while the errors on the measurement of the gain in each cell are below 1\%.

\subsection{Energy threshold}
\label{sec:enerthers}
\begin{figure}[htb!]
\centering
\includegraphics[width=0.48\textwidth]{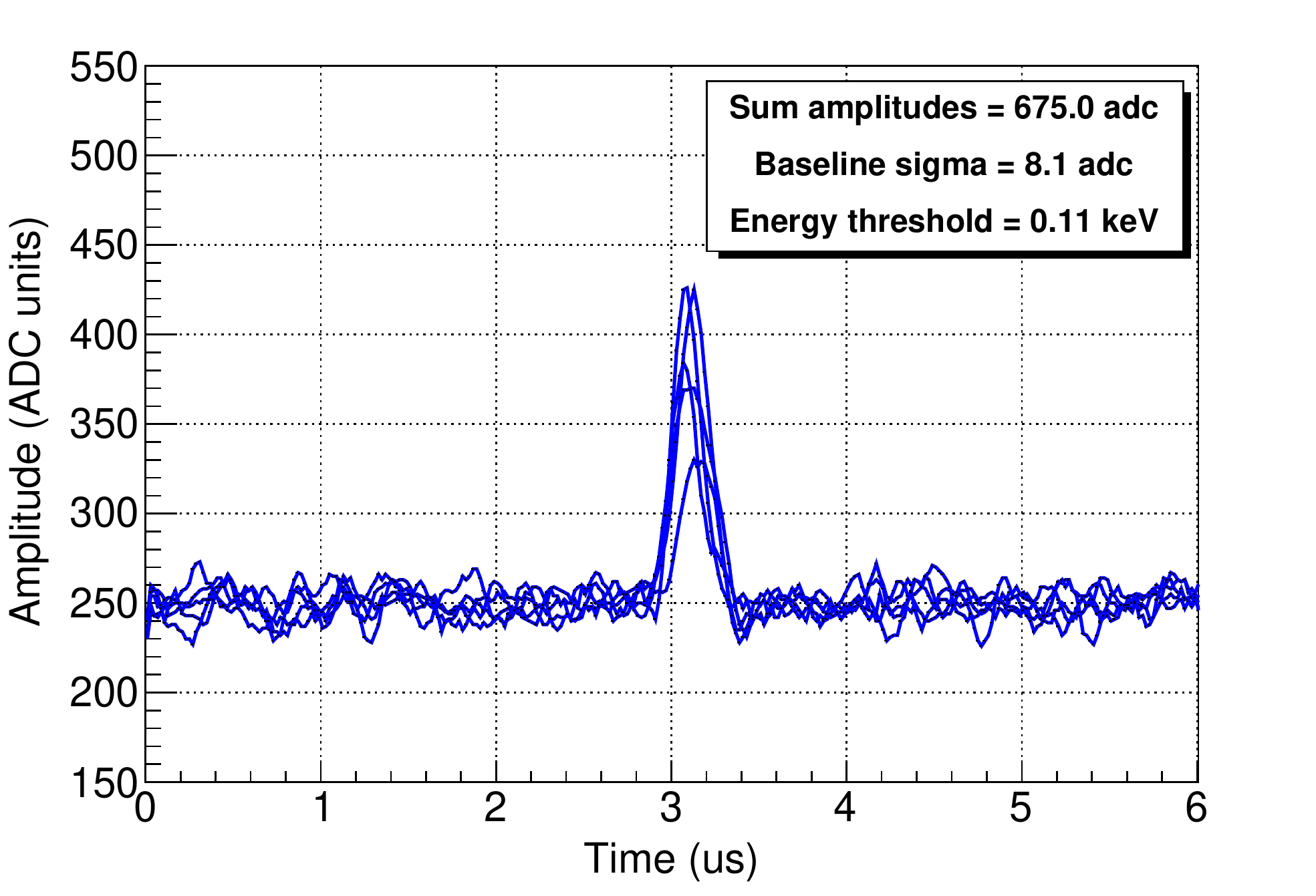}
\caption{The active strip pulses of a 1~keV event,
acquired by the AFTER-based electronics in non-compressed mode.
The energy threshold has been estimated using the pulse baseline's sigma
and the same pulse trigger level (4.0~sigma) as the one of AFTER electronics.}
\label{fig:ThresPulse}
\end{figure}

In the current DAQ implementation, the trigger is built from the mesh signal.
The energy threshold is thus limited by the readout gain
given by the avalanche multiplication of the primary electrons and by the electronic noise of the mesh channel,
that is relatively high due to its high capacitance ($\sim$6~nF).
In the final DAQ implementation planned, based on the AGET chip~\cite{Baron:2011},
the trigger will be generated individually by each single strip signal.
The strip channels enjoy a factor at least $\sim$6 times better in signal-to-noise ratio,
as they show much lower capacitance ($\sim$0.2~nF, including the contributions of the flat cables
and the interface cards).
This fact is illustrated in Fig.~\ref{fig:ThresPulse},
showing the strip signals of a random event of 1~keV
taken in the best noise conditions and using the non-compressed mode of the actual electronics.
From the baseline fluctuations we can estimate that the energy threshold
in the strips could be well at the level of 0.1~keV.
Indeed, the first tests of this DAQ with the IAXO-D0 prototype
have shown an effective energy threshold of 100~eV~\cite{Garza:2015jg}

\begin{figure}[htb!]
\centering
\includegraphics[width=0.48\textwidth]{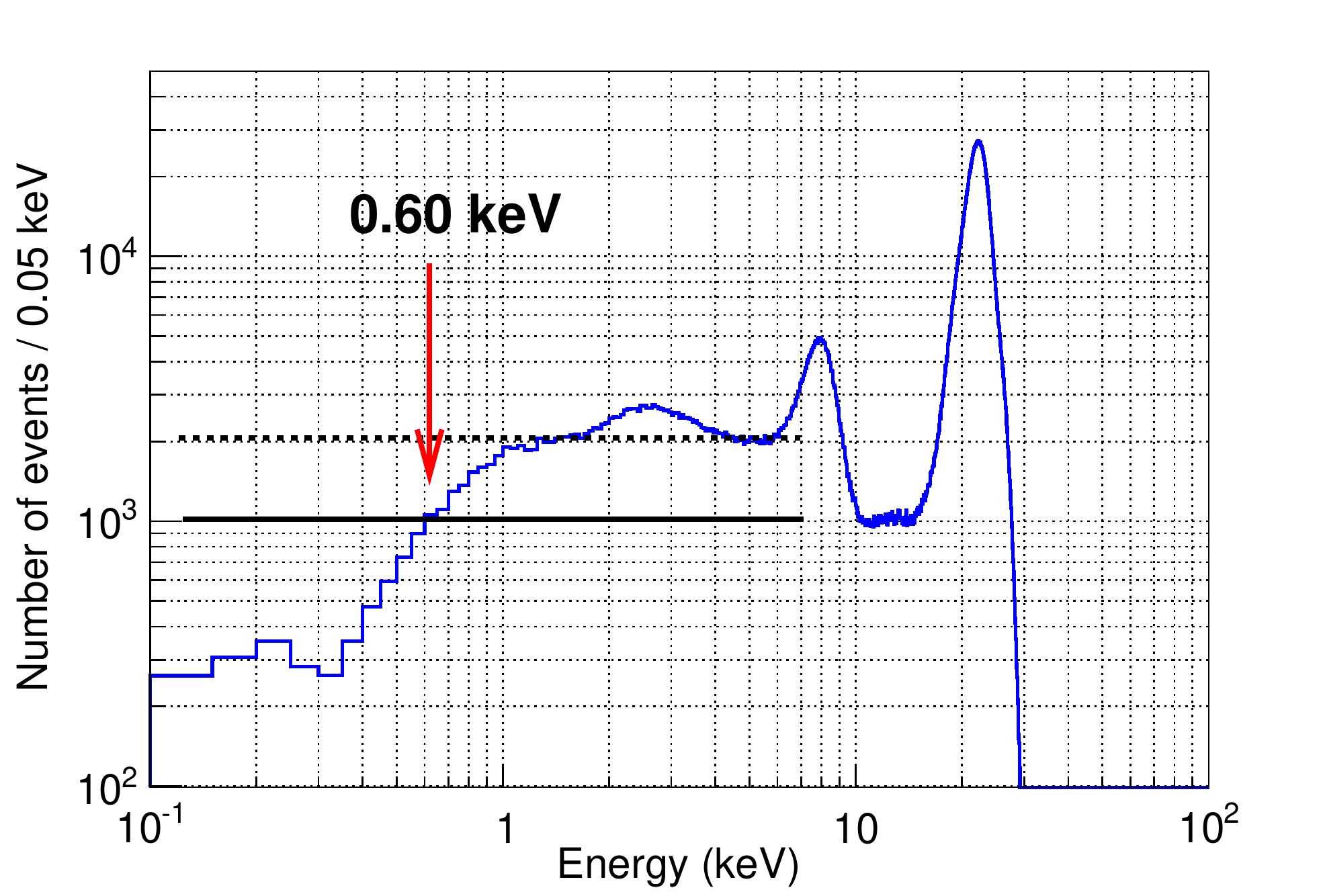}
\caption{Energy spectrum generated by the strip strips of $MM$1 readout plane,
when it was irradiated by a $^{109}$Cd source in Ar+5\%iC$_4$H$_{10}$ at 1.2 bar.
Apart from the K-lines of the source and the iron and copper fluorescence at 6.4~and 8.0~keV,
the argon fluorescence at 3.0~keV is also present. An energy threshold of $0.60\pm0.05$~keV
has been calculated, as described in detail in the text.}
\label{fig:EnerSpecThres}
\end{figure}

Some calibration data-sets were taken in the best conditions of the actual setup
to estimate the energy threshold. The vessel was filled with Ar+5\%iC$_4$H$_{10}$ at 1.2 bar,
there was no MHz-frequency noise, the readout gain was $10^3$
and the energy resolution was 14\%~FWHM at 22.1~keV,
In these conditions, the strip signals of the readout \emph{MM1} were used
to generate the energy spectrum shown in Fig.~\ref{fig:EnerSpecThres}.
The Compton level between 4.0 and 6.0 keV, which is in between the argon and iron fluorescence,
was fitted to a constant value, deriving the dashed black line shown in the figure.
Then, the energy threshold was calculated as the first energy bin,
whose intensity is the half of this Compton level,
marked by a continuous black line.
The calculated energy threshold is $0.60 \pm 0.05$(stat)$\pm 0.30$(sys)~keV,
not far from TREX-DM prospects (0.4~keV).

\section{Radiopurity measurements}
\label{sec:radiopurity} A material screening program was undertaken
to evaluate the bulk radioactivity of all the relevant components of
the detector and surrounding materials used for gas vessel, field
cage, electronics or shielding, to help both in the design of the
set-up and in the construction of the background model of the
experiment. First results were presented in~\cite{Aznar:2013jwa}. In
this section, the techniques applied to carry out these measurements
are described and the results obtained are shown and discussed. The
impact on the background levels of the measured activity in the
components selected for the TREX-DM set-up will be presented at
Sec.~\ref{sec:backmodel}.

The screening program is based mainly on germanium gamma-ray
spectrometry performed deep underground and, complementing these
results, some measurements based on Glow Discharge Mass Spectrometry
(GDMS) were also carried out. GDMS was performed by Evans Analytical
Group in France, providing concentrations of U, Th and K; it must be
noted that having no information on daughter nuclides in the chains,
a possible disequilibrium cannot be detected.

All the germanium measurements were made using a $\sim$1~kg
ultra-low background detector of the University of Zaragoza (named
\emph{Paquito}) and operated at the hall E of the Canfranc
Underground Laboratory (LSC) at a depth of 2450~m.w.e.. This
detector has been used for radiopurity measurements at Canfranc for
several years (details can be found
in~\cite{Cebrian:2010ta,Aznar:2013jwa}). It is a p-type close-end
coaxial High Purity germanium detector, with a crystal of
190~cm$^{3}$ and a copper cryostat. The energy threshold is set at
$\sim$60 keV. It is operated inside a shield made of 10~cm of
archaeological lead plus 15~cm of low activity lead, enclosed in a
plastic bag continuously flushed with boil-off nitrogen to avoid
radon intrusion. The electronic chain for the data acquisition is
based on a linear amplifier\footnote{Canberra 2020.} and an
analog-to-digital converter\footnote{Canberra 8075.}.

The detector background is periodically characterized by taking data
with no sample for periods of time of at least one month; the total
counting rate below 3~MeV is at the level of 5~c/h. Activities of
different sub-series in the natural chains of $^{238}$U, $^{232}$Th
and $^{235}$U as well as of common primordial, cosmogenic or
anthropogenic radionuclides like $^{40}$K, $^{60}$Co and $^{137}$Cs
are typically evaluated by analyzing the most intense gamma lines of
different isotopes\footnote{Typically, the gamma lines analyzed are
1001.0, 295.2, 351.9, 609.3, 1120.3 and 1764.5~keV for $^{238}$U,
338.3, 911.0, 969.0, 238.6, 727.2, 583.2, 860.6 and 2614.5~keV for
$^{232}$Th, 143.8 and 185.7~keV for $^{235}$U, 1460.8~keV for
$^{40}$K, 1173.2 and 1332.5~keV for $^{60}$Co and 661.7~keV for
$^{137}$Cs.}; upper limits are derived if the gross signal does not
statistically differ from the background
signal~\cite{Currie:1968lac,Hurtgen200045}. The detection efficiency
is determined by Monte Carlo simulations based on
Geant4~\cite{Agostinelli:2002hh} for each sample, accounting for
intrinsic efficiency, the geometric factor and self-absorption of
the sample. The simulation environment has been validated by
comparing the measured efficiency curve with a $^{152}$Eu reference
source of known activity (having relevant gamma emissions from 121.8
to 1408.0~keV) with the simulated one \cite{Aznar:2013jwa}; although
agreement between data and simulation is at or below 10\% for all
the gamma lines, a conservative overall uncertainty of 30\% is
considered for the deduced efficiency and properly propagated to the
final results to account for the limited reproduction of samples in
simulation.

A wide range of materials and components related to Micromegas
readout planes and the whole set-up of TREX-DM has been screened,
like the gas vessel, the field cage, the radiation shielding or the
electronic acquisition system. Massive elements and those in contact
with the sensitive volume of the detector are in principle the most
relevant. In the following, the screened samples will be described
and the results presented. The activity values obtained are
summarized in Table~\ref{tab:resrad}; reported errors include both
statistical and efficiency uncertainties.

\footnotesize \onecolumn
\begin{landscape}
\begin{table*}[p]
\centering
\caption{Results obtained for the activity of the natural
chains and some common radioactive isotopes in components and materials intended to be used at the TREX-DM setup.
Values reported for
$^{238}$U and $^{232}$Th correspond to the upper part of the chains
and those of $^{226}$Ra and $^{228}$Th give activities of the lower
parts. Reported errors correspond to $1\sigma$ uncertainties and
upper limits are given at 95\%~CL.}
\label{tab:resrad}
\begin{tabular}{llllllllllll}
\hline
\# & Material,Supplier & Technique & Unit & $^{238}$U & $^{226}$Ra & $^{232}$Th & $^{228}$Th & $^{235}$U & $^{40}$K & $^{60}$Co & $^{137}$Cs\\
\hline
1 & Pb, Mifer  & GDMS &  mBq/kg & $<$1.2 && $<0$.41 &&&  0.31 &&\\
2 & Pb, Mifer &  GDMS & mBq/kg & 0.33 &&0.10&&& 1.2&&\\
3 & Cu, Sanmetal &GDMS &mBq/kg &$<$0.062&&$<$0.020&&&&&\\
4 & Cu, hot rolled, Luvata &  GDMS &mBq/kg& $<$0.012&& $<$0.0041&&& 0.061&&\\
5 & Cu, cold rolled, Luvata  & GDMS &mBq/kg &$<$0.012&& $<$0.0041 &&& 0.091&& \\
6 & Cu, Luvata & Ge &  mBq/kg& &$<$7.4 &  $<$0.8 & $<$4.3 && $<$18 & $<$0.8 & $<$1.2 \\ \hline
7 & Kapton-Cu, LabCircuits & Ge & $\mu$Bq/cm$^{2}$ & $<$160 & $<$14 & $<$12 & $<$8  & $<$2 & $<$40 & $<$2 & $<$2  \\
8 & Teflon, Sanmetal & Ge &  mBq/kg  &$<$157 &   $<$4.1 &   $<$6.6 & $<$4.8 &$<$4.8 &$<$19 &$<$1.2 &   $<$1.4 \\
9 & Teflon tube, RS & Ge & mBq/kg & $<$943&    $<$21& $<$37& $<$31& $<$19& 510$\pm$170 &$<$7.6 &$<$8.0 \\
10 & Stycast, Henkel & Ge & mBq/kg& (3.7$\pm$1.4)10$^{3}$ & 52$\pm$10& 44$\pm$12 & 38$\pm$9  & & (0.32$\pm$0.11)10$^{3}$ &$<$5.5& $<$6.5 \\
11 & Epoxy Hysol, Henkel & Ge & mBq/kg & $<$273 & $<$16 &$<$20& $<$16 & & $<$83& $<$4.2& $<$4.5 \\
12 & SMD resistor, Farnell  & Ge & mBq/pc& 2.3$\pm$1.0 & 0.16$\pm$0.03 & 0.30$\pm$0.06 & 0.30$\pm$0.05  & $<$0.05 & 0.19$\pm$0.08 & $<$0.02 & $<$0.03 \\
13 & SM5D resistor, Finechem   & Ge & mBq/pc& 0.4$\pm$0.2 & 0.022$\pm$0.007 & $<$0.023 & $<$0.016  & 0.012$\pm$0.005 & 0.17$\pm$0.07& $<$0.005 & $<$0.005 \\
14 & CF40 flange, Pfeiffer & Ge & mBq/kg & & 14.3$\pm$2.8 & 9.7$\pm$2.3 & 16.2$\pm$3.9 & 3.2$\pm$1.1 & $<$17 & 11.3$\pm$2.7 & $<$1.6 \\ \hline
15 & Connectors, Samtec  & Ge & mBq/pc  &$<$77& 9.2$\pm$1.1& 19.6$\pm$3.6& 18.5$\pm$2.2 &1.5$\pm$0.4 &12.2$\pm$4.1 &   $<$0.6 &   $<$1.3 \\
16 & Connectors, Panasonic & Ge & mBq/pc & $<$42 & 6.0$\pm$0.9 & 9.5$\pm$1.7 & 9.4$\pm$1.4 & $<$0.95 & 4.1$\pm$1.5 & $<$0.2 & $<$0.8\\
17 & Connectors, Fujipoly & Ge & mBq/pc & $<$25 & 4.45$\pm$0.65 & 1.15$\pm$0.35 & 0.80$\pm$0.19 & & 7.3$\pm$2.6 & $<$0.1 & $<$0.4\\
18 & Flat cable, Somacis & Ge & mBq/pc & $<$370 & 101$\pm$13 & 165$\pm$29 & 164$\pm$23 & & 80$\pm$25 & $<$5 & $<$15 \\
19 & Flat cable (rigid), Somacis & Ge & mBq/pc & $<$1.5 10$^{3}$ & 123$\pm$17 & 225$\pm$40 & 198$\pm$29 & & 112$\pm$40 & $<$5.8 & $<$20 \\
20 & Flat cable (flexible), Somacis & Ge & mBq/pc & $<$102 & $<$3.8 & $<$4.0 & $<$1.4 & $<$1.8 & $<$15 & $<$0.7 & $<$0.8 \\
21 & Flat cable, Somacis & Ge & mBq/pc & $<$45 & $<$1.7 & $<$1.8 & $<$0.61 & $<$0.77 & $<$6.6 & $<$0.3 & $<$0.3 \\
22 & Flat cable, Somacis & Ge & mBq/pc & $<$14 & 0.44$\pm$0.12 & $<$0.33 & $<$0.19 & $<$0.19 & 1.8$\pm$0.7 & $<$0.09 & $<$0.10 \\
23 & RG58 cable, Pro-Power & Ge & mBq/kg & (2.2$\pm$0.9)10$^{3}$ & (0.9$\pm$0.1)10$^{3}$ & 40$\pm$12 & 29$\pm$8 & $<$212 & 108$\pm$43 & $<$9.2 & $<$8.9 \\
24 & Teflon cable, Druflon & Ge & mBq/kg & $<$104 & $<$2.2 & $<$3.7 & $<$ 1.7 & $<$1.4 & 21.6$\pm$7.4 & $<$0.7 & $<$0.8 \\
25 & Teflon cable, Axon & Ge & mBq/kg & $<$650 & $<$24 & $<$15 & $<$9.9 & $<$7.9 & 163$\pm$55 & $<$4.3 & $<$5.1 \\
26 & Kapton tape, Tesa & Ge & mBq/kg & $<$1.7 10$^{3}$ & $<$34 & $<$40 & $<$22 & $<$14 & (0.46$\pm$0.15)10$^{3}$ & $<$10 & $<$10 \\ \hline
27 & FR4 PCB, Somacis & Ge & Bq/kg & 31$\pm$11 & 15.3$\pm$2.1 & 25.5$\pm$4.4 & 22.5$\pm$3.5 & & 15.5$\pm$4.7 & $<$0.16 & \\
28 & PTFE circuit, LabCircuits &  Ge & Bq/kg  & $<$36& 4.7$\pm$0.6 &5.0$\pm$1.1 &6.2$\pm$0.9& $<$0.50 &  4.5$\pm$1.5 &$<$0.16&   $<$0.54 \\
29 & Cuflon, Crane Polyflon & Ge & mBq/kg & $<$103 & $<$3.7 & $<$3.6 & $<$1.4 & $<$1.8 & $<$13 & $<$0.6 & $<$0.7 \\
30 & Classical Micromegas, CAST & Ge & $\mu$Bq/cm$^{2}$ & $<$40 &&4.6$\pm$1.6&& $<$6.2& $<$46 & $<$3.1& \\
31 & Microbulk Micromegas,CAST & Ge & $\mu$Bq/cm$^{2}$ & 26$\pm$14&& $<$9.3&& $<$14 &57$\pm$25& $<$3.1& \\
32 & Kapton-Cu foil, CERN & Ge & $\mu$Bq/cm$^{2}$ & $<$11&& $<$4.6 &&$<$3.1& $<$7.7 &$<$1.6&\\
33 & Cu-kapton-Cu foil, CERN & Ge & $\mu$Bq/cm$^{2}$ & $<$11 && $<$4.6&& $<$3.1& $<$7.7& $<$1.6& \\
34 & Pyralux, Saclay & Ge & $\mu$Bq/cm$^{2}$& $<$19& $<$0.61&$<$0.63& $<$0.72& $<$0.19& 4.6$\pm$1.9& $<$0.10 &$<$0.14 \\
\hline
\end{tabular}
\end{table*}
\end{landscape}
\normalsize \twocolumn

\subsection{Shielding and vessel}
Lead and copper are commonly used to reduce the external gamma
background in passive shielding. Several metal samples from
different suppliers were analyzed by GDMS and activities were
obtained from the measured U, Th and K
concentrations~\cite{Aznar:2013jwa}. Lead samples from the Spanish
company Mifer for two different raw materials were considered (\#1-2
of Table~\ref{tab:resrad})\footnote{GDMS is not sensitive to the
$^{210}$Pb content which is typically used to qualify bricks as low
activity lead.}.

Copper is also used for mechanical and electric components: vessel,
central cathode, Micromegas readout planes, HV feedthroughs or rings in TREX-DM.
Three copper samples with different origins were studied (\#3-5 of Table~\ref{tab:resrad}).
One is ETP (C11000) copper supplied by Sanmetal while the other two were made of OFE (C10100)
copper\footnote{Purity guaranteed at 99.99\%.} from Luvata, having different production
mechanism (hot versus cold rolling). A Luvata copper sample with
681~g was screened with the germanium detector as well (\#6 of table
\ref{tab:resrad}); the upper bounds on activities derived from this
germanium spectrometry measurement were much less stringent than
those from GDMS due to its limited sensitivity. Although the GDMS
measurement of the Luvata copper has given information only on U and Th
concentration, the upper limits derived are at the same level or
even better than the germanium spectrometry results for the NOSV
copper from the Norddeutsche Affinerie (re-branded as Aurubis)
\cite{Laubenstein2004167}. At TREX-DM, Luvata copper was used for the plates while
the other copper components were made of ETP copper from Sanmetal.

\subsection{Field cage}
Materials and components to be used inside the vessel, mainly
related to the field cage have been screened~\cite{Aznar:2013jwa}.

The monolayer PCB made of kapton and copper used at the field cage,
supplied by LabCircuits, was screened finding good radiopurity with
upper limits from a few to tens of $\mu$Bq/cm$^{2}$ (\#7 of
Table~\ref{tab:resrad}); a sample with a surface of 260.15~cm$^{2}$
was considered. A cylinder of teflon (945~g) supplied by Sanmetal
was measured and an acceptable radiopurity was found deriving upper
limits at the level of mBq/kg (\#8 of Table~\ref{tab:resrad}). This
material is very widely used, due to its physical, mechanical,
dielectric and optical properties. All teflon components at TREX-DM
are from this supplier.

A tube supplied by RS and used in the calibration system described above, to move radioactive
sources in and out of the detector, was measured (\#9 of Table~\ref{tab:resrad}).
It was made of 1-mm-thick teflon and had a diameter of 1~cm;
the mass of the sample was 91~g. The high content of $^{40}$K at a level of one half of Bq/kg advised against its final use and other tubes are being analyzed.

The radiopurity of two types of adhesives to be used to glue
kapton elements was analyzed. One was Stycast 2850 FT (a two
component, thermally conductive epoxy encapsulant) used
with the catalyst 24LV, both from Henkel.
The other was Hysol RE2039 (an epoxy resin also from Henkel
having exceptional resistance to impact and thermal shock) used together with the hardener HD3561.
Massive samples of 551~g for Stycast and 245~g for Hysol were prepared
following the provider specifications. Results are quoted in rows \#10-11 of Table~\ref{tab:resrad}.
High activities of tenths or even a few Bq/kg were measured for $^{40}$K and $^{238}$U for
the stycast sample, which prevents its use, while for the Hysol
epoxy no contaminant could be quantified and therefore it has been used at TREX-DM.
It is worth noting that soldering has been completely avoided inside the vessel.

Resistors are used in the TREX-DM field cage. Surface Mount Device (SMD) resistors supplied by Farnell
(50 pieces) and by Finechem (100 pieces) were screened (\#12-13 of Table~\ref{tab:resrad}). Activity values obtained for
Finechem resistors are up to one order of magnitude lower than for the Farnell ones for some isotopes.
For this reason, Finechem resistors were used at TREX-DM.

Radiopurity information for the CF40 flange in the vessel for pumping was also obtained.
The screened piece, from Pfeiffer, was made of 304L stainless steel having a mass of 347~g.
The activity from the radioactive chains and $^{60}$Co was quantified (\#14 of Table~\ref{tab:resrad}).

\subsection{Electronics}
\label{sec:radiopurityelectronics}
First results of the screening of
different components related to the acquisition system of TREX-DM
(some connectors and circuits) were already presented
in~\cite{Aznar:2013jwa}. More components have been recently analyzed
and results are detailed here.

Various types of connectors have been screened. Narrow pitch
connectors for board-to-board from the Panasonic P5K series and other
ones supplied by Samtec were initially considered
\cite{Aznar:2013jwa} (\#15-16 of Table~\ref{tab:resrad}). The number of
pieces in the samples was 15~(0.67~g/pc) for Panasonic connectors
and 10~(2.2~g/pc) for the Samtec ones. Both types show activities
of several mBq/pc for isotopes in $^{232}$Th and the lower
part of $^{238}$U chains and for $^{40}$K, as found also in
\cite{Alvarez:2014kvs} for similar connectors. All of them are made of
Liquid Crystal Polymer (LCP), thus the activity measured is attributed
to this material. As it will be shown in Sec.~\ref{sec:backmodel},
this activity at connectors would dominate the background level,
and therefore this kind of connectors must be avoided or properly shielded.
Five connectors made of silicone (Fujipoly Gold 8000 connectors type C, 1.14~g/pc)
were also screened, having lower activity of $^{226}$Ra and
specially of $^{232}$Th (\#17 of Table~\ref{tab:resrad});
its use in TREX-DM is foreseen in the future.

Very radiopure, flexible, flat cables made of kapton and copper have been developed in collaboration with Somacis,
performing a careful selection of the materials included. Three different designs of flat cables have been screened;
their dimensions and masses are indicated in Table~\ref{tab:limandas} together with the number of units screened in each case.
In the first design, the cables consisted of a flexible band ended by two rigid boards and large activities were found at the
screening of several units (\#18 of Table~\ref{tab:resrad}); to investigate their origin,
one of the cables was cut and the flexible band and the two rigid heads were separately screened.
Only upper limits were set for the flexible part, while activities of the same order than for the whole cable were found
for the rigid heads (\#19-20 of Table~\ref{tab:resrad}) pointing to materials there
to be the main source of radioactive contamination.
The specific activities quantified for these cables are at the level of 10~Bq/kg,
typical of glass fiber; it seems that the glass fiber-reinforced
materials at base plates of circuit boards can be a source of radioactive contamination~\cite{Heusser:1995wd}.
Two cables produced with a second design as totally flexible cables were screened (\#21 of Table~\ref{tab:resrad}),
finding results compatible with those obtained for the flexible band of the first design
(upper limits are about a factor of 2 lower because two units were analyzed for the new design).
This measurement was useful to fix the allowed materials and procedures in the cables manufacture.
The screening of the final design to be used in TREX-DM was performed for 12 units and activities of $^{40}$K and $^{226}$Ra
were quantified, while upper limits were set for the other common radioisotopes (\#22 of Table~\ref{tab:resrad}).
The results are comparable with previous measurements
and the inclusion in the sample of a larger number of cable units, being in addition more massive,
has allowed to quantify some isotopes and to reduce the upper limits for the rest of nuclides.

\begin{table}
\centering
\caption{Main features of the samples of flat cables made of kapton and copper by Somacis
and screened by germanium spectrometry to analyze and improve their radiopurity.
Dimensions and masses given correspond to one cable unit.}
\label{tab:limandas}
\begin{tabular}{lccccc}
\hline
Design & Units & Length & Width & Thickness & Mass \\
& & (cm) & (cm) & (mm) & (g) \\
\hline
First & 8 &  &  & & 33.2 \\ 
Heads & 1 & 3.3 & 6 & 1.65  & 14.0  \\ 
Band & 1 & 50.5 & 5 & $\sim$0.4 & 19.6 \\ 
Second & 2 & 57.5 & 5-6 & $\sim$0.4 & 24.7 \\ 
Final & 12 & 57.6 & 6.4 & $\sim$0.7 & 54.6 \\ \hline 
\end{tabular}
\end{table}

Several kinds of high voltage and signal cables have been analyzed.
A sample of coaxial cable RG58BU with jacket made of black PVC from
Pro-Power was screened; it was 20-m-long having a mass of 723.4~g.
Large activities, at the level of Bq/kg for $^{238}$U, were found
(\#23 of Table~\ref{tab:resrad}). A sample of the cable AWG 18/19/30
$\times$ 10.0~CR from Druflon Electronics was also measured (\#24 of
table \ref{tab:resrad}). It has Silver Plated Copper wires (19
wires, diameter of 0.225~mm each) with a teflon jacket with outer
diameter of 0.25~inches; the sample was 10.65-m-long with a mass of
780.4~g. This Druflon cable is used to connect the field cage last
ring to HV feedthrough. A sample of the coaxial low noise cable SML
50 SCA from Axon Cable S.A.S. was screened too (\#25 of
Table~\ref{tab:resrad}). The conductor is made of Silver Plated
Copper Alloy, the dielectric of extruded PTFE, the screen of Silver
Plated Copper and the outer sheath of taped PTFE. The length and
mass of the sample were 43.76~m and 125.4~g and the cable has 1.1~mm
as maximum diameter. The Axon cable is used to extract the mesh
signal from the vessel. Only $^{40}$K activity was quantified for
these two cables made basically of copper and teflon. Although the
values of the activity per mass are lower for the Druflon cable, the
Axon cable has a better radiopurity per unit length.

Finally, the kapton tape (Tesa 52408-00008-00) used throughout the set-up
was screened. It is reported to have a polyimide backing with a
silicone adhesive. The sample, with a mass of 49.1~g, was 33-m-long,
19-mm-wide and 65-$\mu$m-thick; only $^{40}$K activity was
quantified (\#26 of Table~\ref{tab:resrad}).

\subsection{Micromegas readout planes}
\label{sec:RadMicro}
Different options can be taken into consideration for PCBs as base
material for a Micromegas detector. A 187.4-g sample of the PCB at
the Micromegas produced by Somacis and used for the moment in
TREX-DM was screened; it is made of FR4/phenolic for core and
pre-impregnated reinforced fabric together with copper and resin.
Very large specific activities of tens of Bq/kg were found for the
common radioisotopes (\#27 of Table~\ref{tab:resrad}); as mentioned
before, this was expected for glass-fiber reinforced
materials~\cite{Heusser:1995wd}. In addition, FR4 should be
disregarded not only because of high radioactivity, but also for an
unacceptable high rate of outgassing. Kapton (or cirlex) and PTFE
are in principle radiopure, as shown in the screening of the PCB
made of kapton and copper supplied by LabCircuits and used at the
field cage (\#7 of Table~\ref{tab:resrad}). However, a 49-g circuit
made of ceramic-filled PTFE composite also from LabCircuits (\#28 of
Table~\ref{tab:resrad}) presented very high activities of Bq/kg for
the natural chains and $^{40}$K, precluding its use. Good
radiopurity was found for cuflon samples from Crane Polyflon,
setting upper limits at the mBq/kg level (\#29 of
Table~\ref{tab:resrad}); a sample taken from a 1.57-mm-thick panel,
made of PTFE sandwiched by two 35 $\mu$m-thick copper sheets, and
with a mass of 705.9~g was screened. However, the use of cuflon for
Micromegas has been disregarded due to the difficulty to fix the
mesh and also because bonding films to prepare multilayer PCBs have
been shown to have unacceptable activity~\cite{Alvarez:2014kvs}.

The radiopurity of Micromegas readout planes (without base material)
was first analyzed in depth in~\cite{Cebrian:2010ta}. Main results
obtained in this work are reproduced here for the sake of
completeness. On the one hand, two samples (\#30-31 of
Table~\ref{tab:resrad}) were part of fully functional microbulk
Micromegas readouts: a full microbulk readout plane formerly used in
the CAST experiment and a classical Micromegas structure without
mesh. Both of them had a diameter of 11 cm.  The second sample
represents an earlier stage in the manufacturing process than the
full microbulk structure of the first sample, in which chemical
baths have been applied to etch the kapton pillars and the mesh
structure. On the other hand, two more samples (\#32-33 of
Table~\ref{tab:resrad}) were screened corresponding just to the raw
foils used in the fabrication of microbulk readouts, consisting of
kapton metalized with copper on one or both sides. Several circular
wafers of the same diameter as the real detectors (11 cm) were
considered in this case. The raw materials (kapton and copper,
mainly) were confirmed to be very radiopure, since no contamination
was quantified. Altogether, the obtained results proved that
Micromegas readouts of the microbulk type are manufactured with
radiopurity levels comparable to the cleanest detector components in
low background experiments.

A new activity measurement for the Cu-kapton-Cu foil was carried out
profiting the great capabilities of the BiPo detector~\cite{Gomez:2013dka}
operating at the LSC.
It is a large planar detector developed to measure mainly the SuperNEMO double beta source foils
with sensitivity to few $\mu$Bq/kg of $^{208}$Tl and $^{214}$Bi
(two isotopes produced in the decays of the natural chains of $^{232}$Th and $^{238}$U),
thus surpassing by almost two orders of magnitude the sensitivity of standard gamma spectroscopy.
Preliminary results~\cite{Gomez:2015hg} have shown that the activities of both isotopes
in the Cu-kapton-Cu foil are near the detector's sensitivity, i.e, $\sim$0.1~$\mu$Bq/cm$^2$.

Together with the kapton and copper foils, a stainless steel mesh
and pyralux are included in the Micromegas produced by Somacis and IRFU/SEDI for
TREX-DM. The screening of the stainless steel mesh is scheduled for
the very next future and pyralux has been already analyzed. Pyralux is
used in the construction of bulk Micromegas
~\cite{Giomataris:2004aa}; it is a photoresistive film placed
between the anode plane and the mesh, subsequently etched to produce
the pillars. A sample of pyralux sheets with a total surface of
4800~cm$^{2}$ and a mass of 65~g from Saclay was screened (\#34 of
Table~\ref{tab:resrad}); only $^{40}$K was quantified and upper
limits were set for all the other common radioisotopes.

Following all these results, a microbulk version of the TREX-DM readout planes
will be built for a physics run at LSC.
This new readout is described in Sec.~\ref{sec:con}.
Apart from that, other readouts based on bulk techniques will be built too,
following the fabrication techniques used for flat cables made of copper and
kapton (see Sec.~\ref{sec:radiopurityelectronics}).


\section{Background model of TREX-DM at LSC}
\label{sec:backmodel}
As a required element to estimate the sensitivity of TREX-DM to low-mass WIMPs, we have created
a first background model of the experiment, as it were installed and in operation at the LSC.
This model is based on the screening program of all materials used
in the setup (described in Sec.~\ref{sec:radiopurity})
and the simulation of the detector response.
The section has been divided in four parts:
the simulation of the detector's response is described in Sec.~\ref{sec:BackDetResponse},
which is followed by a validation of this simulation with real data in Sec.~\ref{sec:ValidSimulation};
then the main contributions to the background model are detailed in Sec.~\ref{sec:BackContibutions};
an analysis based on x-ray cluster features is then proposed
and a first estimation of background levels is made in Sec.~\ref{sec:BackXraysAna}.

We have considered two light gas mixtures at 10 bar: Ar+2\%iC$_4$H$_{10}$ and Ne+2\%iC$_4$H$_{10}$,
which are good candidates to detect WIMPs of masses below 20 GeV
and give a total active mass of 0.300 and 0.160~kg respectively.
The background levels quoted in the following are referred
to a Range of Interest (RoI) of 2-7~keVee\footnote{Electron equivalent energy.},
which is equivalent to 5.2-16.3~keVnr\footnote{Nuclear equivalent energy.} for argon-based mixtures
and 5.5-17.1~keVnr for neon-based ones.
The upper limit is low enough to avoid the contribution of most of K-fluorescence lines of the surrounding materials,
while the lower one has been set to minimize the uncertainties
in the simulation of the detector's response and the analysis.
The calculated levels will be later used to assess the different contributions
to background model and to calculate the sensitivity of TREX-DM to low-mass WIMPs.
In this approach, the background spectrum is supposed to be flat at low keV energies.
Nevertheless, the final analysis of TREX-DM experiment should quantitatively describe
its background spectrum, as in other Dark Matter experiments~\cite{Cebrian:2012yda}.

\subsection{Simulation of the detector response}
\label{sec:BackDetResponse}
\begin{figure}[htb!]
\centering
\includegraphics[width=60mm]{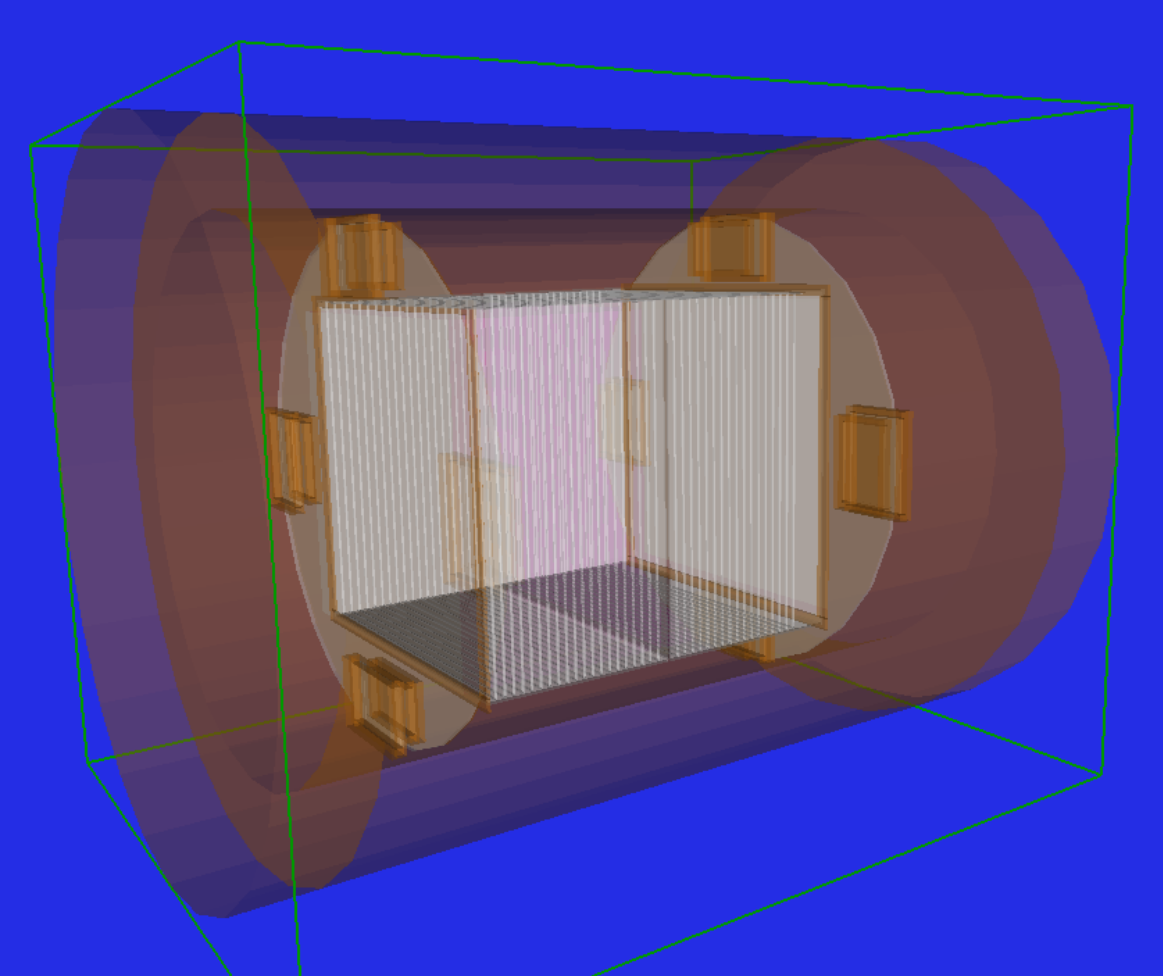}
\caption{A view of the TREX-DM geometry implemented in Geant4. The cylindrical copper vessel (orange volumes)
contains a circular base with four shielded boxes (dark gray surface with four yellow boxes), 
two active volumes (in light gray), the field cage and degrador (white walls) and a central cathode.}
\label{fig:Geant4}
\end{figure}

The simulation of the detector response can be divided into two blocks. The first one covers all the physical processes
involved in the passage of gamma-rays and charged particles through matter,
and is mainly based on the version 4.10 of Geant4~\cite{Agostinelli:2002hh}.
For this purpose, a model of TREX-DM set-up has been created, as shown in Fig.~\ref{fig:Geant4}.
It includes the gas, the cathode, the field-cage, the Micromegas readout planes, the support bases,
the connectors and their shielding pieces.
For computational reasons, some details like small screws or cables have been omitted
and some parts have been simplified.
For instance, each readout plane is a pile of material layers whose thickness match with the real ones.
The low energy models based on Livermore data libraries have been implemented
for interactions of alpha, beta and gamma particles.
These models are accurate for energies between 250~eV-100~GeV
and can be applied down to 100~eV with a reduced accuracy~\cite{Cirrone2010315}.
Apart from that, fluorescence, Auger electrons and atomic de-excitation initiated by other electromagnetic processes
have been explicitly included for energies over 100~eV~\cite{Francis20131}.
In the case of muons, we have only considered electromagnetic processes,
while for neutron-induced recoils, we have used the NeutronHP model.
To accelerate the simulation,
we have used the Decay0 code~\cite{Ponkratenko:2000um} as generator of initial events, instead of the Geant4 Radioactive
Decay Module. Decay0 generates the particles from the decay of radioactive nuclides of many known unstable isotopes.

The second block simulates all physical processes of a TPC:
the generation of electrons in the gas, the diffusion effects during the drift to the readout plane,
the charge amplification at the Micromegas readout and the generation of signals both at mesh and strips.
It is based on the REST program~\cite{Tomas:1557181}, with some minor changes due to the two-volumes geometry
and the AFTER-based electronics. The resulting data has the same format as the DAQ data,
so as both real and simulated data may be analyzed by the same routines.
We describe step by step the complete simulation chain:
\begin{itemize}
 \item \textbf{Primary electrons}: The number of electrons ($n_e$) generated by an energy deposit ($E$)
 follows a distribution, which is empirically described
 by the average energy needed to produce an electron-ion pair ($W$)
 and the Fano factor ($F$), which accounts for the primary charge fluctuations.
 With these parameters: $n_e = E / W$ and $\sigma^2 = F \times n_e$.
 The W-value is about 20-30~eV for noble gases and hydrocarbons
 (see Table~\ref{tab:GasPar} for those used in the simulation),
 while the Fano factor lies between 0.15-0.2, i.e., the distribution of $n_e$ is not Poissonian.
 For computational reasons, we have combined the primary charge fluctuations
 with the amplification ones in a later step.
 \item \textbf{Quenching factor}: For the specific case of nuclear recoils,
 we have considered the conservative parametrization~\cite{Smith:1988kw} given by
 \begin{equation}
 \label{eq:quench1}
	Q(E_R)=\frac{g(E_R)}{1+g(E_R)}
	\end{equation}
 \noindent where $E_R$ is the event energy expressed in keVee and
 the function $g(E_R)$ is parametrized in terms of the atom number ($Z$) and mass ($A$) as

 \begin{equation}\label{eq:quenchParam}
	g(E_R) \simeq 0.66 \left( \frac{Z^{5/18}}{A^{1/2}} \right)  E_R^{1/6}(\mbox{keV}) 
	\end{equation}

 This parametrization is more conservative than the Lindhard model for k=0.157~\cite{Lindhard:2010jl}
 and the value of $\sim$0.3 measured for scintillation light in liquid argon in~\cite{Creus:2015fqa} at 1~keVnr.
 \item \textbf{Diffusion effects}: Each primary electron is projected to the XY plane and the time line
 following two gaussian distributions, whose widths are calculated by the distance to the readouts and the gas parameters
 (drift velocity, longitudinal and transversal diffusion coefficients) generated by Magboltz~\cite{Biagi:1999nwa}.
 The gas parameters are detailed in Table~\ref{tab:GasPar}.
 \item \textbf{Charge amplification}: The Micromegas readout amplifies the primary charge but it also
 introduces a fluctuation due to the avalanche formation.
 This variation depends on the gas and on the readout~\cite{Schindler201078}.
 In this model, the avalanche fluctuations ($f$) have been combined
 with primary ones ($F$), so that the energy resolution (\% FWHM) follows the expression
 \begin{equation}
  R = 2.35 \ \sqrt{(F + f) \ \frac{E}{W} + \left(\frac{ENC}{g} \ \frac{E}{W}\right)^2 + \sigma^2_{surf}}
 \end{equation}
 where $g$ is the readout gain (or charge amplification), $ENC$ is the equivalent noise charge
 and $\sigma_{surf}$ accounts for surface fluctuations.
 In a first approximation, noise and surface effects have been discarded,
 so that the energy resolution scales with $1/\sqrt{E}$ from a reference value.
 This simplification may not be applied for surface fluctuations of the actual readouts as they represent a 10\%.
 However, surface fluctuations of future readouts should be a minimum factor 2 lower, as already shown in \cite{Aznar:2015iia}.

 About the gain and energy resolution, the values detailed in Table~\ref{tab:ValidParam} have been used
 in the validation.
 In the background model, a gain of $10^3$ and an energy resolution of 13\%~FWHM at 5.9~keV have been considered.
 This resolution is the best obtained by a fully equipped Micromegas readout plane~\cite{Aznar:2015iia}.
 \item \textbf{X-Y readout}: The detector readout is divided in pads, which are alternatively interconnected
 in X and Y directions. This specific feature is simulated dividing the amplified charge between X and Y planes by:
 $Q_X = d_Y / (d_X + d_Y) \times Q$ and $Q_Y = d_X / (d_X + d_Y) \times Q$, where $d_X$ and $d_Y$ are respectively 
 the minimum distance to pixels connected to X and Y directions.
 \item \textbf{Electronics response}: Each X and Y charge create a pulse, whose amplitude and widths are calculated
 considering the AFTER-based features~\cite{Baron:2008zza,Baron:2010zz}:
 a sampling time of 10~ns, a shaping time of 100~ns and a transfer function of 10~mV/fC.
 The electronics noise has been partially modellized by setting a strip energy threshold
 in the cluster analysis. Further details are given in Sec.~\ref{sec:ValidSimulation}.
\end{itemize}

\begin{table}
\centering
\caption{Summary of the gas parameters used in the simulation.
W-values of argon and neon have been used for its corresponding mixtures with isobutene,
disregarding the contribution of isobutane (iC$_4$H$_{10}$) due to its low concentration.
These values are based on measurements and have been extracted from \cite{Christophorou:1971}.
Velocity and diffusion coefficients have been calculated using Magboltz~\cite{Biagi:1999nwa}
and considering a reduced drift field of $100$~V/cm/bar.}
\begin{tabular*}{\columnwidth}{@{\extracolsep{\fill}}cccccc@{}}
\hline
Gas   & Pressure     & $W$  & Velocity  & \multicolumn{2}{c}{Diff. ($\mu$m cm$^{-0.5}$)}\\
      &  bar         & eV   & cm/$\mu$s & Long.  & Trans.\\
\hline
Ar+2\%iso &  2.0 & 26.3 & 3.33      & 298.5  & 494.6\\
          & 10.0 &   "  &  "        & 133.5  & 221.2\\
Ar+5\%iso &  1.2 &   "  & 3.45      & 364.0  & 450.0\\
Ne+2\%iso & 10.0 & 36.4 & 2.18      & 107.2  & 168.2\\
\hline
\end{tabular*}
\label{tab:GasPar}
\end{table}

\subsection{Validation of the simulation}
\label{sec:ValidSimulation}
The expected signals in TREX-DM are nuclear recoils with energies below 20~keV. These events will create short tracks
of a few microns length, which will then induce two compact group of active strips or \textit{clusters}
at both $XZ$ and $YZ$ directions. Their widths will be short and will be mainly defined by diffusion effects.
Only at higher energies (or at mbar pressures), the cluster features may be slightly different
for electrons and neutrons due to the longer tracks of the former~\cite{Billard:2012dy}.
For this reason, we have applied the analysis used in CAST Micromegas detectors~\cite{Aune:2013pna},
based on the cluster features of low energy x-rays,
to separate point-like events from complex topologies, that may be generated by high energy gammas or cosmic muons.
In the case of CAST detectors, a $^{55}$Fe source (5.9~keV) was used as a reference.
For TREX-DM detector, the K- and L-lines of a $^{109}$Cd source have been used instead.

In a first step, the cluster limits in $X$- and $Y$-direction are calculated
by looking for two consecutive strip with induced pulses greater than a strip threshold.
This threshold is related to the strip noise conditions and
has been set to the values of Table~\ref{tab:ValidParam} in the validation of the simulation
and to 0.05~keV in the background model.
The calculated limits remove the contribution of noisy strips to the calculation of event energy and cluster widths.

Once the cluster limits have been defined, the cluster width in each direction is calculated by
\begin{equation}
\sigma_a = \sqrt{\frac{\sum_j q_j \times (a_j - \overline{a})^2}{\sum_j q_j}}
\end{equation}

\noindent
where $a = X$ or $Y$, $q_j$ is the pulse integral, $a_j$ is the pulse spatial position (either in $X$ or $Y$),
$\overline{a}$ is the mean cluster position and the index $j$ runs over the set of event pulses
whose spatial position is in between the cluster limits previously set.
From these two variables, two cluster observables are defined:
the $XY$ width, $\sigma_{XY} = \sqrt{\sigma_X^2 + \sigma_Y^2}$, which is mainly determined by the event topology
and transversal diffusion;
and the width balance, $\Delta\sigma_{XY} = (\sigma_Y - \sigma_X) / (\sigma_X + \sigma_Y)$, which only depends on
energy as charge fluctuations between the two readout projections increase at low energy as less charge is shared.

The last observable is the width in $Z$ direction ($\sigma_Z$),
which is calculated using the pulses of both $XZ$ and $YZ$ planes as
\begin{equation}
\sigma_z = v_{drift} \times \sigma_t = v_{drift} \times \sqrt{\frac{\sum_j q_j \times (t_j - \overline{t})^2}{\sum_j q_j}}
\end{equation}

\noindent
where $v_{drift}$ is the electron drift velocity,
$q_j$ is the pulse integral, $t_j$ is the temporal position of the pulse maximum
and the index $j$ runs over the set of event pulses
whose spatial position is in between the cluster limits previously set.
This observable includes information both on the intrinsic event's topology and the longitudinal diffusion.

These observables have been used to validate the complete simulation chain,
by comparing their distributions to those of real data.
We have used two data-sets acquired by TREX-DM detector,
when it was irradiated by a $^{109}$Cd source (x-rays of 22.1 and 24.9~keV)
situated at a calibration point and the vessel was filled by two different gases: Ar+2\%iC$_4$H$_{10}$ at 2 bar
and Ar+5\%iC$_4$H$_{10}$ at 1.2 bar. The detector conditions are specified in Table~\ref{tab:ValidParam}.
In the detector geometry, a calibration tube of 1 mm-thickness has been implemented, but not the metallic source container.

\begin{table}
\centering
\caption{Detector conditions of the two-data sets used in the validation of the simulation chain.}
\begin{tabular*}{\columnwidth}{@{\extracolsep{\fill}}ccccc@{}}
\hline
Gas   & Pressure     & Gain  & Ener. res. & Strip thres.\\
      &  bar         &       & \% FWHM & keV\\
\hline
Ar+2\%iC$_4$H$_{10}$ &  2.0 & $10^3$ & 24.0 & 0.36\\
Ar+5\%iC$_4$H$_{10}$ &  1.2 & $10^3$ & 15.0 & 0.12\\
\hline
\end{tabular*}
\label{tab:ValidParam}
\end{table}

\begin{figure*}[htb!]
\centering
\includegraphics[width=0.48\textwidth]{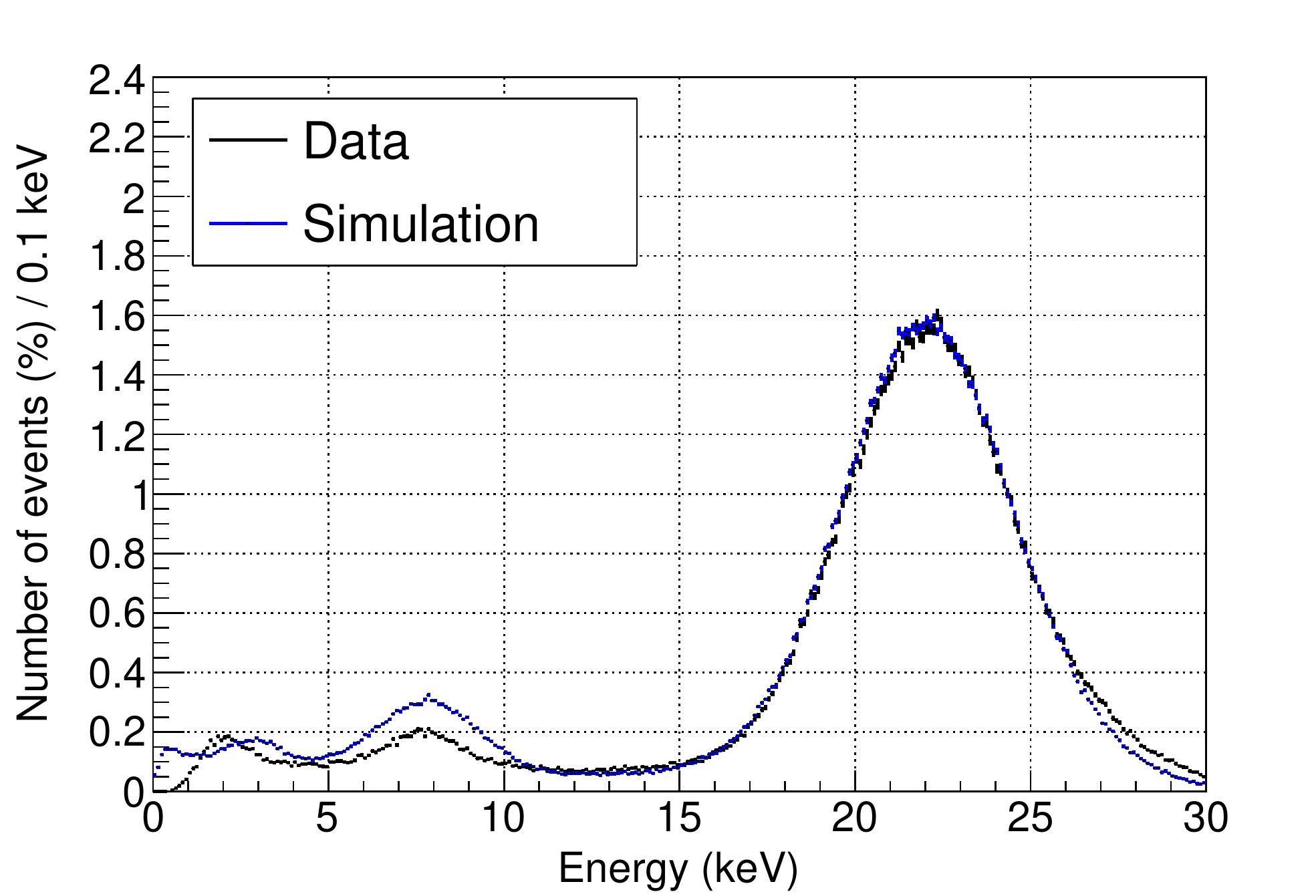}
\includegraphics[width=0.48\textwidth]{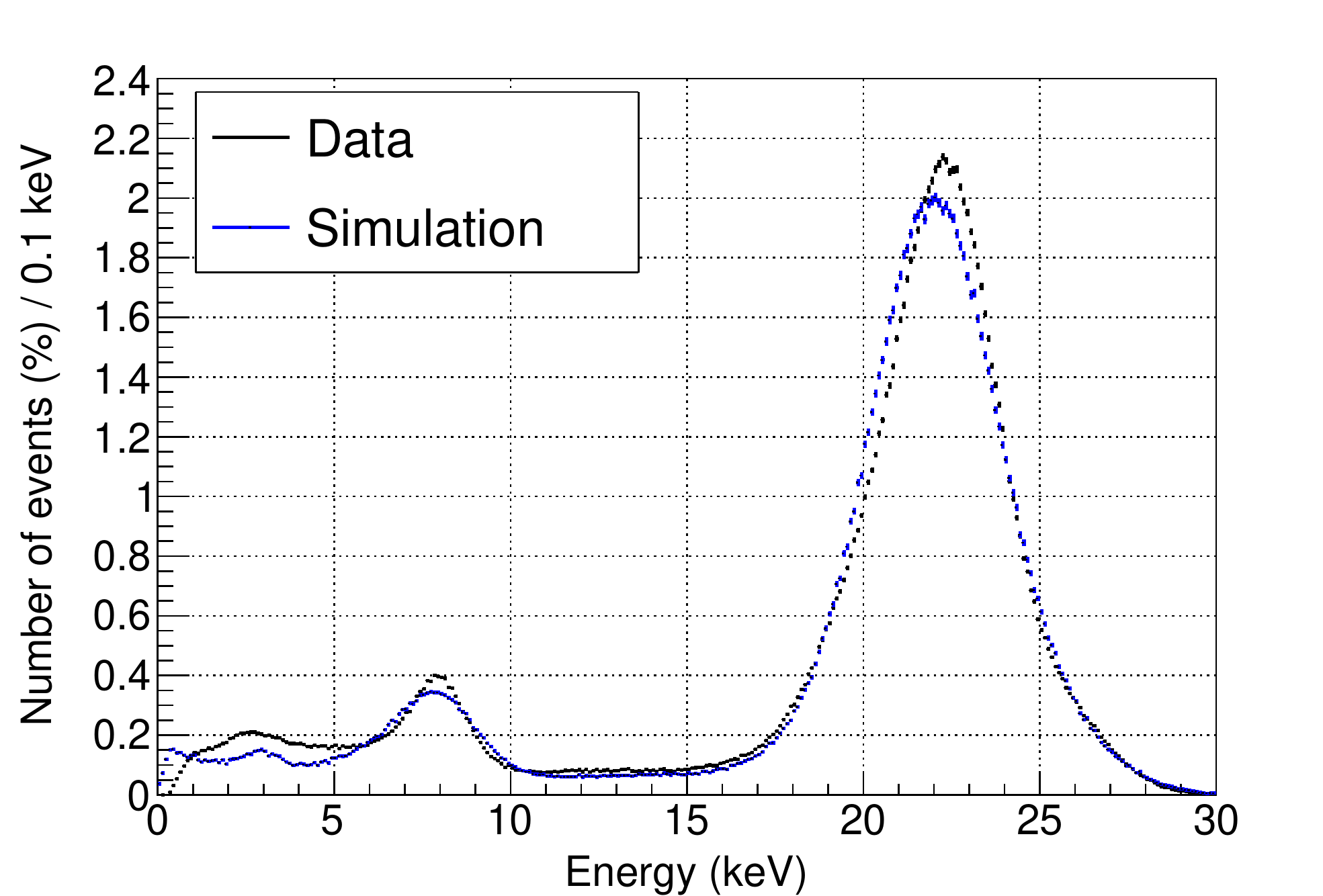}
\includegraphics[width=0.48\textwidth]{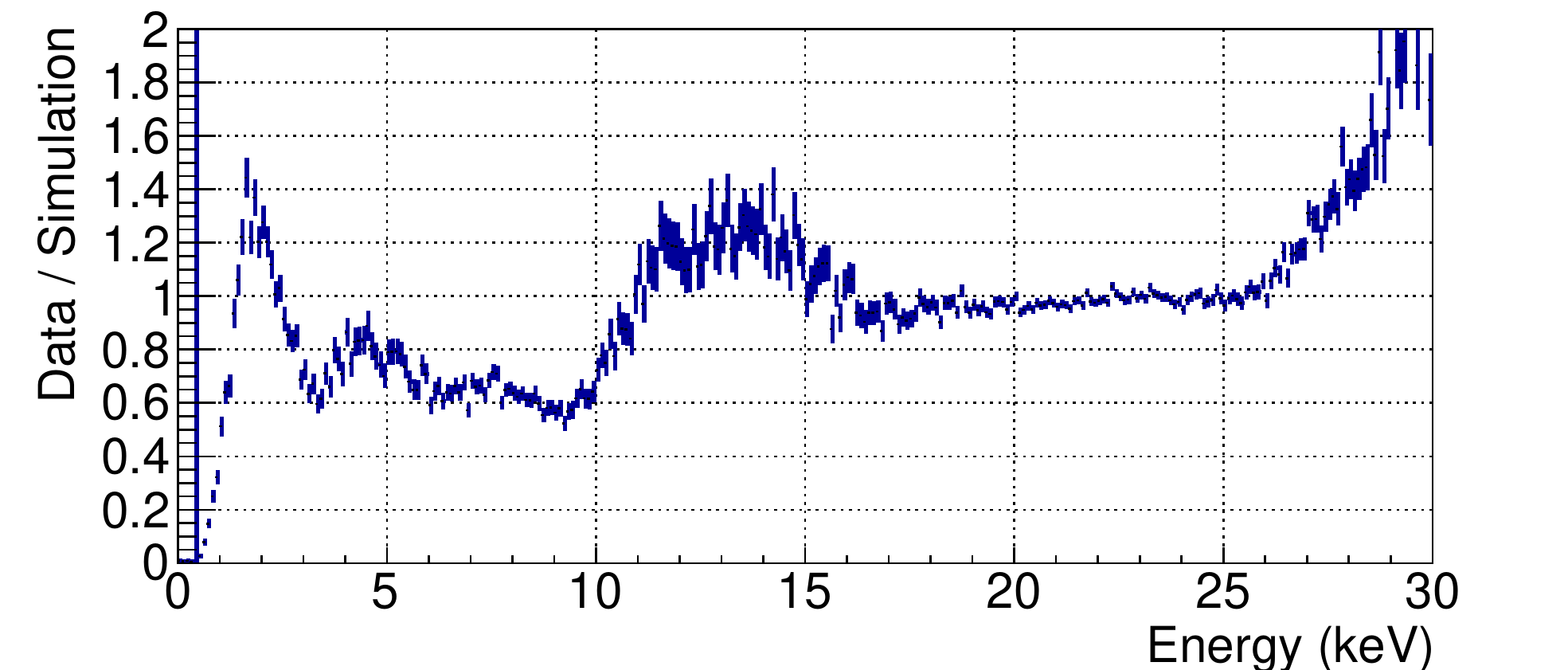}
\includegraphics[width=0.48\textwidth]{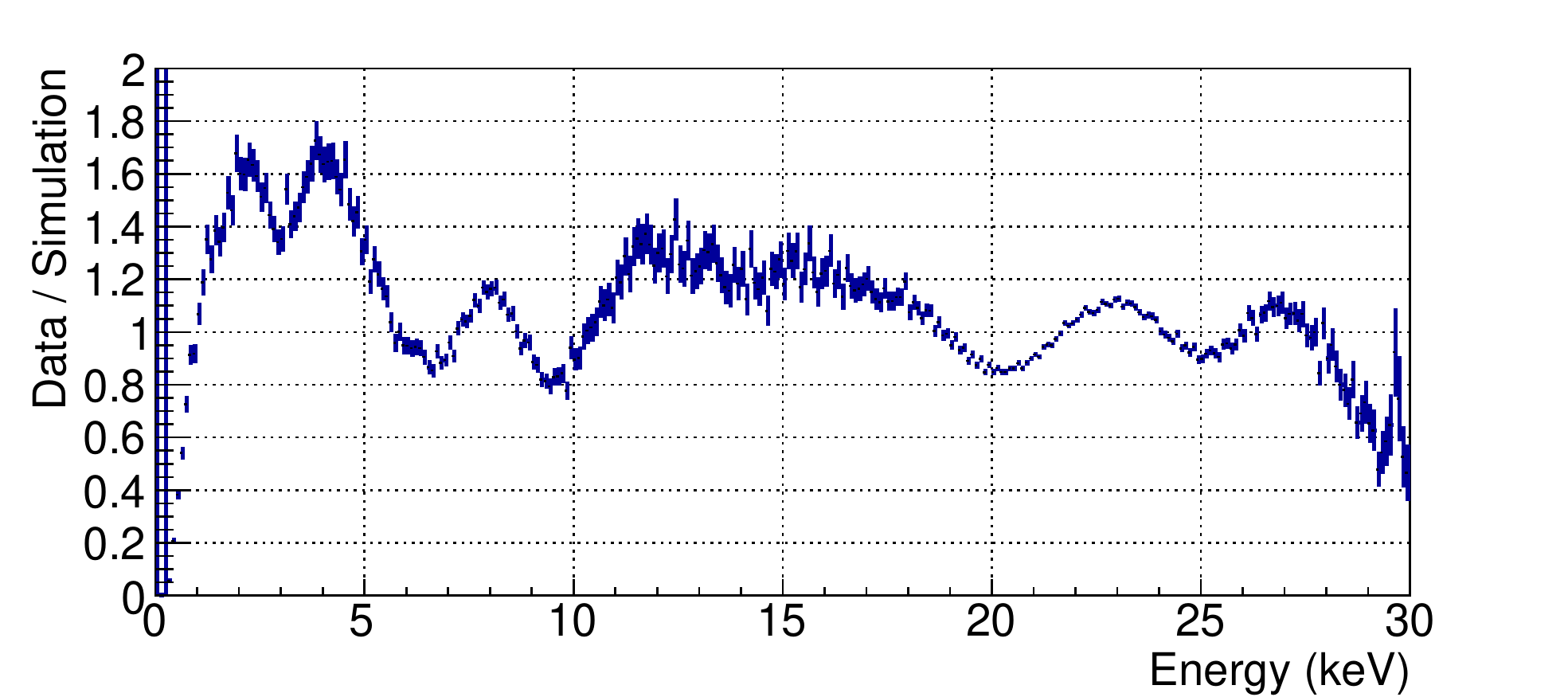}
\caption{Energy spectra of real data (black line) and Monte Carlo simulation (blue line)
generated by the strip signals when one of the active volumes is irradiated by a $^{109}$Cd source
situated at a calibration point. The vessel was filled with Ar+2\%iC$_4$H$_{10}$ at 2 bar (left)
or Ar+5\%iC$_4$H$_{10}$ at 1.2 bar (right).
All energy spectra have been normalized to the total number of events for the comparison.
The statistical error of each energy bin has been graphically represented by an error bar.
At both spectra, the K$_\alpha$ (22.1~keV) and K$_{\beta}$ (24.9~keV) x-ray lines
generated by the source are present, as well as their corresponding escape peaks,
located at 19.1 and 21.9~keV.
The argon, iron and copper K-fluorescences, respectively induced by the source at the argon gas,
the Micromegas readout planes and the central cathode, are also present at 3.0, 6.4 and 8.0~keV.}
\label{fig:EnergyValid}
\end{figure*}

The comparison between the real and the simulated energy spectra is made in Fig.~\ref{fig:EnergyValid}.
The level of agreement is reasonable: the Monte Carlo simulation reproduces
the energy and the intensities of x-ray lines of the $^{109}$Cd source;
the main differences appear at energies below 10~keV, up to 40\% in some cases,
where material fluorescence is important.
In the 5-10~keV range, these disagreement can be attributed to the simplified model of the readout planes,
which may affect the intensities of iron (6.4 keV) and copper (8.0 keV) fluorescence.
For energies below 5~keV, the divergences may be explained by some simplifications in the simulation,
like surface fluctuations and the noise level.

\begin{figure*}[htb!]
\centering
\includegraphics[width=55mm]{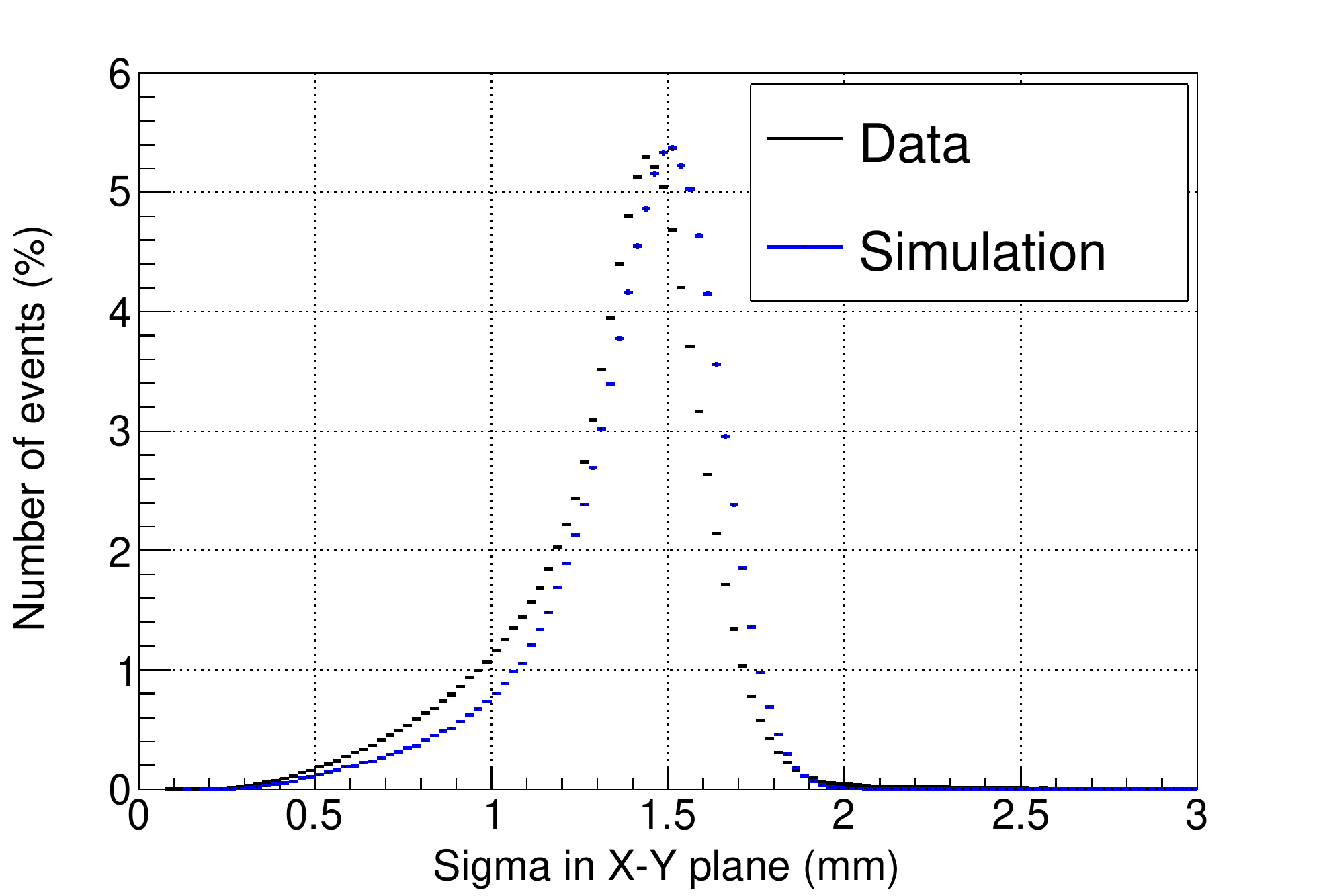}
\includegraphics[width=55mm]{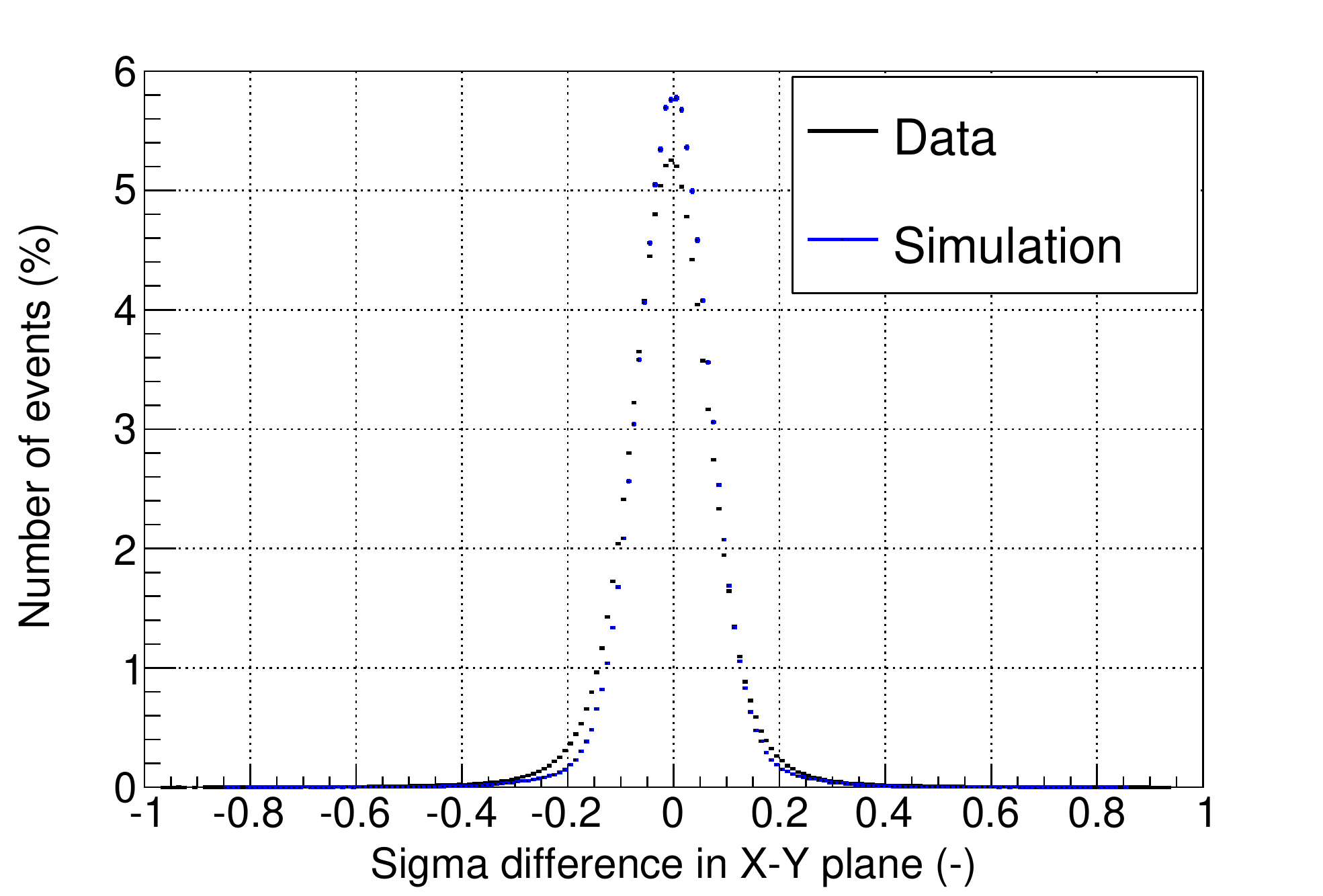}
\includegraphics[width=55mm]{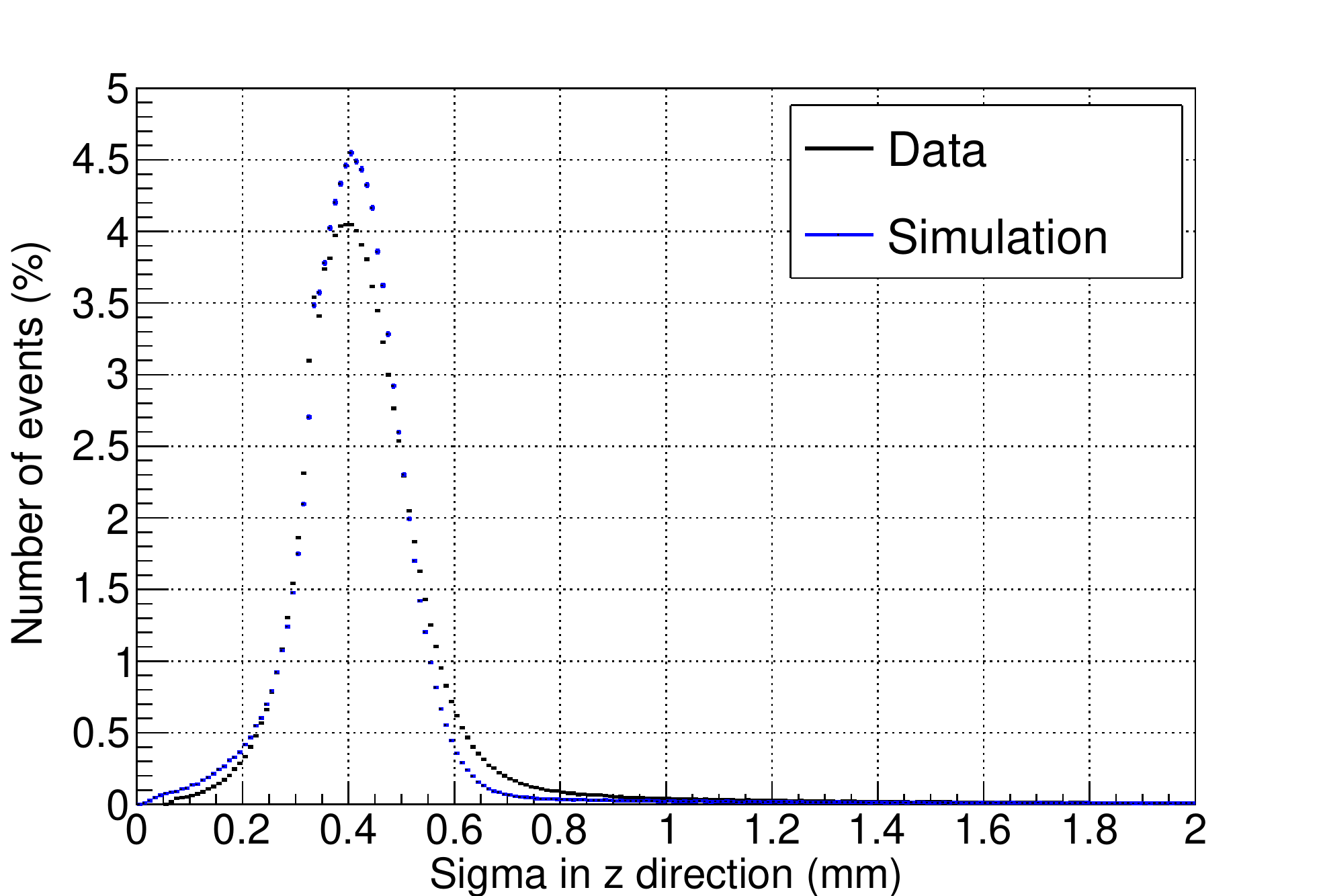}
\includegraphics[width=55mm]{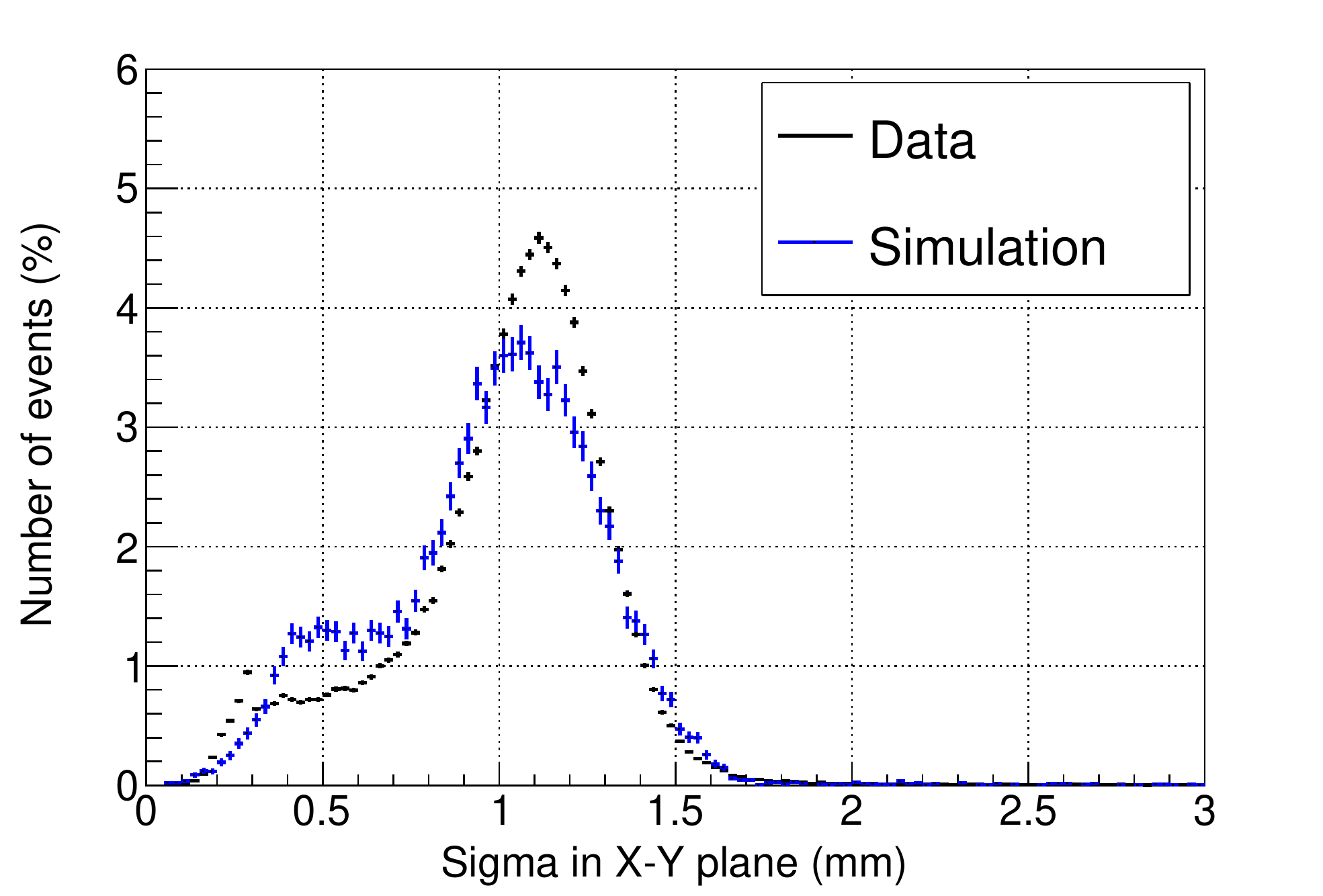}
\includegraphics[width=55mm]{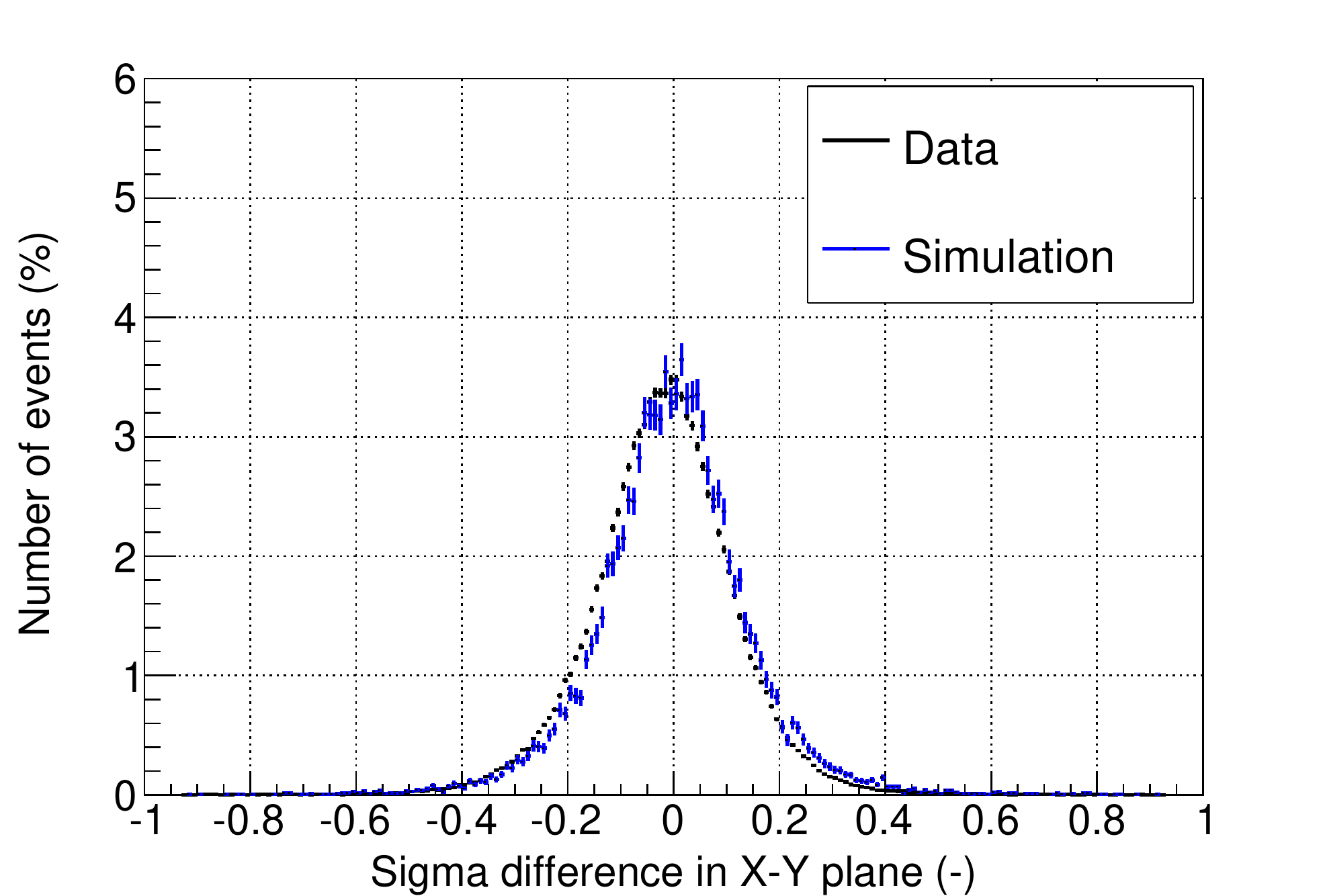}
\includegraphics[width=55mm]{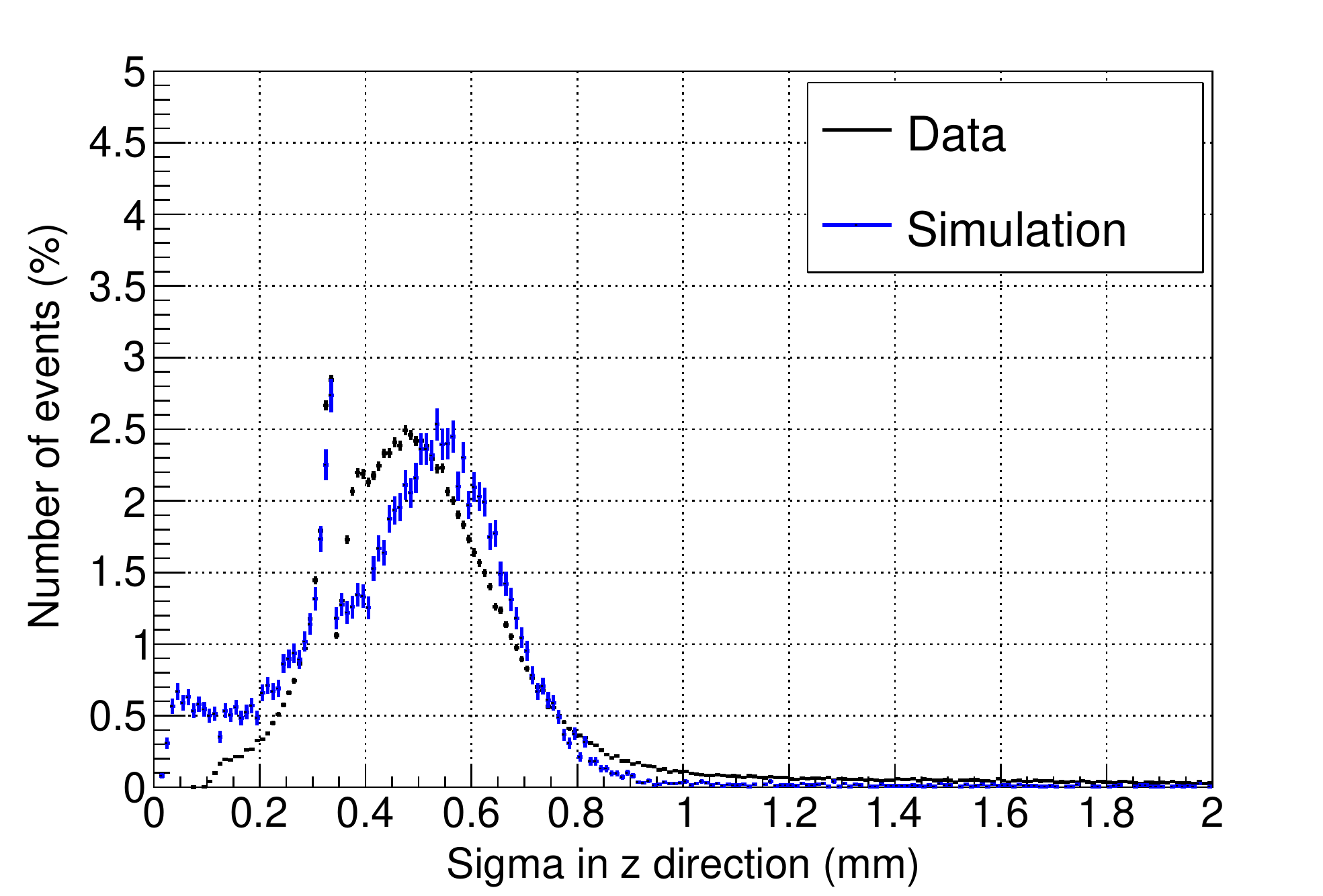}
\includegraphics[width=55mm]{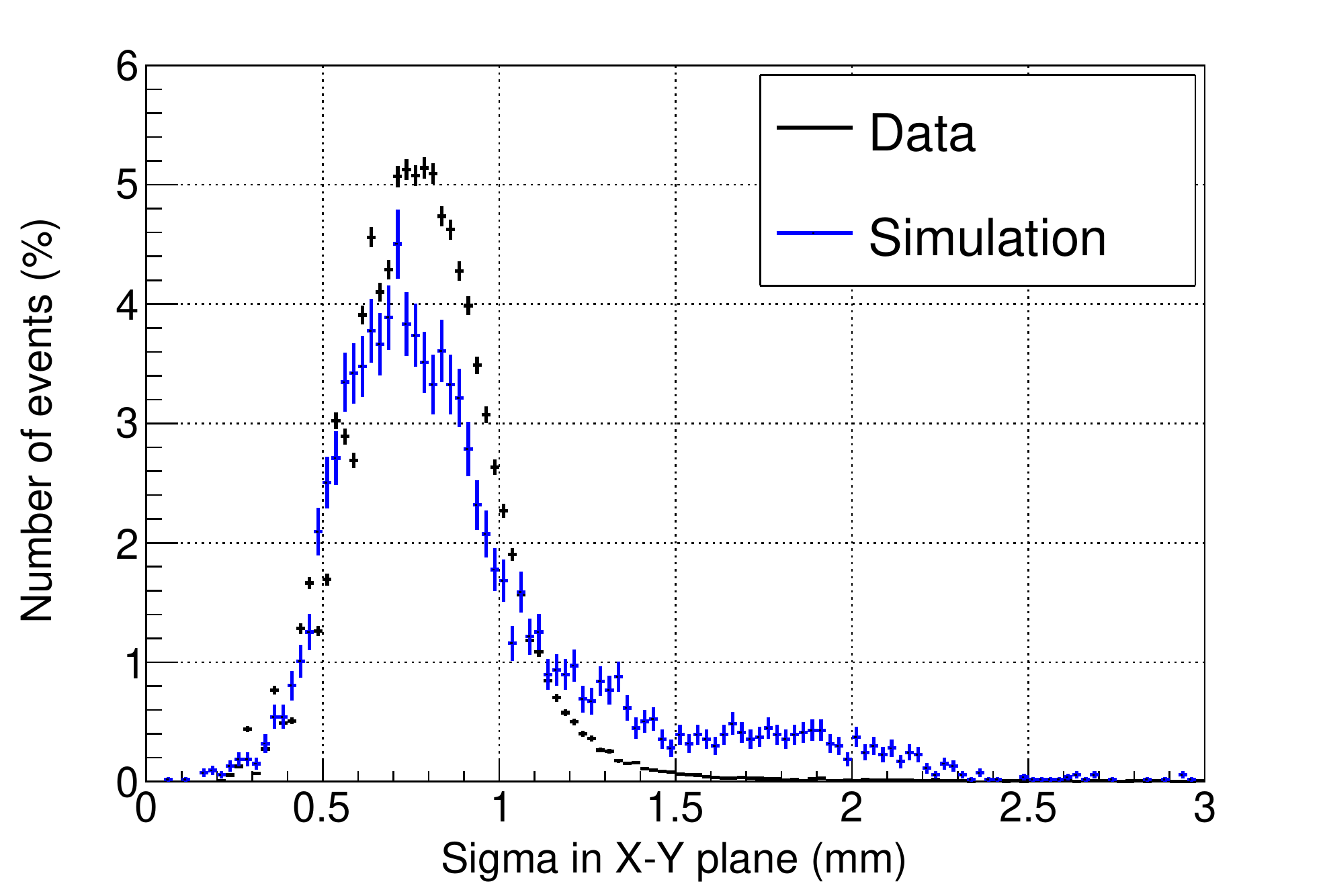}
\includegraphics[width=55mm]{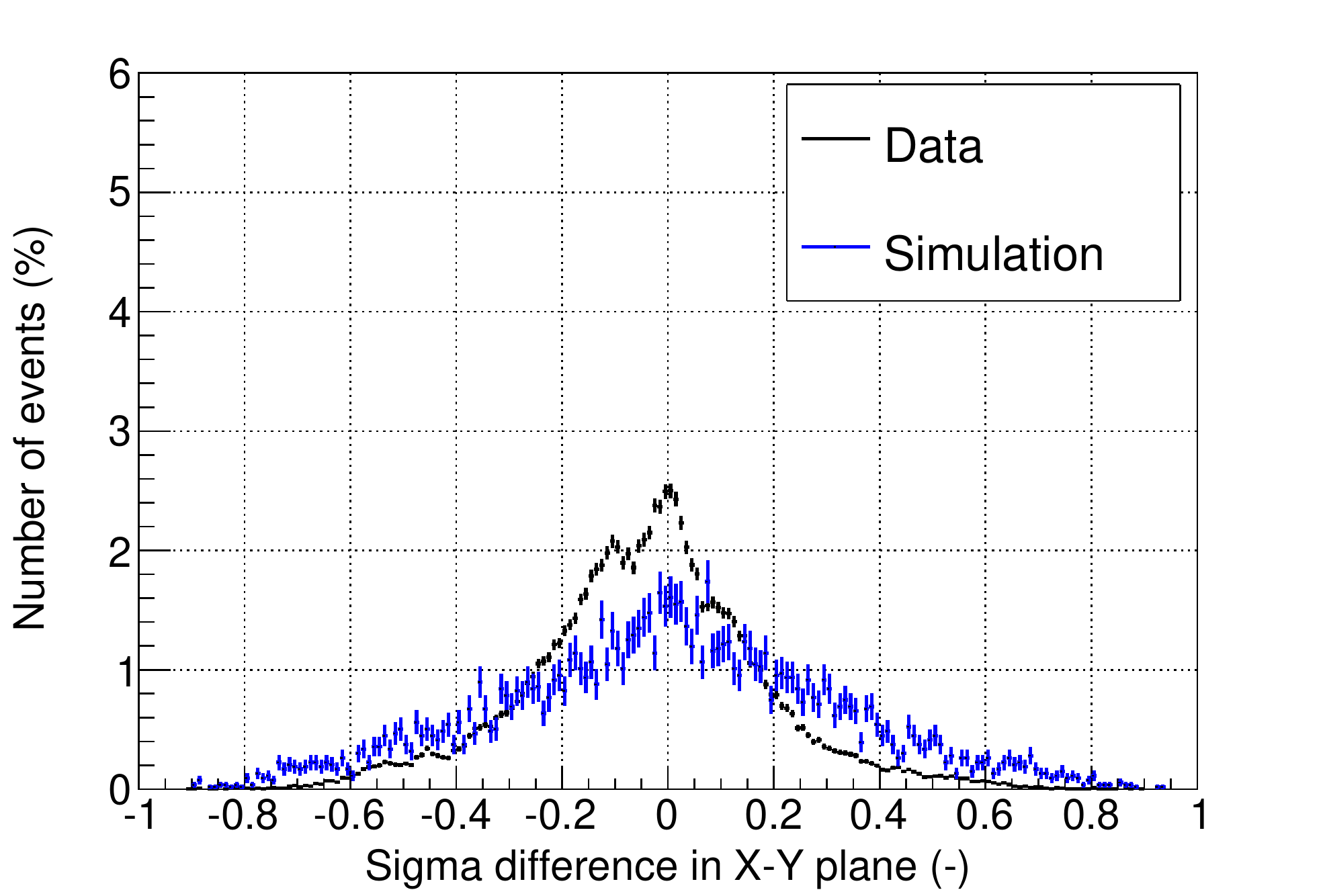}
\includegraphics[width=55mm]{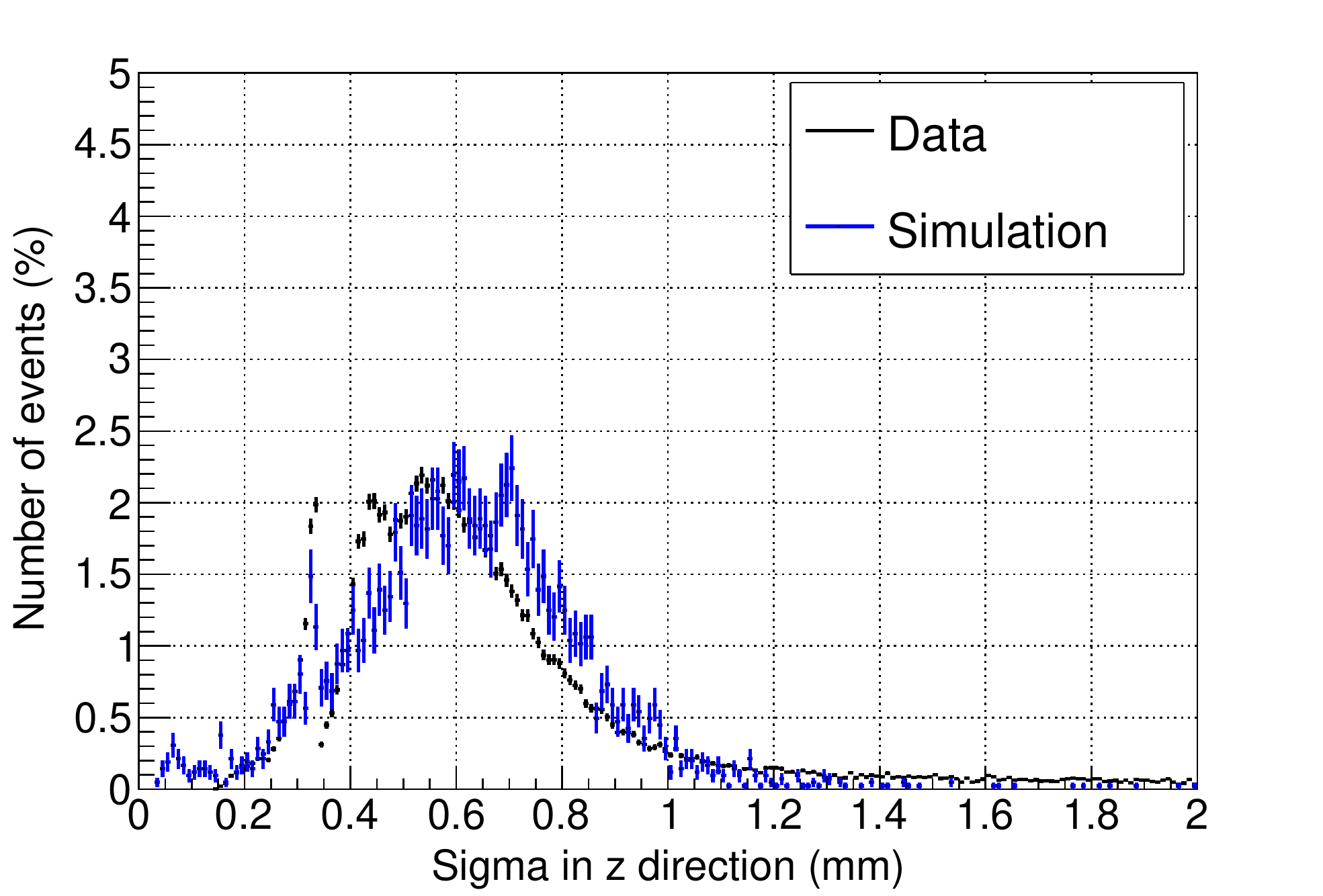}
\caption{Comparison between real data (black line) and Monte Carlo simulation (blue line)
for three analysis observables: the $XY$ width, $\sigma_{XY}$ (left);
the balance of cluster widths, $\Delta\sigma_{XY}$ (center); and the $Z$ width, $\sigma_Z$ (right);
and three energy ranges: 16-28 keV (top); 5-10 keV (center); and 2-4 keV (bottom).
Data was acquired when a $^{109}$Cd source was situated at a calibration point of TREX-DM
and the detector was filled with Ar+2\%iC$_4$H$_{10}$ at 2 bar.
The statistical error of each bin has been graphically represented by an error bar.}
\label{fig:ValidAr2iso}
\end{figure*}

\begin{figure*}[htb!]
\centering
\includegraphics[width=55mm]{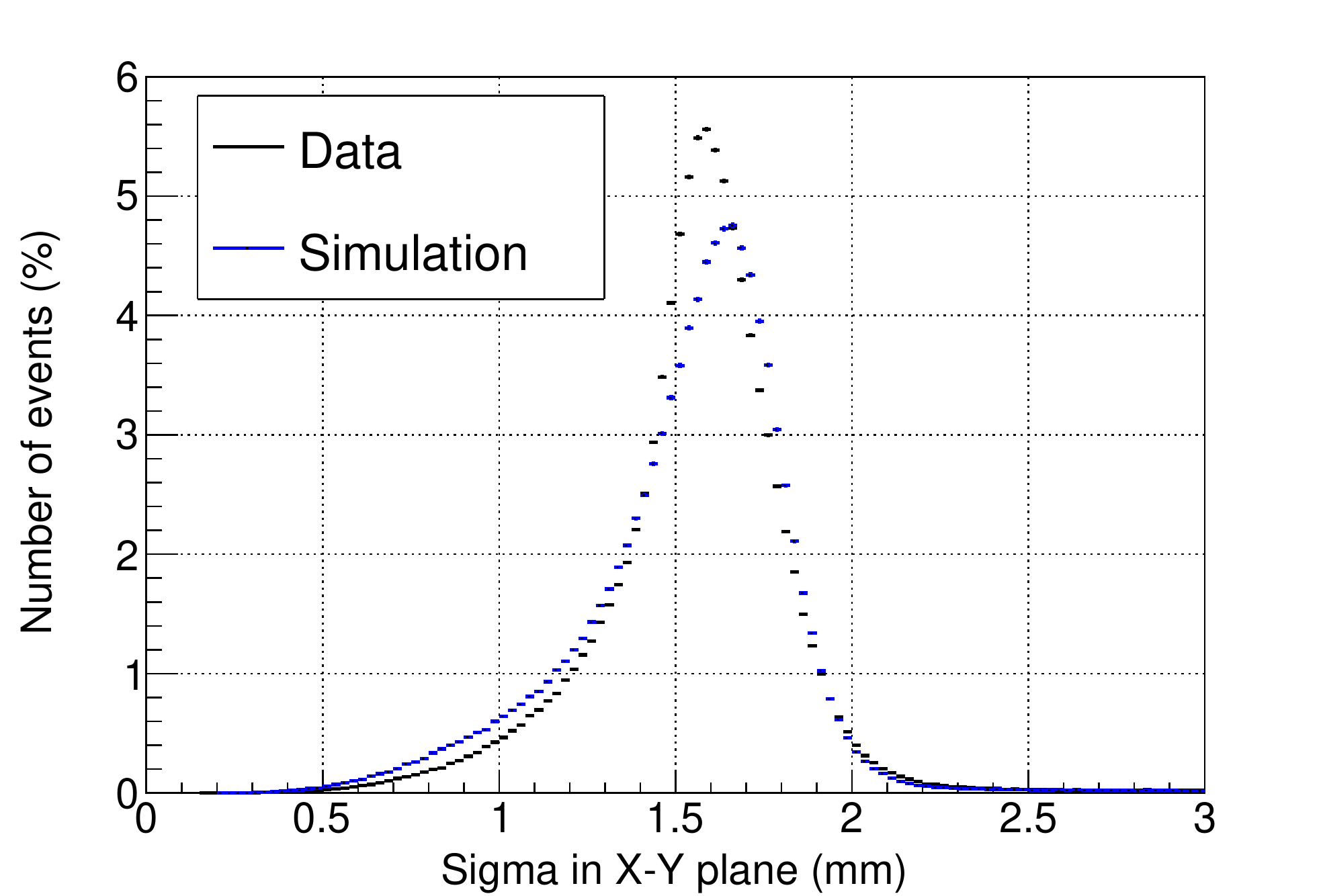}
\includegraphics[width=55mm]{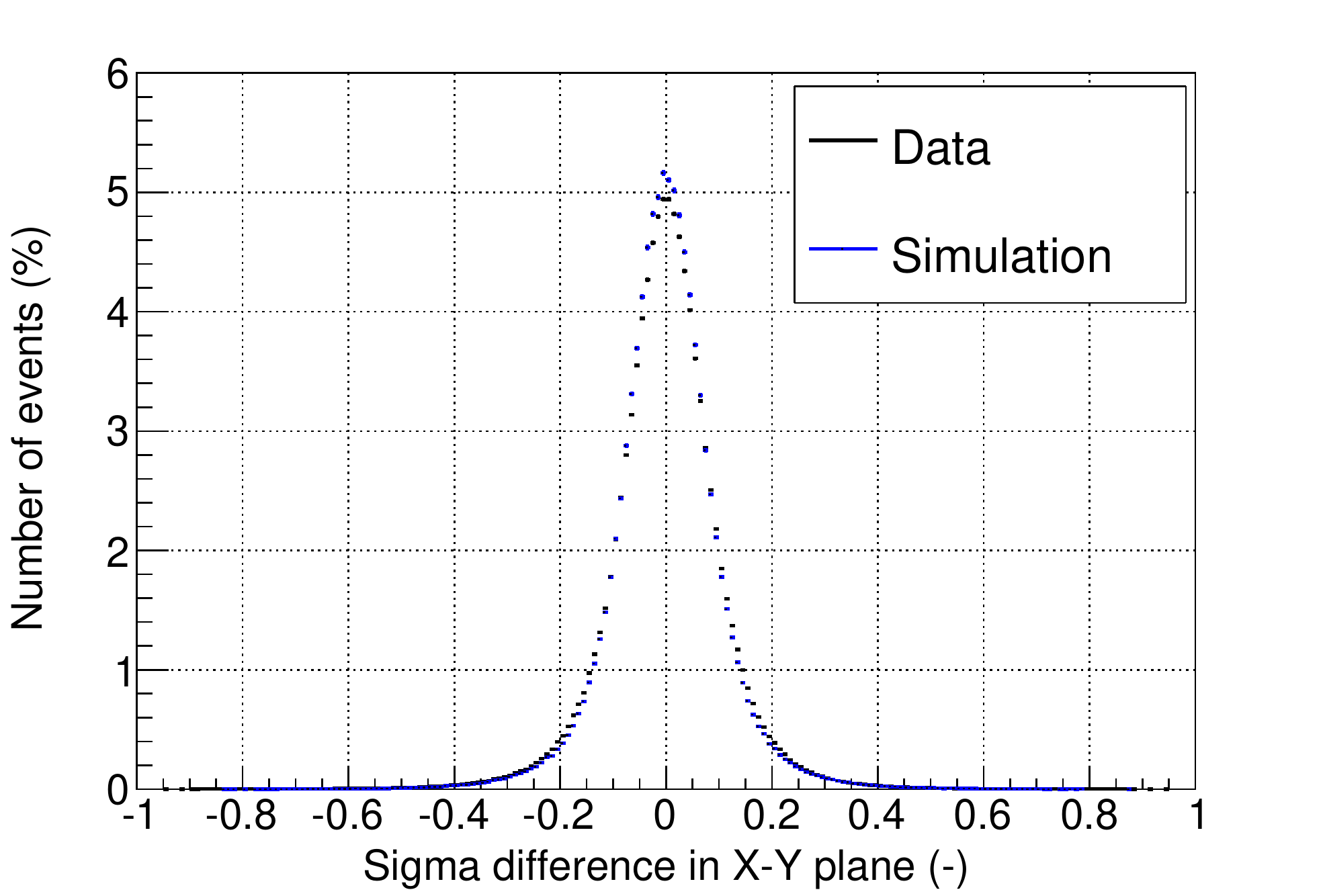}
\includegraphics[width=55mm]{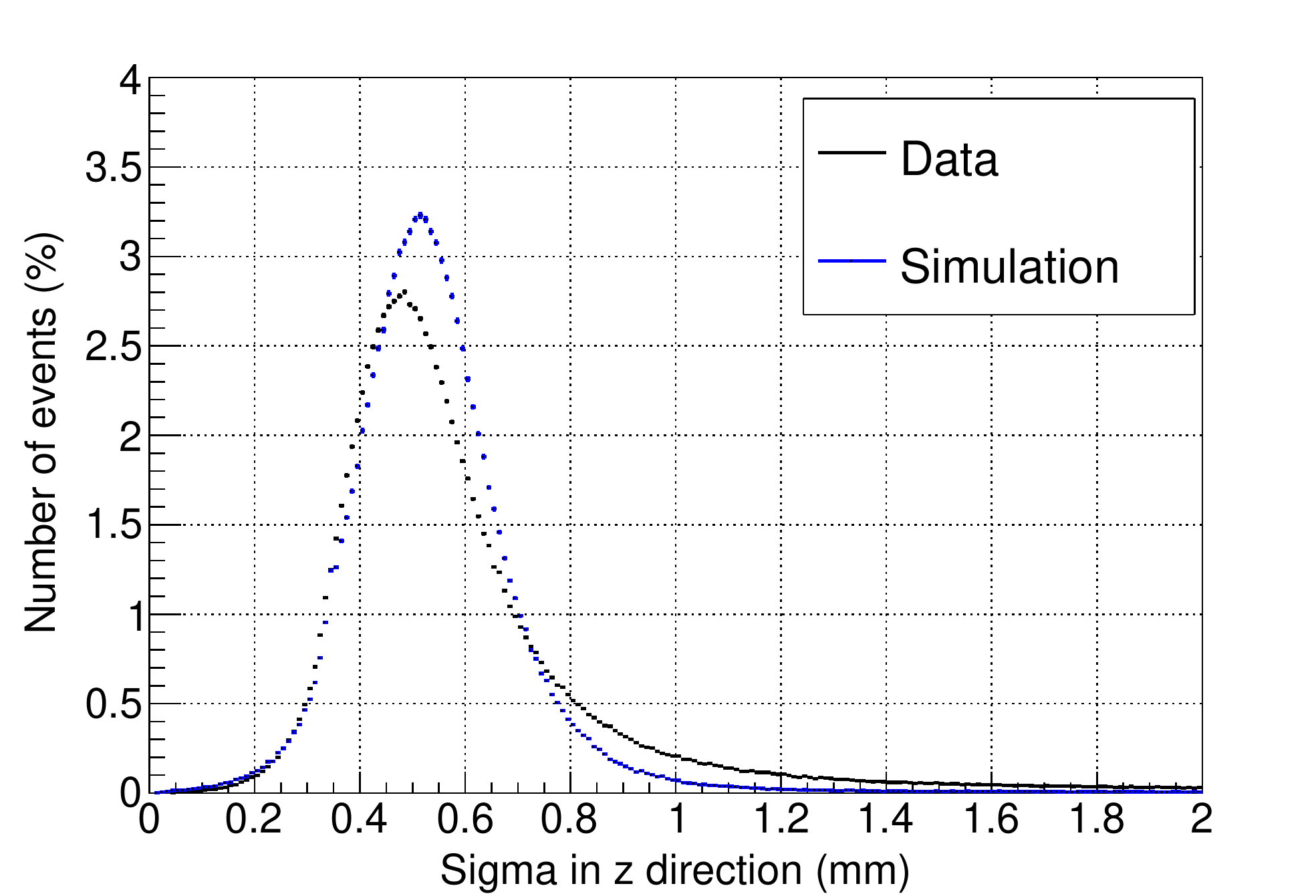}
\includegraphics[width=55mm]{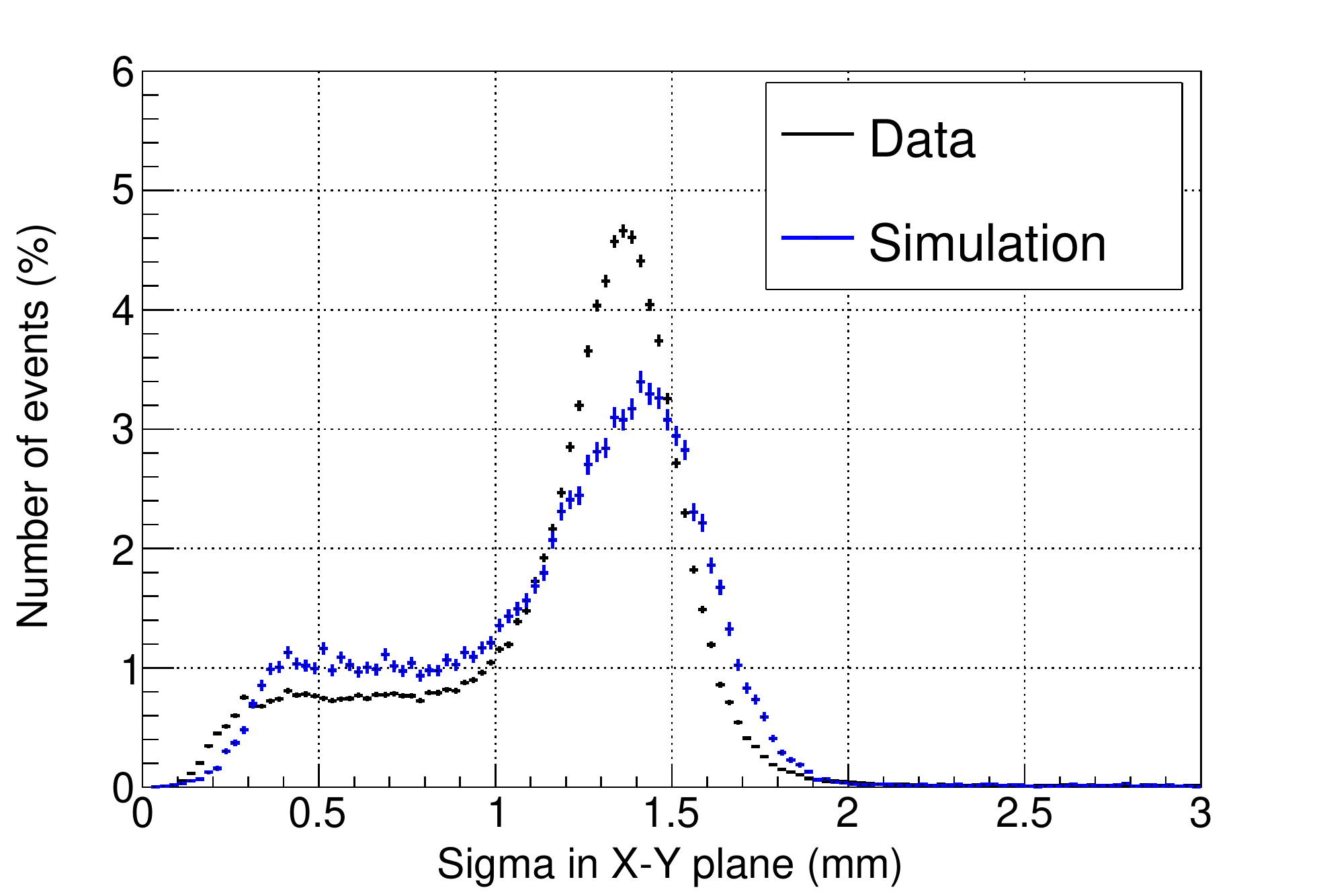}
\includegraphics[width=55mm]{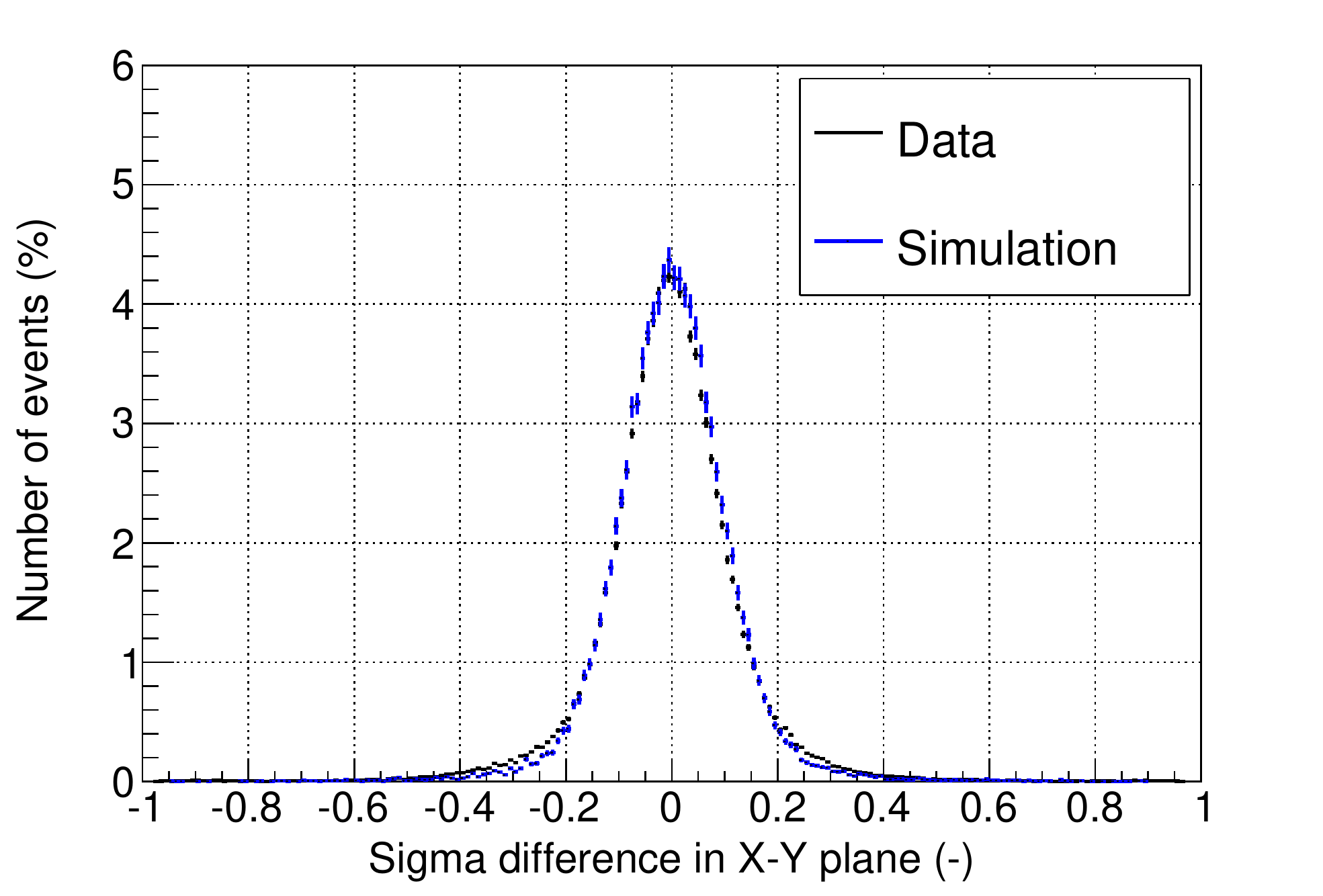}
\includegraphics[width=55mm]{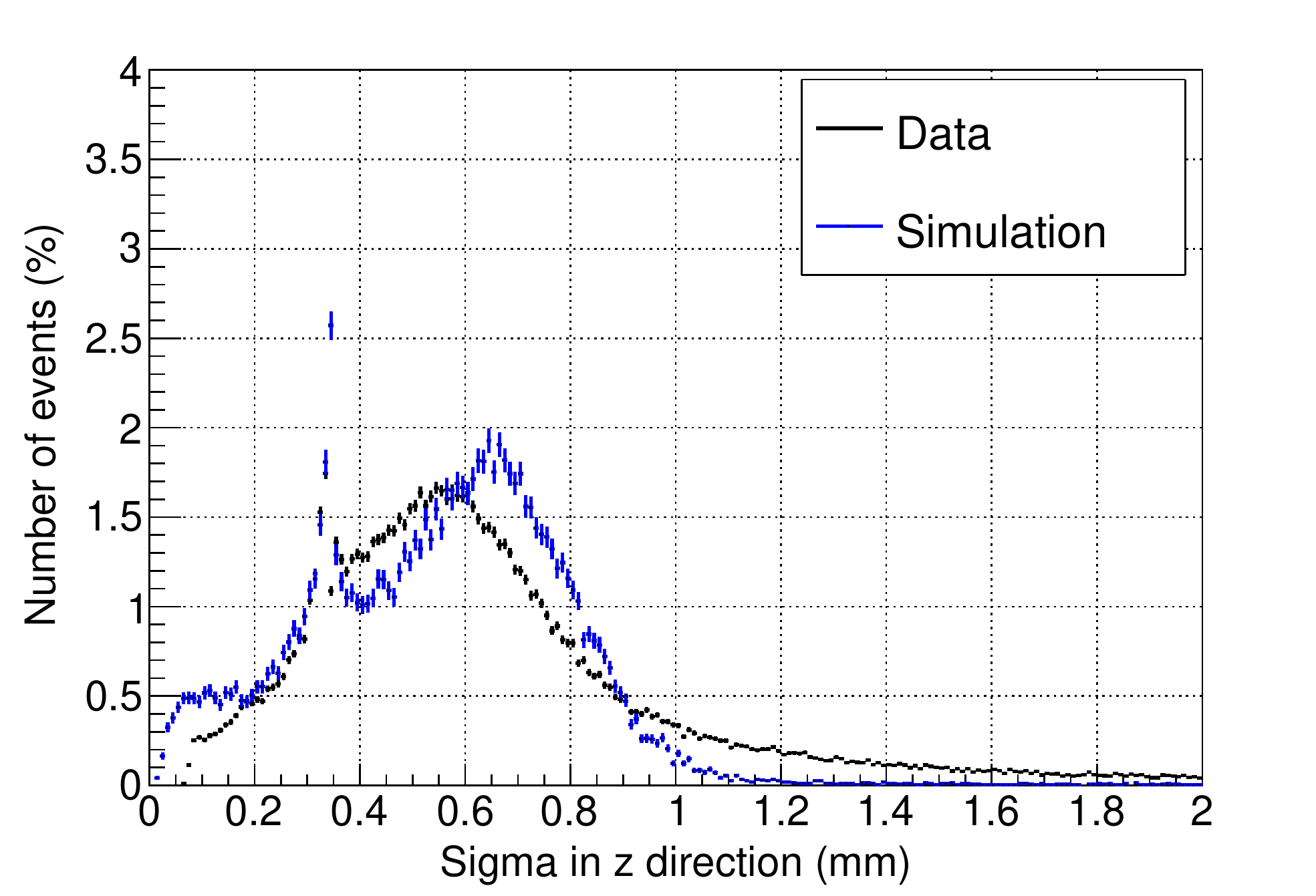}
\includegraphics[width=55mm]{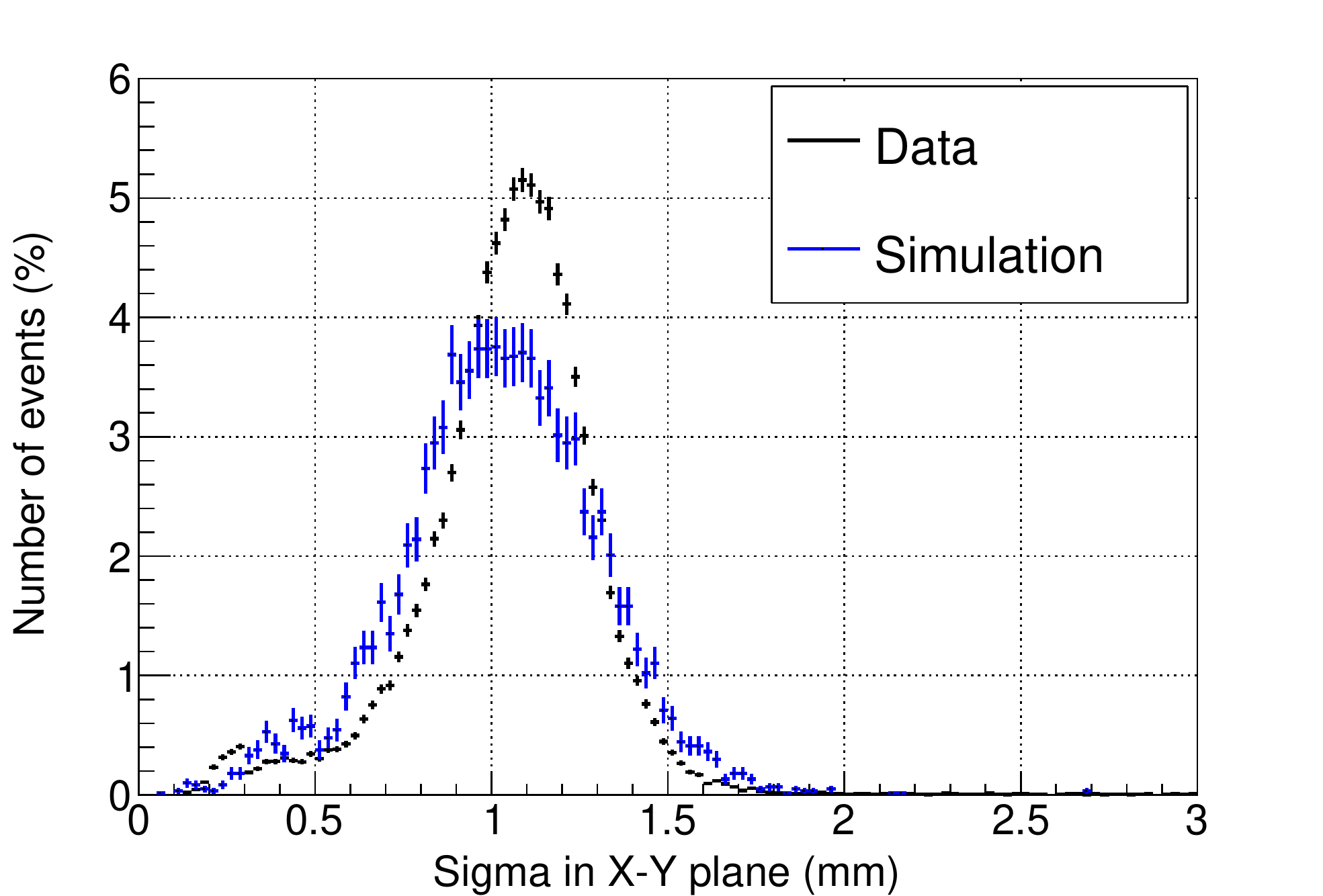}
\includegraphics[width=55mm]{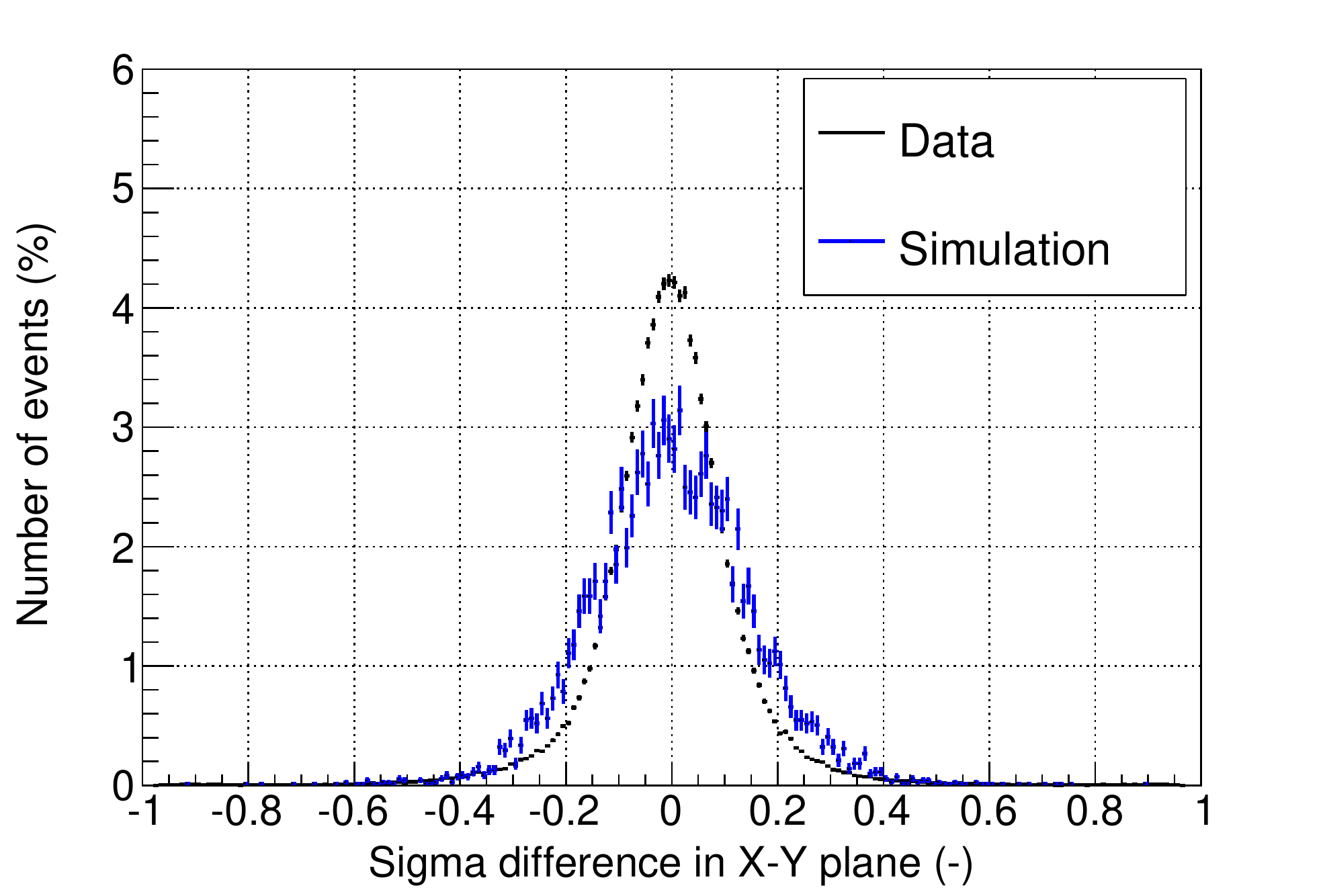}
\includegraphics[width=55mm]{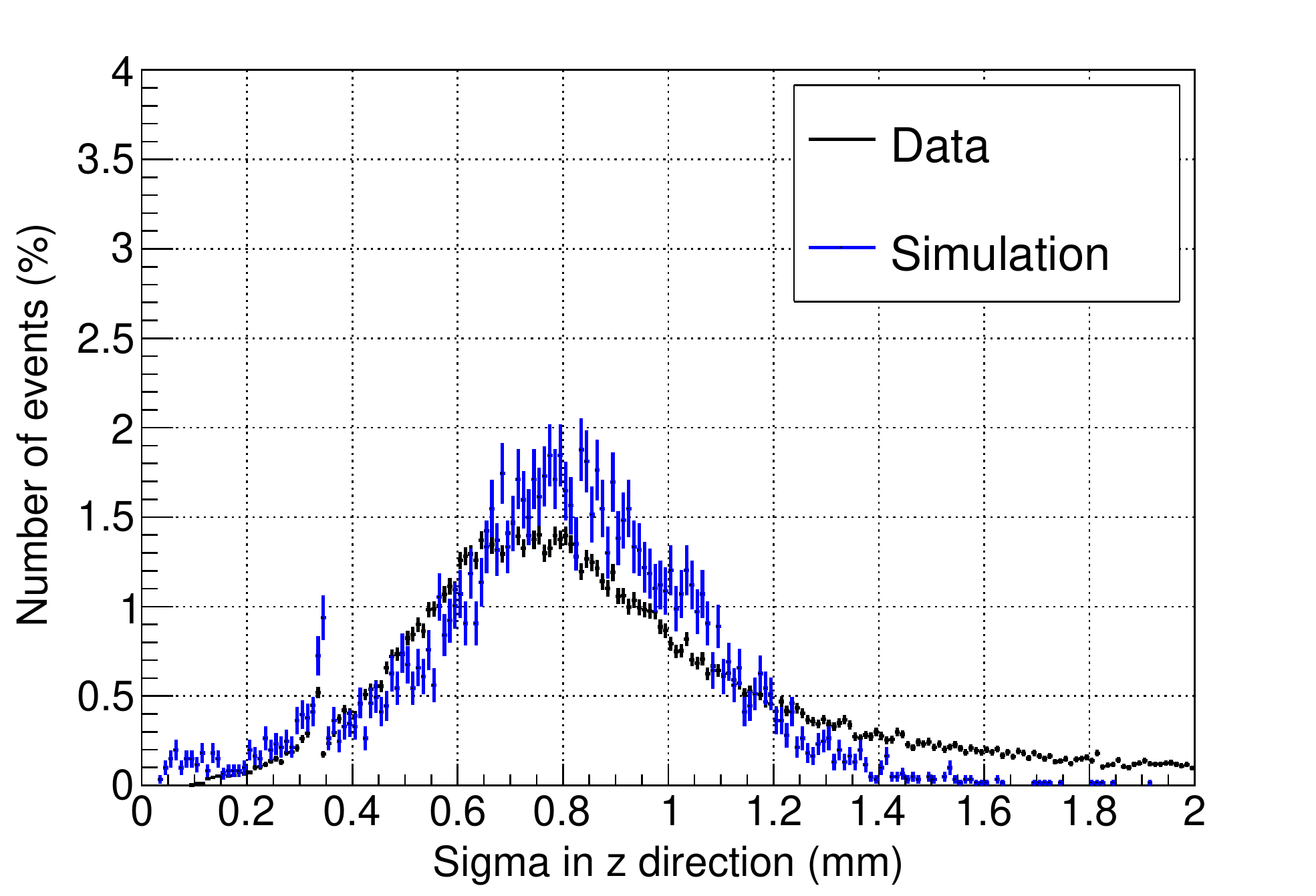}
\caption{Comparison between real data (black line) and Monte Carlo simulation (blue line)
for three analysis observables: the $XY$ width, $\sigma_{XY}$ (left);
the balance of cluster widths, $\Delta\sigma_{XY}$ (center); and the $Z$ width, $\sigma_Z$ (right);
and three energy ranges: 16-28 keV (top); 5-10 keV (center); and 2-4 keV (bottom).
Data was acquired when a $^{109}$Cd source was situated at a calibration point of TREX-DM
and the detector was filled with Ar+5\%iC$_4$H$_{10}$ at 1.2 bar.
The statistical error of each bin has been graphically represented by an error bar.}
\label{fig:ValidAr5iso}
\end{figure*}

The comparison of the observables (the cluster widths in $XY$-plane
and $Z$-direction, $\sigma_{XY}$ and $\sigma_Z$,
and the width balance, $\Delta\sigma_{XY}$) between real and simulated events
is made in Figs.~\ref{fig:ValidAr2iso} and Figs.~\ref{fig:ValidAr5iso}
for three energy ranges: 16-28 keV, 5-10 keV and 2-4 keV.
These energy ranges correspond to the $K$-lines of the source and the main fluorescence lines in the chamber.
There are some differences between distributions which can be attributed to the geometry in Geant4
and some simplification in the simulation chain.
These differences should be reduced in future upgrades of the simulation code.
Nevertheless, we have reproduced by simulations the dependence of observables with energy,
which can be explained by diffusion effects and a threshold effect in the strip electronics.
In general, the width by diffusion responds to the spatial distribution of the events in the conversion volume,
as the range of primary electrons is too small compared to diffusion effects.
Copper fluorescence (at 8.0 keV) is roughly expected everywhere in the detector,
but with more intensity close to the cathode.
These photons are absorbed near the cathode due to their short mean free path,
the electrons suffer the diffusion effects along all the drift and
as a consequence, their cluster should be wide in $XY$.
In contrast, iron fluorescence (at 6.4 keV) is only emitted from the readouts,
and their clusters are therefore narrow.
These two contributions are clearly present in the $\sigma_{XY}$ distribution in 5-10 keV range.
The decrease of the $XY$-width at low energies is due to the threshold effect
in the strip electronics, that effectively cuts the low energy tails of the electron clouds.
The width in $Z$-direction ($\sigma_{Z}$) shows the inverse dependence with energies,
i.e., clusters are wider at low energies as it is correlated to the number of primary electrons.
Finally, the balance of cluster widths ($\Delta\sigma_{XY}$) shows
a wider distribution for low energy events,
as the relative charge differences between each direction increase.

\subsection{Simulated contributions in this first background model}
\label{sec:BackContibutions}
In this first model, we have simulated the radioactive isotopes of the main internal components,
and we have scaled the results by the measured activities described in Sec.~\ref{sec:radiopurity}.
If only an upper limit was set, this value was used in the scaling.
In some cases, we have considered a radiopure alternative,
like in the case of the Micromegas readout planes.
For this component, we have imposed a secular equilibrium of both $^{232}$Th and $^{238}$U chains
to estimate the activities of the different isotopes from those of $^{208}$Tl and $^{214}$Bi,
while we have kept the values reported for $^{40}$K and $^{60}$Co in \cite{Cebrian:2010ta}.
In the case of teflon, we have used the activities reported by EXO-200,
as the values in Table~\ref{tab:resrad} (\#8 and \#9) are just upper limits.
Finally, for the specific case of argon-based mixtures,
we have considered the isotope $^{39}$Ar,
which decays by beta-emission (Q = 565~keV) and has a long half-life (239~yr).
It is produced at surface level by cosmogenic activation
and the best way to avoid it is extracting argon for underground sources.
The lowest activities have been obtained by DarkSide collaboration
using this technique~\cite{Xu:2015hja}.
The components and the activities included in this first background model of TREX-DM
are detailed in Table~\ref{tab:BackLevels}.

\begin{table*}
\centering
\caption{Activities and estimated background levels (in keV$^{-1}$ kg$^{-1}$ day$^{-1}$) in the RoI (2-7~keV)
of the different components of the TREX-DM experiment for an argon- and neon-isobutane mixture at 10 bar,
using the analysis described in the text.
The numbers with \# at reference refer to Table~\ref{tab:resrad}.
Upper limits of activities are given at 95\% C.L.
In the specific case of connectors, the $^{238}$U limit has been used for the upper part of the chain,
while the $^{226}$Rn value has been used for the lower part.
The statistical error of these values is 5\%,
while the systematic error includes a 30\% uncertainty associated to the measurement of the component's activity,
a 60\% due to the simulation of the detector response and a 25\% for the fiducial efficiency of the analysis.
For each component, the isotopes that gives the main contribution to background level have been specified for discussion purposes.}
\label{tab:BackLevels}
\begin{tabular}{cc|cccccc|cc|cc}
\hline
                &           & \multicolumn{6}{c}{Radioactive isotopes}       & \multicolumn{2}{|c|}{Background level} & Main\\
Component       & Ref. & Unit                 & $^{232}$Th & $^{238}$U & $^{40}$K & $^{60}$Co & Others             & Argon      & Neon & contr.\\
\hline
Muon flux       & \cite{Luzon:2006sh} & s$^{-1}$ m$^{-2}$    &            &           &          &           & $5 \times 10^{-3}$ & 0.019 & 0.029 & -\\
\hline
Vessel          & \#4& $\mu$Bq/kg & $< 4$      &$< 12$     & $< 61$   &           &                              & $<$ 0.079  & $<$ 0.093 & $^{238}$U\\
Connectors      & \#17& mBq /pc              & 1.2        &  $< 25$   & 7.3      & $< 0.1$   & $^{226}$Rn: 4.5    & 0.61       & 0.90 & $^{232}$Th,$^{238}$U\\
Field cage      & \cite{Leonard:2007uv}& $\mu$Bq/kg           & $< 1.2$    & $< 9.7$   & $<10.0$  &           &                    & $<$ 0.00096& $<$ 0.0012& $^{238}$U\\
Cathode         & \#4& $\mu$Bq/kg           & $< 4$      &$< 12$     & $< 61$   &           &                    & $<$ 0.0042 & $<$ 0.0046& $^{232}$Th,$^{238}$U\\
Readouts        &\cite{Cebrian:2010ta,Gomez:2015hg}& nBq/cm$^2$           & $< 120$    & $< 110$   & $6 \times 10^4$ & $< 3000$ &             & 3.35       & 3.34 & $^{40}$K, $^{60}$Co\\
\hline
Target          & \cite{Agnes:2015ftt}& mBq/kg               &            &           &          &           & $^{39}$Ar: 0.73    & 0.084      & -   & -\\
\hline
\multicolumn{8}{c|}{Total background level}                                                                 & 4.15       & 4.43\\
\hline
\end{tabular}
\end{table*}

This first background model does not include some inner components like the cabling,
the calibration tube and the pieces used to shield the connectors. Their activities
can be considered small in comparison to other inner components. Regarding external
components, we have made a rough estimation of the contribution of the AGET-based
electronics (based on \#27 of Table~\ref{tab:resrad}), the lead shielding (\#2 of Table~\ref{tab:resrad}),
the environmental gamma flux~\cite{Luzon:2006sh} and the LSC rock-induced neutrons~\cite{Carmona:2004qk}.
Their contribution to background level will be below $10^{-1}$ counts keV$^{-1}$ kg$^{-1}$ day$^{-1}$
if the external shielding is composed of a lead layer of 20~cm thickness and a polyethylene layer of 40~cm thickness.
These contributions and others like cosmogenics, muon-induced neutrons in the surrounding rock or
radon emanation should be simulated using a detailed geometry of the TREX-DM experiment at LSC.

\subsection{Analysis and results}
\label{sec:BackXraysAna}
An analysis has been developed to perform event-by-event signal identification and background rejection
using the topological information provided by the readout planes.
It is based on cluster features of a given x-ray source, as the expected WIMP-induced recoil
signals are point-like events, whose width is mainly determined by diffusion.
The x-ray analysis is composed of two parts.
In the first one, a veto area of 5~mm thickness at the borders of each readout plane is used to reject background events,
with a small reduction of the signal efficiency (91.8\%). This veto is specially useful in the case of cosmic
muons, as its acceptance effienciency\footnote{Defined as the ratio of events in the RoI after
and before the application of the selection criteria.} is 7-8\% for events in the RoI are kept,
as shown in Table~\ref{tab:RejecEff}.
It is also powerful for events coming from the readout surface,
as alphas or high energy electrons are easily rejected.
In the rest of the cases, the rejection power of this selection cut is modest.

\begin{table}
\centering
\caption{Mean acceptance efficiencies (\%) in the RoI (2-7~keV)
of the veto area and the cut defined by
cluster features of $^{109}$Cd x-ray lines,
for the different components of the TREX-DM experiment,
supposing an argon- and neon-isobutane mixture at 10 bar.
The efficiency of the cluster cut has been calculated over the events
that have survived the veto area cut.
The acceptance efficiency may vary for the different simulated
isotopes of each component.}
\label{tab:RejecEff}
\begin{tabular}{c|cc|cc}
\hline
                & \multicolumn{2}{|c|}{Argon} & \multicolumn{2}{c}{Neon}\\
Element       & Veto area & Cluster & Veto area & Cluster \\
\hline
Muons         & 7.8 & 68.9 & 6.9 & 48.5\\
\hline
Vessel        & 78.3 & 86.2 & 74.7 & 78.8\\
Connect.      & 81.1 & 72.7 & 71.8 & 78.0\\
Field cage    & 65.3 & 78.5 & 65.7 & 78.7\\
Cathode       & 67.6 & 81.7 & 64.5 & 73.2\\
Readouts      & 55.0 & 67.3 & 49.3 & 38.1\\
\hline
Target        & 91.8 & 72.4 & - & - \\
\hline
\end{tabular}
\end{table}

\begin{figure}[htb!]
\centering
\includegraphics[width=0.48\textwidth]{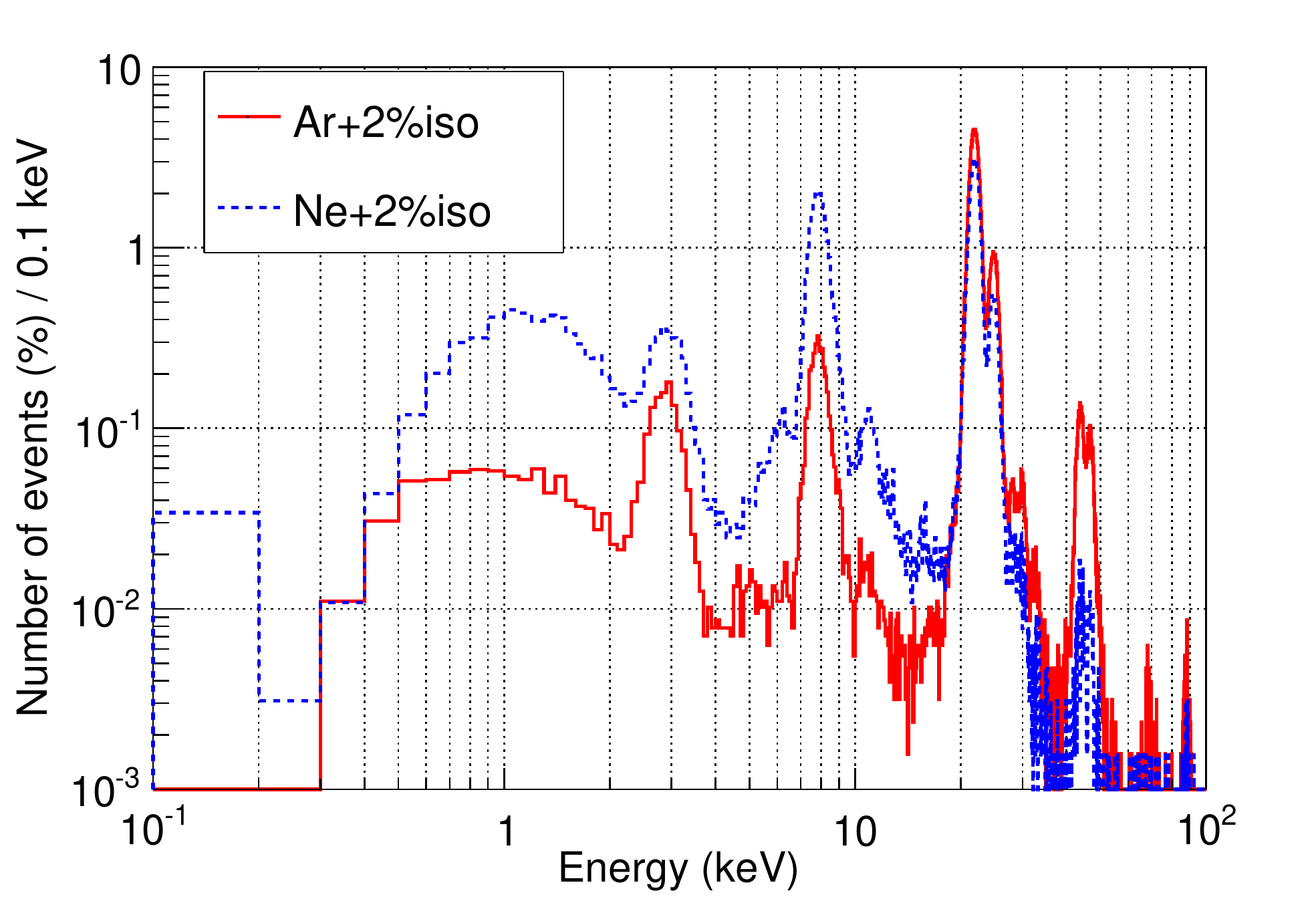}
\caption{Simulated energy spectra generated by a $^{109}$Cd source
situated inside the vessel but outside the calibration plastic tube,
when it is filled with Ar+2\%iC$_4$H$_{10}$ (red line) and Ne+2\%iC$_4$H$_{10}$ (dashed blue line) at 10 bar.
The two energy spectra have been normalized to the total number of events for the comparison.
The source generates two intense lines at 22.1~keV (K$_\alpha$) and 24.9~keV (K$_{\beta}$)
and two other ones at 2.98~keV (L$_{\alpha}$) and 3.15~keV (L$_{\beta}$),
which cannot be separated due the energy resolution of the detector.
The iron and copper K-fluorescences, induced by the source at the Micromegas readout plane and the central cathode,
are also present at 6.4 and 8.0~keV, respectively. In the case of the argon target,
there is an extra contribution at 3.0~keV line by the argon K-fluorescence (at 2.96~keV).}
\label{fig:SpecArNe}
\end{figure}

\begin{figure*}[htb!]
\centering
\includegraphics[width=0.48\textwidth]{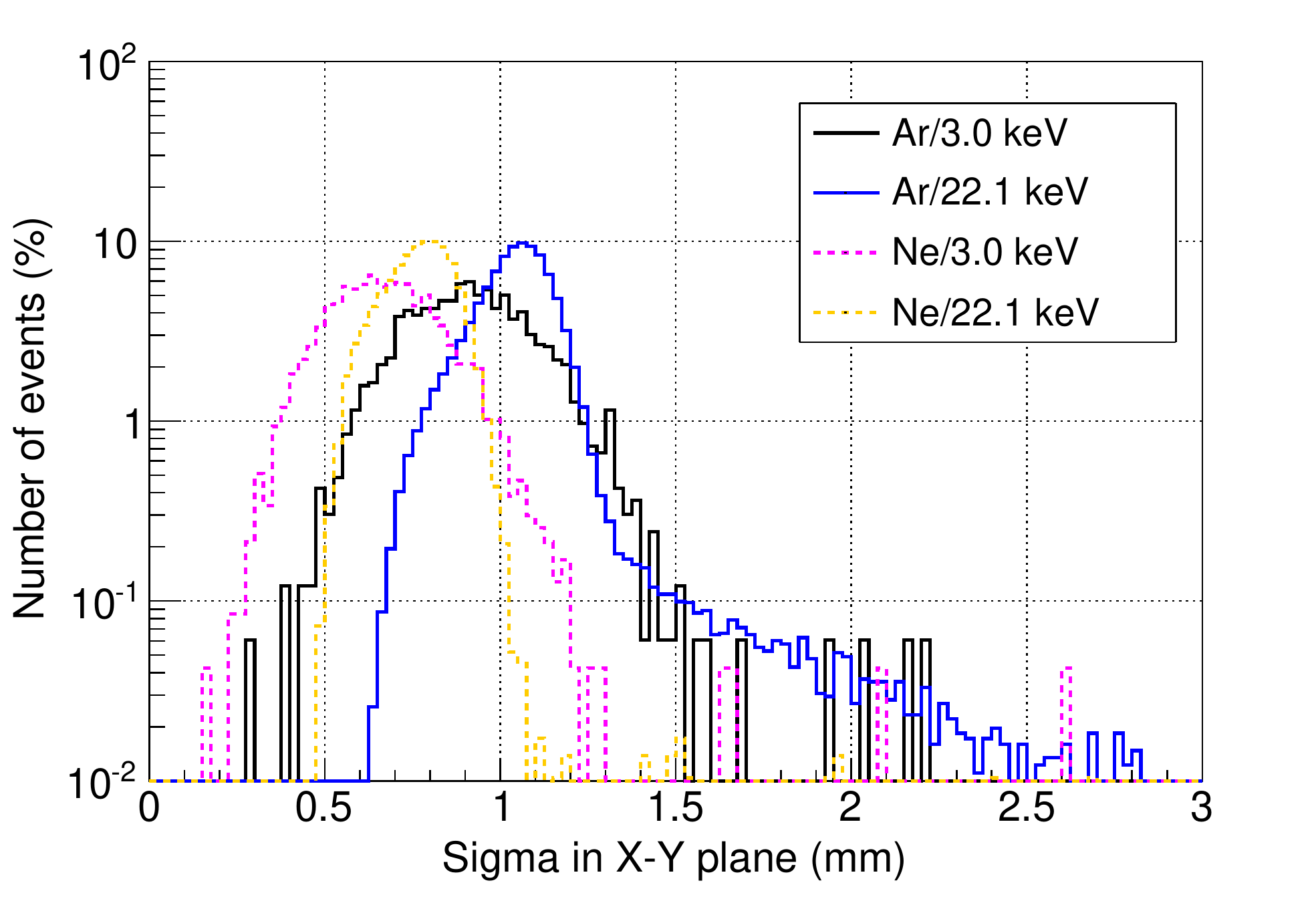}
\includegraphics[width=0.48\textwidth]{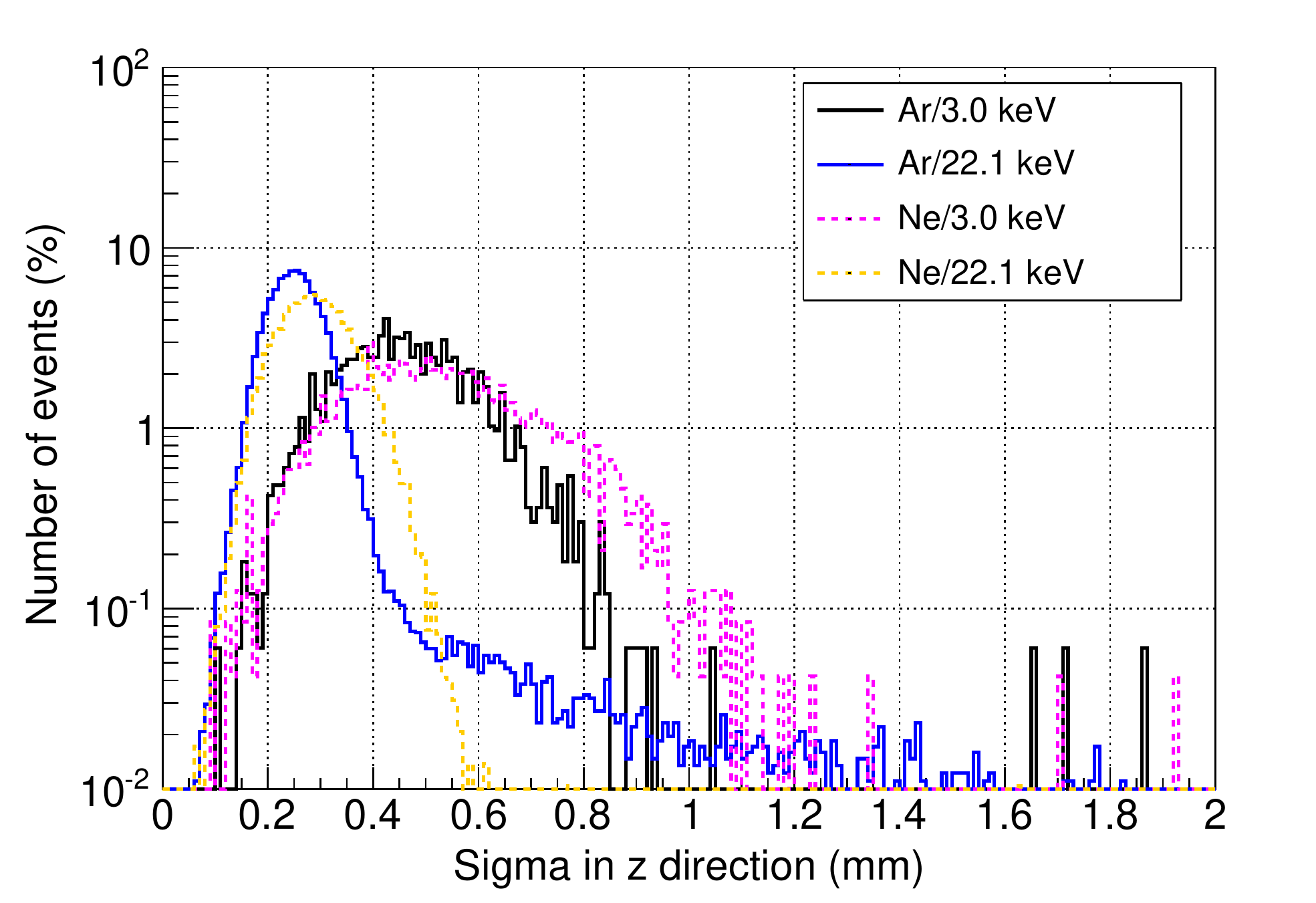}
\caption{Distribution histograms of the $XY$ width (left) and $Z$ width (right) for the x-ray events at 3.0 and 22.1~keV
generated by the simulation of a $^{109}$Cd source situated inside the vessel but outside the calibration plastic tube,
when it is filled with Ar+2\%iC$_4$H$_{10}$ (black and blue lines)
and Ne+2\%iC$_4$H$_{10}$ (dashed magenta and orange lines) at 10 bar.}
\label{fig:DistObs}
\end{figure*}

The second part of the analysis is based on the simulation of a $^{109}$Cd source situated
inside the vessel but outside the calibration plastic tube.
As shown in Fig.~\ref{fig:SpecArNe}, we have in this way access to an extra x-ray at 3.0~keV,
which is in the RoI.
These x-rays are generated by the L$_{\alpha}$ (at 2.98~keV) and L$_{\beta}$ (at 3.15~keV) lines of the source
and are blocked by the calibration tube in the actual setup.
The 3.0~keV and the 22.1~keV x-ray lines are used to generate the distribution histograms ($P^j_i$)
of the three observables defined in Sec.~\ref{sec:ValidSimulation}:
the widths in $XY$ ($\sigma_{XY}$) and $Z$ directions ($\sigma_Z$), and the the width balance ($\Delta\sigma_{XY}$);
which are shown in Fig.~\ref{fig:DistObs}.
Each distribution define the probability that an observable takes a specific value for simulated signal events.
The two x-ray lines may be absorbed at any position of the active volume
but there is a dependence with the z-position, i.e., x-rays are mainly absorbed near the cathode plane.
This dependence creates a fiducial efficiency, as wider clusters are expected for events absorbed near the cathode.
By comparing the analysis selection efficiencies for readout planes and the other parts (in Table~\ref{tab:RejecEff}),
we deduce that this effect is less than 25\%.
In the x-ray analysis, we have discarded the use of the iron (at 6.4~keV) and copper K-fluorescences (at 8.0~keV),
which are in between the other lines,  as the z-position dependence is more important:
most of the copper fluorescence come from the central cathode
and its events show larger widths; while iron fluorescence is induced at the Micromegas readout
and its clusters are narrower.

The distribution histograms ($P^j_i$), shown in Fig.~\ref{fig:DistObs},
are used to define two likelihood ratios $\mathscr{F}^j$ of the form
\begin{equation}
\label{eq:likelihood}
\mathscr{F}^j = -\log{\mathscr{L}^j} = - \sum^3_{i = 1} \log\left(\frac{P^j_i}{1-P^j_i}\right)
\end{equation}

\noindent
for the two x-ray lines: the first ratio is defined by the 3~keV line and is applied for energies up to 10~keV,
while the second one is defined by the 22.1~keV line and is applied from 10 to 100~keV.
For each function $\mathscr{F}^j$, an upper acceptance limit $q^j(90\%)$ is calculated
by setting an analysis efficiency of 90\%, equivalent to the veto cut.
This means that for each x-ray line, 90\% of its events
show cluster features whose corresponding ratio is below the acceptance limits. The specific
values used in this analysis are detailed for each gas mixture in Table~\ref{tab:AnaXrays}.
As shown in Table~\ref{tab:RejecEff}, the acceptance efficiency of this cut is modest:
values 70-80\% are obtained, just slightly better than the analysis efficiency 90\%.
This analysis should be optimized in future by including the dependence of cluster widths
with energy and z-position.

\begin{table}
\centering
\caption{Summary of the cluster-based analysis parameters: x-ray lines, selection and application energy ranges
and upper acceptance limits in argon- and neon-isobutane mixtures.}
\begin{tabular*}{\columnwidth}{@{\extracolsep{\fill}}ccccc@{}}
\hline
Line  & \multicolumn{2}{c}{Energy range (keV)} & \multicolumn{2}{c}{Upper limit} \\
(keV)       & Selection & Aplication    & Argon & Neon \\
\hline
 3.0        & 2.0-4.0   & 0.0-10.0      & 13.25 & 13.75\\
22.1        & 21.0-23.0 & 10.0-100.0    & 11.75 & 11.05\\
\hline
\end{tabular*}
\label{tab:AnaXrays}
\end{table}

\begin{figure*}[htb!]
\centering
\includegraphics[width=0.48\textwidth]{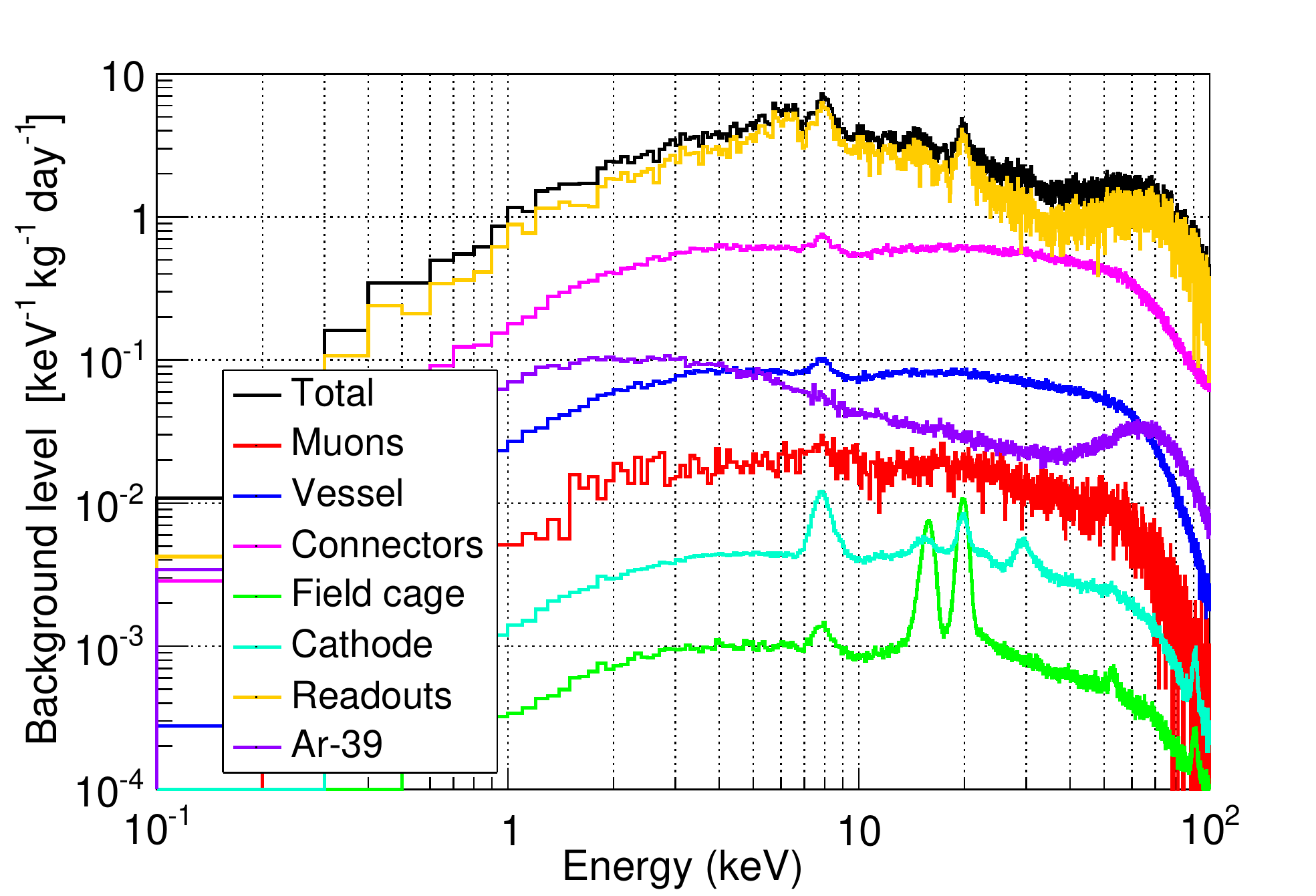}
\includegraphics[width=0.48\textwidth]{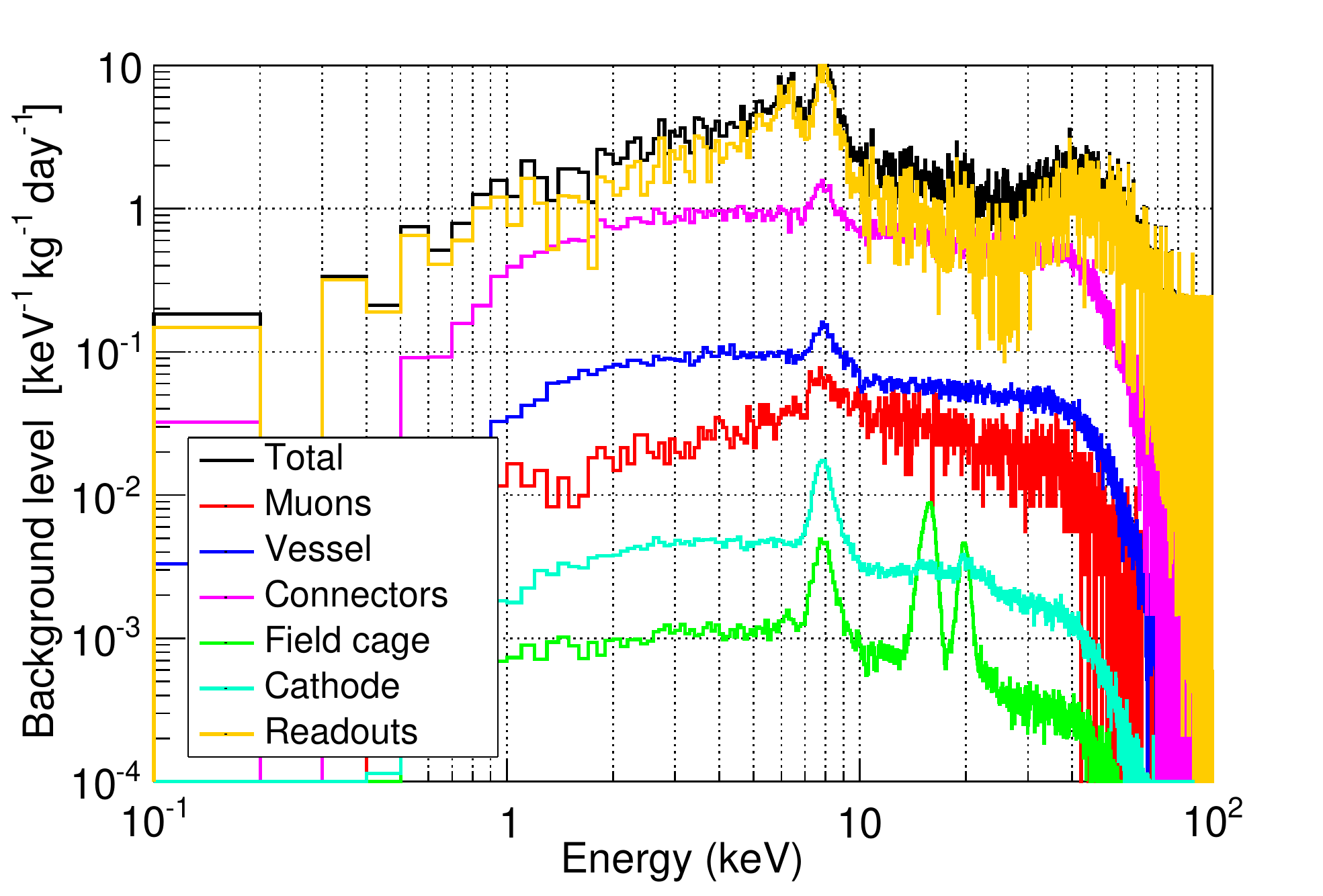}
\caption{Simulated background spectrum expected in TREX-DM experiment (black line) during a physics run at LSC
if operated in Ar+2\%iC$_4$H$_{10}$ (left) or Ne+2\%iC$_4$H$_{10}$ at 10 bar.
The contribution of the different simulated components is also plotted: external muon flux (red line),
vessel contamination (blue line), connectors (magenta line), field cage (green line), central cathode (cyan line),
Micromegas readout planes (orange line) and $^{39}$Ar isotope for the argon case (violet line).}
\label{fig:BackSpecArNe}
\end{figure*}

Once the likelihood ratios and the acceptance limits have been defined,
the observables of all events in the simulation-sets are calculated.
For each component and isotope, an energy spectrum with the events that survive the selection criteria
is then generated and scaled by the isotope activity, the mass of the component and the total active mass.
Finally, the spectra are summed for each component, which results in the background spectra
for the argon- and neon-isobutane shown in Fig.~\ref{fig:BackSpecArNe}.

Each background spectrum has a flat and continuous component in a wide range of energies,
generated by gamma events that have suffered a Compton process.
This flat spectrum decays at high energy due to an efficiency loss.
At low energies, clusters show a shorter $XY$-width
and a larger $Z$-width in comparison to 3~keV x-rays clusters (see Fig.~\ref{fig:DistObs}), which causes a signal loss.
These differences are due to the energy dependences discussed in Sec.~\ref{sec:BackDetResponse}.
Apart from that, there are two intense peaks at 6.4 and 8.0~keV,
which respectively correspond to the iron and copper K-fluorescences.
These events are induced at the Micromegas readout plane and the central cathode by gammas.
Finally, the contributions of the cathode, the Micromegas readout and the field cage show other lines
between 10 and 20~keV, mainly generated by the x-ray lines of $^{228}$Ac (at energies of 13.0, 16.2 and 19.0~keV),
$^{212}$Bi (at 14.6~keV) and $^{214}$Pb (at 12.9~keV).

The estimated background level in the RoI (2-7~keV) and its different contributions are detailed in Table~\ref{tab:BackLevels}.
For the argon-(neon-) isobutane mixture, the total background level
is 4.15~(4.43)~counts~keV$^{-1}$~kg$^{-1}$~day$^{-1}$.
The statistical error of these values is 5\%,
while the systematic error includes a 30\% uncertainty associated to the measurement of the component's activity,
a 60\% due to the simulation of the detector response and a 25\% for the fiducial efficiency of the analysis.
For both gases, the main contribution (81\% and 75\% of background events, respectively)
is due to the readout planes, followed by the connectors (15\% and 20\%) and the vessel (2\%).
In the case of argon, the contribution by the $^{39}$Ar isotope is similar to the vessel one.

According to this estimation, we can conclude that a background level of $1-10$ counts~keV$^{-1}$~kg$^{-1}$~day$^{-1}$
is feasible if the final configuration of TREX-DM experiment follows these conclusions:
\begin{itemize}
  \item \textbf{Readout planes}: the proposed strategy of building them only of copper and kapton,
	using either the microbulk techology or a radiopure version of the actual bulk ones,
	will give a contribution to background level of 3.35 (3.34)~counts~keV$^{-1}$~kg$^{-1}$~day$^{-1}$.
	For bulk technology, it is already a big step as the actual bulk readouts
	are dirty in terms of radiopurity ($10^4$ worse).
	
	The limiting activity in background prospects is due to $^{40}$K and, in a factor 5 lower, to $^{60}$Co.
	For this reason, to further reduce this contribution, the activity of $^{40}$K should be measured
	with better sensitivity. It may improve, as it happens for $^{238}$U from \cite{Cebrian:2010ta} to \cite{Gomez:2015hg},
	as the first quantification was near the sensitivity limits of the germanium detector. If it is not the case,
	its origin should be found and the readout construction technique should be improved in radiopurity terms.
	
	\item \textbf{Connectors}: the actual strategy of shielding them with a 0.5 cm-thick layer of copper and a 0.5 cm-thick layer of lead
	gives an estimated contribution of 0.61 (0.90)~counts~keV$^{-1}$~kg$^{-1}$~day$^{-1}$.
  To reach lower values, they should be better shielded or put further away from the active volume.
	In the proposed design for LSC, they will put behind the copper basements and will be shielded by a 6 cm-thick layer of copper.
	
  \item \textbf{Gas}: the actual strategy of using argon extracted from underground sources
  gives an estimated contribution of 0.084~counts~keV$^{-1}$~kg$^{-1}$~day$^{-1}$.
	The use of atmospheric argon should be discarded as the contribution of $^{39}$Ar to background
	may increase a factor $\sim 10^3$~\cite{Agnes:2015ftt}.
\end{itemize}


\section{Sensitivity to low-mass WIMPs}
\label{sec:sensitivity}

\begin{figure}[htb!]
\centering
\includegraphics[width=0.48\textwidth]{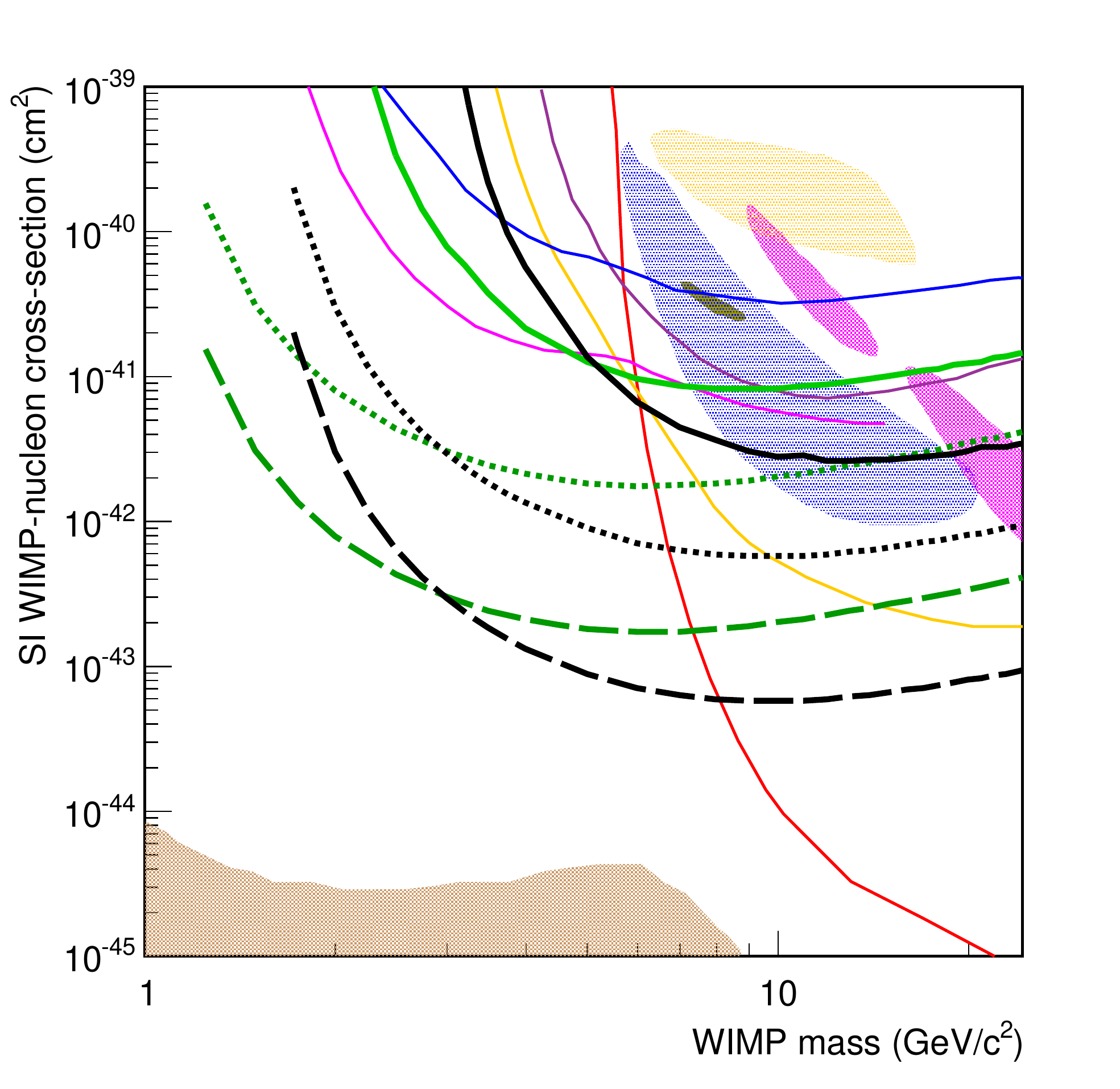}
\caption{90\% confidence level projected sensitivity of TREX-DM
assuming an exposure of 1~kg$\cdot$y in argon (\emph{black thick lines})
and neon (\emph{green thick lines}) with a conservative (\emph{solid})
and realistic (\emph{dotted}) assumptions on the background levels
of 10 and 1~keV$^{-1}$kg$^{-1}$day$^{-1}$, respectively,
and an energy threshold of 0.4 keVee for the first scenario and 0.1~keVee for the latest.
The dashed lines represent the sensitivity of an upgraded detector with 0.1~keVee threshold,
0.1~keV$^{-1}$kg$^{-1}$d$^{-1}$ and 10~kg$\cdot$y exposure.
Closed contours shown are CDMS II Si~\cite{Agnese:2013rvf} (\emph{blue}, 90\% C.L.),
CoGeNT~\cite{Aalseth:2012if} (\emph{dark gray}, 90\% C.L.),
CRESST-II~\cite{Angloher:2011uu} (\emph{magenta}, 95\% C.L.),
and DAMA/LIBRA~\cite{Bernabei:2013rb} (\emph{tan}, 90\% C.L.).
For comparison we also show 90\% C.L. exclusion limits from SuperCDMS~\cite{Agnese:2014aze} (orange),
CDMSlite~\cite{Agnese:2015nto} (\emph{magenta}), LUX~\cite{Akerib:2013tjd} (\emph{red}),
and CDEX1~\cite{Yue:2014qdu} (purple) and CRESST-II 2015~\cite{Angloher:2015ewa} (\emph{blue}).
The brown shaded region corresponds to the sensitivity limit
imposed by the solar neutrino coherent scattering background~\cite{Billard:2013qya}.}
\label{fig:exclusion}
\end{figure}

TREX-DM could be sensitive to a relevant fraction of the low-mass WIMP parameter space. 
Figure~\ref{fig:exclusion} shows 90\% confidence level projected sensitivity of TREX-DM
assuming a total exposure of 1~kg$\cdot$yr in argon (\emph{black thick lines})
and neon-based (\emph{green thick lines}) gas mixtures,
under two assumptions on a flat-shaped background level (10 and 1~keV$^{-1}$kg$^{-1}$day$^{-1}$, respectively)
and for an energy threshold of 0.4 keVee in the first scenario (solid lines)
and 0.1~keVee in the latest (dotted lines).
The dashed lines represent the sensitivity of a future detector for 0.1~keVee threshold,
0.1~keV$^{-1}$kg$^{-1}$day$^{-1}$ and 10~kg$\cdot$yr exposure.

The projected exclusion curves have been derived using a binned Poisson method \cite{Savage:2008er}
with background subtraction.
This simple method works relatively well in case of large background levels, like ours.
The Poissonian probability $p$ of observing $N$ or more events, where $N = s + b$,
being $s$ and $b$ the signal and background events, is $p = \sum_{\rm{k=s+b}}^{\infty} \frac{e^{-b} b^k}{k!}$,
from which we can derive an exclusion contour at $1 - \alpha$ confidence level
by looping on the scattering cross-section $\sigma_N$, for each WIMP mass,
until $p < \alpha$, being $\alpha$ set at 0.1.
As the quenching factor of neither gaseous argon nor neon has been measured yet,
we have considered the parametrization described in Eq.~\ref{eq:quench1} and \ref{eq:quenchParam}.
According to this model, our energy threshold prospects of 0.4 and 0.1~keVee expressed as nuclear recoil energy
would be 2 and 0.6~keVnr, respectively.

In the calculation we have used a standard WIMP halo model
with Maxwell-Boltzmann velocity distribution,
though this model is known to be an oversimplification~\cite{Mao:2012hf},
and standard values of the astrophysical parameters:
local dark matter density $\rho_{0} = 0.3$ GeV/c$^{2}$,
local velocity $v_{0} = $ 220 km/s,
laboratory velocity $v_{\mbox{lab}} = $ 232 km/s and $v_{\mbox{esc}} = $ 544 km/s.
We have also assumed that the WIMP couples identically to neutrons and protons,
though different coupling values are generically available~\cite{Feng:2011vu}. 

It is shown that under these hypotheses the experiment could reach higher sensitivity
to low-mass WIMPs ($m_{\chi}< 8$~GeV) than many of the current experiments,
and could exclude the ``region of interest'' invoked by some positive interpretations of some Dark Matter experiments.

\section{Conclusions and outlook}
\label{sec:con}
New detection techniques, focused on the use of light target nuclei together
with low energy thresholds, are needed to explore the low-mass range of the WIMP parameter space.
Recent advances in radiopure Micromegas readout planes for gaseous TPCs and in electronics
are improving the low-background prospects and scalability of Micromegas-based TPCs.
If we add to these features the tracking capabilities and the low intrinsic energy threshold,
they are a good detection option for the search of low-mass WIMPs.
In this context, we present TREX-DM, a prototype built to test this concept.
It is designed to host an active detection mass of $\sim$0.300~kg of Ar at 10 bar,
or alternatively $\sim$0.160~kg of Ne at 10 bar and fully built with radiopure materials.

The experiment consists of a copper vessel divided into two active volumes, each of them equipped with a field cage
and a bulk Micromegas readout plane. Signals are extracted from the vessel by flat cables
and are read by an AFTER-based electronics. Each side is calibrated at four different points by a $^{109}$Cd source.
The experiment has been successfully built and commissioned
and the first calibration data in Ar+2\%iC$_4$H$_{10}$ have been described in detail.
The role of the quencher quantity will be further studied in the near future.
A better performance has been observed with a 5\% isobutane at atmospheric pressure~\cite{Iguaz:2011yc,Iguaz:2014koa}
but so much quencher may degrade the detector performance at high pressure.
Neon-based mixtures will be also studied,
which are expected to show higher gains and a better energy resolution,
as theoretically shown in~\cite{Schindler201078} and practically shown in~\cite{Iguaz:2012ur,Iguaz:2014koa}.

Several changes are planned for a physics run at the Canfranc Underground Laboratory (LSC),
mainly at the external support and shielding, the gas, the calibration system, the readout plane and the electronics.
The actual aluminum support structure should be replaced by a copper-based one,
cleaner in terms of radiopurity.
The structure should also be compatible with a lead shielding
to reduce the effect of the external gamma flux and
a polyethylene shielding to remove neutrons.
Other systems will also be affected like the gas and vacuum systems,
which should be made of copper near the vessel.

The new gas system is being designed to work in either open or close loop;
and to recover the gas using cryogenic nitrogen. By this way, precious gases could be used in future.
The gas should no contain significant amounts of radioactive isotopes.
Very light gases like neon do not have any but for instance,
natural argon contains an unacceptable amount of $^{39}$Ar
which could increase the background level of the experiment.
The DarkSide collaboration has proven that argon from underground sources has negligible levels,
and its use in large scales is feasible.
Either neon or underground argon will be used in the final setup.

The calibration system will be automatize to minimize the number of openings of the shielding
and an extra x-ray line at lower energies will be included.
Several options are being studied: the fluorescence of neon at 0.85 keV,
the use of a movable $^{55}$Fe source (5.9~keV x-rays) installed at one of the two free ports of the vessel
or the dilution of $^{37}$Ar (0.25 and 2.6~keV x-rays) in the gas.

About the readout plane, two materials must be replaced by clean ones,
in terms of radiopurity: FR4 PCB, present at the readout plane,
and Liquid Crystal Polymer, present at the connectors. Both changes are technically feasible in the near future:
a microbulk Micromegas readout built only out of kapton and copper, and connectors made of silicone.
The microbulk plane will be glued on a radiopure copper support, to give mechanical strength to the readout,
while the routing of the signal channels will be extracted via a flexible card that is the continuation
of the same kapton-copper foil. This cable brings the signals far enough from the readout,
so as connectors could be additionally shielded far from the active volume.

Finally, a new electronics, based on the AGET chip, will be implemented.
Its trigger will be generated individually by each single strip signal,
which will reduce the energy threshold down to 0.1~keVee.
In the best noise conditions of the actual setup,
an energy threshold of 0.60~keVee was measured for a readout gain of $10^3$.
The final setup should keep at least the same noise level and reach the same gain.
There are good prospects for microbulk technology to reach operational gains much higher than $10^3$
in either argon or neon at 10~bar, as shown in \cite{pacotesis} for argon-isobutane mixtures
and quencher percentages of 0.5-2\%.

During the design and construction of TREX-DM,
a material screening program (mainly based on germanium gamma-ray spectrometry)
was undertaken to evaluate the radioactivity of all the relevant components of the detector and surrounding materials.
These results have been used to build a first background model of the experiment,
in combination with the full simulation of the detector's response and an analysis optimized to
discriminate point-like events from complex topologies. Based on this first model,
the background level of this detection concept has been estimated in 1-10~counts~keV$^{-1}$ kg$^{-1}$ day$^{-1}$ for energies in 2-7~keVee.
Supposing a flat-shape background for lower energies
and an energy threshold of 0.4~keVee or below, TREX-DM could reach higher sensitivity
to low-mass WIMPs than many of the current mainstream experiments,
and could exclude the \emph{region of interest} invoked by some positive interpretations of some Dark Matter experiments.


\begin{acknowledgements}
We acknowledge the Micromegas workshop of IRFU/SEDI for bulking our readout planes and the
\emph{Servicio General de Apoyo a la Investigaci\'on-SAI} of the University of Zaragoza for the fabrication
of many mechanical components.
We also thank D.~Calvet from IRFU/SEDI for his help with the AFTER electronics.
We acknowledge the support from the European Commission under the European Research Council
T-REX Starting Grant ref. ERC-2009-StG-240054 of the IDEAS program of the 7th EU Framework Program,
the Spanish Ministry of Economy and Competitiveness (MINECO) under grants FPA2011-24058 and FPA2013-41085-P
and the University of Zaragoza under grant JIUZ-2014-CIE-02.
F.I. acknowledges the support from the \emph{Juan de la Cierva} program
and T.D. from the \emph{Ram\'on y Cajal} program of MINECO.
\end{acknowledgements}

\bibliographystyle{spphys}
\bibliography{TREXDM_Rev_01}
\end{document}